# Comparing electricity generation technologies based on multiple criteria scores from an expert group

Euan Mearns*[1] and Didier Sornette[1,2]
ETH Zurich
[1]Department of Management, Technology and Economics
Scheuchzerstrasse 7 , CH-8092 Zurich,  Switzerland
[2]Institute of Risk Analysis, Prediction and Management (Risks-X), Academy for Advanced Interdisciplinary Studies, Southern University of Science and Technology (SUSTech), Shenzhen, 518055, China
emearns@ethz.ch      dsornette@ethz.ch

* Corresponding author.

Highlights:
- 13 common electricity generation technologies were evaluated using multi criteria decision analysis (MCDA) where values were assigned by the expert judgment of a professional group.
- A hierarchical scheme of 12 criteria organized into 5 categories of health, environment, grid, economics and resources was used to provide a holistic measure of energy quality.
- The three leading technologies to emerge are nuclear power, combined cycle gas and hydroelectric power. The three trailing technologies are solar PV, biomass and tidal lagoon.
- Our findings are consistent with the baseline cost approach of the 2004-2009 EU funded NEEDS project but contrast sharply with the MCDA survey of the same project that found CaTe solar PV and solar thermal power to be the most promising technologies for central Europe.

**Abstract**: Multi criteria decision analysis (MCDA) has been used to provide a holistic evaluation of the quality of 13 electricity generation technologies in use today. A group of 19 energy experts cast scores on a scale of 1 to 10 using 12 quality criteria – based around the pillars of sustainability – society, environment and economy - with the aim of quantifying each criterion for each technology. The total mean score is employed as a holistic measure of system quality. The top three technologies to emerge in rank order are nuclear, combined cycle gas and hydroelectric. The bottom three are solar PV, biomass and tidal lagoon. All seven new renewable technologies fared badly, perceived to be expensive, unreliable, and not as environmentally friendly as is often assumed. We validate our approach by 1) comparing scores for pairs of criteria where we expect a correlation to exist; 2) comparing our qualitative scores with quantitative data; and; 3) comparing our qualitative scores with NEEDS project baseline costs. In many cases, $R^2$>0.8 suggests that the structured hierarchy of our approach has led to scores that may be used in a semi-quantitative way. Adopting the results of this survey would lead to a very different set of energy policy priorities in the OECD and throughout the world. (210 words)

Abstract

# 1. Introduction

The global electricity generation system is in process of being transformed, based mainly on the summary opinion of the International Panel on Climate Change (IPCC) who have now stated that, if global mean temperatures rise more than 1.5°C above pre-industrial levels, dangerous climate change may occur (IPCC special report 8th October 2018 [1]).

The IPCC tends to attribute most observed climate change (i.e. temperature rise) to greenhouse gas emissions in particular $CO_2$ produced mainly from combusting fossil fuels [2]. Given near total failure to reduce emissions in the 30 years from 1990 to 2020 (Figure 1), it is perhaps naïve to assume that the next 30 years will be any more successful, as drastic changes with momentous ramifications would be required. The previous assessment was to limit warming to 2°C [2] and lowering the threshold has raised the bar making, what so far has been unattainable, even more so.

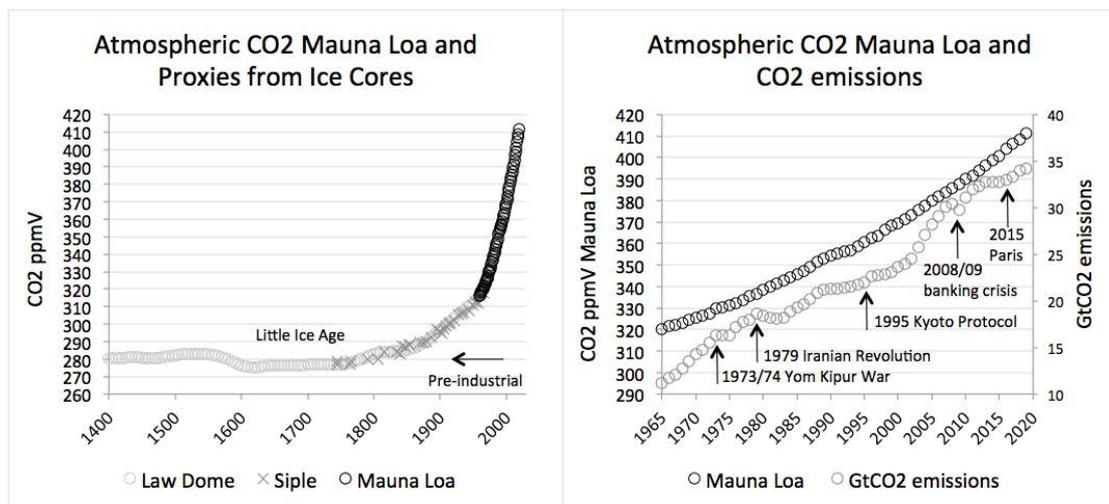

Figure. 1 Left – reconstruction of atmospheric $CO_2$ based on ice cores from Law Dome [3] and Siple [4] (Antarctica) spliced with annual averages of actual atmospheric measurements made at Mauna Loa (Hawaii) [5]. Right – annual average $CO_2$ from Mauna Loa (Hawaii) [5] compared with human emissions as reported by BP [6]. The only detectable declines in emissions occur in 1973 and 1979 linked to spikes in oil price that led to a decline in oil consumption and in 2008/09 linked to decline in energy consumption resulting from the global financial crisis. Climate treaties in 1995 and 2015 did not result in discernible declines in $CO_2$ emissions although it can still be argued that emissions might have risen more than recorded without those treaties. Irregularities in the emissions trend are not visible in the atmospheric $CO_2$ trend.

Fossil fuels provide easy to access high quality energy and industrial society is founded on their consumption [7]. Some argue that it is nigh impossible for a prosperous civilized society to function without fossil fuels, especially oil [8] while others claim to show using computer models that a 100% renewables electricity supply can easily be achieved [9, 10]. A compromise position is to use carbon capture and storage to remove $CO_2$ from the flue gas of power stations [11], thus enabling the continued use of fossil fuels. Others advocate nuclear power as the most viable low carbon source of electricity [12].

Recognising the failure of past mandatory agreements to deliver $CO_2$ reductions, the latest treaty agreed in Paris opted for voluntary reductions where individual nations set their own targets [13]. However, summing the volunteered reductions does not come close to the emissions reductions required to reach the climate goals [14] and the Paris accord seems on a route to policy failure. Notwithstanding the reality of almost assured failure, many OECD countries have committed to draconian cuts in emissions with targets for net zero emissions by 2050. For example, the UK has passed a law mandating net zero emissions by 2050 [15], predicated on the observation that emissions have fallen by 42% since 1990. The ease with which this remarkable achievement has come about has encouraged ministers to believe that this can be easily replicated going forward. This overlooks the fact that the 42% decline has been achieved through sleight of hand by conveniently ignoring the emissions embedded in imports that have grown in lock step with industrial decline in the UK [16]. Including embedded emissions and the decline since 1990 is almost halved to 23% (Appendix A).

One thing to emerge is that most OECD countries are committed to significant $CO_2$ reductions even if they will make no significant difference to the global $CO_2$ budget. $CO_2$ reduction has grown to totally dominate the energy policies of many countries. This one-dimensional approach has led to a number of bizarre outcomes such as transporting wood pellets from N America to Europe for industrial – scale power generation [17], converting food to bio-fuel for the transport sector [18], covering fields with solar panels and hillsides with wind turbines. Many environmentalists will welcome the abandonment of these practices. However, not all outcomes of the current strategy are necessarily bad. For example, coal fired power stations are latently dirty, and it is probably a good thing that many have closed. And there is perhaps an altruistic niche for local grids powered by solar photovoltaics and batteries in some parts of the world, for example Africa, where a little electricity can provide giant steps towards rural development.

In the past, electricity generating systems were developed by engineers and economists, following a system's approach, taking due cognisance of prevailing laws and regulations. It was against this backdrop that we recognized a need to develop a holistic approach to classify the quality of a range of electricity generating technologies, new and old. If the one-dimensional $CO_2$ reduction policy were to be abandoned, what should take its place? We strongly support the idea that future power generation should be sustainable and a holistic classification should be built around the three pillars of sustainability, namely environment, economy and society. However, these three categories alone are insufficient to characterize an effective power supply as discussed below.

This paper describes the results of a poll of 19 energy experts conducted on-line in 2018. 13 electricity generation technologies, old and new, were evaluated against 12 criteria using expert judgment within a framework of multi criteria decision analysis (MCDA). The twelve criteria chosen may give rise to some debate. We have little doubt that basing an evaluation on 12 criteria is better than using only 1. Where possible, we validate the judgments of our expert group by comparing scores for criteria with real-world data, for example cost of electricity with reported levelised cost of electricity (LCOE) for the various technologies.

Section 2 describes the design of the project, first delineating our objectives, and presenting the choice of electricity technologies in competition. We then introduce the MCDA hierarchy and the way we acquired the data and the selection of participants. The obtained data is then presented. Section 3 documents that our sample of experts is sufficiently large as shown from a convergence test. We also present different weighting methods and find that the un-weighted scores for the competing electricity technologies are robust. Section 4 describes several procedures we followed to validate our results. Section 5 concludes with policy implications. Appendix A presents a synthesis of UK energy policies as a case study to put in perspective with respect to our results on ranking energy technologies. Appendix B discusses the weaknesses of the NEEDS MCDA approach in comparison with our method. Appendix C presents the detailed method used to validate the EMETS MCDA scores.

## 2. Project Design

## 2.1 Objectives

How do we compare diverse electricity generating technologies? For example, how do we compare solar photovoltaic (PV) with coal-fired electricity and wind with nuclear power? How do we evaluate which is best and which is worst? This is like comparing apples with potatoes with barley. The EU majority funded (€7.6 of 11.7 million) project called New Energy Externalities Development for Sustainability (NEEDS) [19, 20] set out to answer these difficult questions and employed a poll-based multi criteria decision analysis (MCDA) method to rank future electricity technologies for Europe in the year 2050. We were drawn to the strengths and adaptability of this approach but found weaknesses in the way it was applied in NEEDS that are summarised in Appendix B (supplementary materials). For example, many of the 40 NEEDS questions were impossibly difficult to answer, especially for non-experts. The door was left open for answers to be based on calculation or expert judgment and it is not clear which methodology prevailed. Many questions were technology specific and repetitive, imparting a strong bias to the results. And we question if it is actually possible for non-experts to judge complex electricity generation technologies 40 years into the future?

We strongly agreed with the fundamentals of the multi-criteria approach and with the simplicity of asking experts to rank technologies by applying expert judgment and assigning scores to the various criteria per technology. We therefore developed an MCDA approach founded on expert judgment designed to provide a holistic measure of technology quality. An overarching objective was to find the best, or a mix of the best, electricity supply options that may sustain the whole of humanity into the future on a sustainable basis.

## 2.2 Choice of electricity technologies

The 13 technologies selected for evaluation are listed in the right hand column of Figure 2. These technologies essentially chose themselves as they include all of the most commonly used electricity generating technologies in use today. Some experimental technologies are included such as wave and tidal stream that have been deployed [21, 22] but not at large scale. The only technology not currently deployed on our list is tidal lagoon. This was included since the UK Swansea Bay scheme was under government evaluation at the time our survey was conducted [23].

| EMETS category | EMETS criteria |    | EMETS technologies |
|---|---|---|---|
| Health | Fatalities | 1 | Coal |
|  | Chronic illness | 2 | Gas CCGT |
| Environment | Environmental costs | 3 | Biomass |
|  | $CO_2$ intensity | 4 | Nuclear |
|  | Footprint | 5 | Wind |
| Grid | Costs to grid | 6 | Hydro |
|  | Benefits to grid | 7 | Solar PV |
| Economics | Taxes raised | 8 | Diesel |
|  | Subsidies paid | 9 | Solar thermal |
|  | Cost of energy | 10 | Tidal lagoon |
| Resources | ERoEI | 11 | Geothermal |
|  | Resource availability | 12 | Wave |
|  |  | 13 | Tidal stream |

Figure 2 The Energy Matters Energy Technology Survey (EMETS) hierarchy showing 5 categories, 12 Criteria and, to the right, the 13 electricity technologies that were evaluated. The rays depict how each technology was ranked against 12 criteria, here illustrated for the case of Coal.

2.3 The MCDA hierarchy and design

All knowledge is hierarchical, from the general to the specific. Similarly, any classification of energy sources needs to be organized according to a hierarchy going from main themes to more specific attributes within each theme. This led us to propose a MCDA hierarchy.

The starting point to design our hierarchy was the top-level structure of the NEEDS project that included Environment, Economy and Society – the three pillars of sustainable development [19, 24]. We agreed that Environment and Economy must be included and that the Health components of social values are also essential. However, we also recognized that these categories alone, and the criteria beneath them, were not sufficient to fully characterize the quality of an electricity technology. Certain fundamental elements were missing.

The evolution of Homo sapiens is defined by our mastery of fire that provided early humans with heat and light, protection from wild animals and the ability to cook meat. This mastery of thermal energy sets us apart from all other species. The early use of fire may have been haphazard, based upon dry lightning strikes starting bush fires that were then nurtured into campfires that had to be kept alight. Early humans (Homo Erectus), 1 million years ago, however, learned how to make fire by frictional heating of dry wood or through using a spark from flints and from that point humans had the luxury of lighting a fire, i.e. using energy, when they wanted or needed it [25], for example when it was cold, when predators were about or to cook food. Our ability to access energy when it was needed is written into our social DNA.

This fundamental social need of human populations is manifest in how modern power grids balance supply with demand exactly for every second of every day. The system is built around our ability to control generation. An electricity grid may be viewed as a commons [26], paid for by all who use electricity, and absolutely vital to modern life. It is analogous to our system of roads, water supply, sewers, and telecommunications networks. These are all commons where there is a communal responsibility to maintain and to not abuse or degrade the system.

The intermittent nature of new renewable energies (solar, wind and tidal), therefore, presents a natural barrier to their successful integration on to modern power grids since they do not satisfy one of our basic needs and that is to provide energy on demand. Renewables advocates dismiss this fundamental barrier, claiming that it is easily solved with a range of interventions such as battery storage, interconnectors, imports and back up from dispatchable sources such as CCGTs [see for example 9 and 10]. This may be true but these authors also maintain it is cheap and more reliable, contradicted by the rising cost of electricity, the ever present risk of blackouts and the on-going requirement for renewables subsidies.

We therefore include Grid as a category with criteria of grid costs and grid benefits (Figure 3). This provides a clear distinction between electricity systems that pro-actively benefit the maintenance of grid voltage and frequency and those that actively work to degrade the system by adding voltage and frequency noise. Having two grid criteria may seem like double accounting. We argue that it is necessary to account for the dissymmetry between (engineering) benefits and (dollar) costs.

One further vital ingredient missing from the traditional three pillars approach is the future availability of energy resources that is a hotly contested issue split into two main camps. One group (the peak oilers) have argued that resources such as crude oil are finite in nature and will one day run out, compromising the viability of future generations [for example 27, 28, 29]. The second group (the market economists) have argued that the availability of resources is linked to price and that scarcity will always lead to higher price (and ultimately to substitution) that will then make more resources available [30, 31]. During the incredible run in oil prices between 2002 and 2008, the peak oilers were claiming victory and warning of dire consequences for humanity lest oil should run scarce [8, 33]. And then sustained high prices led to the "shale revolution" bringing oil and gas independence to the USA that once was heavily reliant on imports. The market economists, therefore, now have the upper hand causing one branch of the peak oilers to change the tenor of their argument. If oil and gas resources are indeed found to be more plentiful than previously assumed, then, it is argued, they must be left in the ground to avoid catastrophic climate change [34, 35].

Setting these arguments to one side, there is no doubt that the size of resource and availability of reserves are of vital importance. What use is an ideal energy source if the resource is only large enough to power civilization for 6 months? We therefore add Resources as a category to our quality classification scheme. Under this category we have two criteria 1) Resource availability and 2) Energy Return on Energy Invested (ERoEI). The latter is a measure of how much energy we have to invest to produce a unit of energy and is a fundamental criterion to determine the quality of energy resources [36, 37}.

Finally, we have a problem fitting social issues into our hierarchy as a category in its own right. We find that social issues transcend many of the other categories. For example, air and water pollution are social issues; despoiling landscape with wind turbines; fracking for gas in populated areas; unreliable power supplies and high cost of electricity are all social issues. Therefore we do not include social issues as a category in its own right, but find that one hugely important social issue, namely health impacts of power generation is missing and we include this as a category in its own right. Under the Health category, we have two criteria, namely Fatalities and Chronic illness, which helps, among others, to distinguish the short-term impacts from the longer ones.

In defining criteria, we were sensitive to the need to capture all the information required to comprehensively describe the quality of individual technologies but, in the interest of simplicity, we wanted to use as few criteria as possible to achieve this. We also recognized the importance of having universal criteria, i.e. ones that could be applied to every technology. For example, fatalities, cost of energy and $CO_2$ intensity can all be calculated and quantified for each of the 13 technologies evaluated. This is in contrast to the NEEDS hierarchy where many criteria were technology specific and repetitive that inevitably imparted a bias to these results [19, 20 see also Appendix B].

We also recognized the importance of grain size in our categories and criteria. Our system was scored at the criterion level and we set out to try and ensure that each criterion had approximately equal weight. The success of this approach may be reflected in the fact that when we came to apply weights to scores, this made little to no difference to the outcomes (Sections 3.2 and 3.4).

It is recognized that certain groups may feel there are important omissions in our structure, for example "social acceptance". We considered including soft social measures such as this but resisted since social acceptance is not at the same hierarchical level as, for example, cost of energy or the health impacts. Indeed, social acceptance is a perception, and is thus more subjective and prone to the fashion of the time, while cost of energy or health impacts are evidence-based and, in our understanding, should be used to shape social acceptance. In other words, we do not put social acceptance at the same level, in this following a logic of the arrow of causality going from facts and evidence to perception. For example, some members of the public may object to nuclear power. But if confronted with regular blackouts and escalating electricity price in a decade's time, their opinion might change. Metrics such as social acceptance occur at a level beneath criteria in our hierarchy, as it derives from them.

We also set out to minimize duplication between criteria. Ideally, duplication should be eliminated, but this is not always possible. For example, low cost energy may support higher taxation while high cost energy may require subsidies, ultimately supported by the low cost energy sources that power economic growth. Cost may also be reflected in ERoEI and resource availability. High ERoEI energies tend to be lowest cost and scarce resources tend to have higher costs and vice versa, etc.

### 2.4 Data acquisition: The Energy Matters Energy Technology Survey (EMETS)

> *The whole problem with the world is that fools and fanatics are always so certain of themselves, and wiser people so full of doubts.*
> *Bertrand Russell [72]*

We faced the challenge of how to populate our matrix of 12 criteria multiplied by 13 technologies with data (156 bits of information in total). We wanted the outcomes to be universal, i.e. at a global level. It is theoretically possible to compile this information in a quantitative way but it is an enormous task. For example, we do not know how many roofers fall to their death installing PV at the global scale annually, but we do know the answer is likely to be "quite a few", but not in hundreds or thousands, and that solar PV will likely have a higher mortality rate than nuclear power where it is well established that fatalities in civilian nuclear power is extremely low [12] (perhaps contrary to the public perception in some countries).

We opted, therefore, to experiment with MCDA methodology using expert judgment to quantify variables by asking a group of experts to assign scores. We audit and validated this approach in three ways (section 4):

1. By looking for pair-wise correlations between the MCDA scores for different technologies where a correlation was to be expected, for example between cost of electricity and ERoEI and between fatalities and illness.

2. By comparing the MCDA scores with real data where this was possible. For example, by comparing the scores for cost and calculated levelised cost of electricity.

3. By comparing the MCDA scores with the NEEDS project baseline costs.

One of us (EM) had run an energy blog called Energy Matters [36] since September 2013 (the blog is still on-line but currently moth-balled). The blog aspired to peer review quality content, but bypassing the time-cost of the peer review process [39]. Peer review takes place instantaneously on-line via the views expressed by commenters. Commenters are heavily moderated where those expressing monotonic extreme views (commonly known as trolls), often not grounded in physics or economics, are excluded. This fostered an amiable discussion, where all views backed up by reasoned argumentation were allowed.

The blog had a couple of hundred regular commenters drawn from diverse backgrounds including academia, industry and finance. We sought to capitalise on this resource of expert opinion via a game where commenters were invited to score the 13 electricity generating technologies against the 12 criteria using a scale of 1 to 10, where 1 = good and 10 = bad. We hoped that between 20 and 40 regular commenters might participate, but in the end we got 19 participants who scored most rounds (Figure 3). This to some extent may reflect the reluctance of experts to express views on hugely complex issues where they perceived their expertise was lacking.

The survey was conducted on-line in the comments section of the blog where participants simply listed the criteria and their scores. Participants were also free to provide commentary on how they reached their results.

On any given day, only one technology was scored. The order and date the technologies were introduced to the blog are listed below (all 2018):

Coal fired (19 March)
Gas CCGT (20 March)
Biomass (March 22)
Nuclear (March 23)
Wind (March 27)
Hydroelectric (March 28)
Solar photovoltaic (March 29)
Diesel generator (2 April)
Solar thermal (3 April)
Tidal lagoon (April 5)
Geothermal (April 6)
Wave (April 10)
Tidal stream (April 11)
Weighting Criteria (April 20)

| Participant | Profession / qualifications | Nationality / country of residence | Coal | Gas | Biomass | Diesel | Nuclear | Hydro | Wind | Solar PV | Solar thermal | Wave | Tidal lagoon | Tidal stream | Geothermal |
|---|---|---|---|---|---|---|---|---|---|---|---|---|---|---|---|
| 1 | Geology / geochemistry PhD | British living in UK | ■ | ■ | ■ | ■ | ■ | ■ | ■ | ■ | ■ | ■ | ■ | ■ | ■ |
| 2 | Professor physics / economics / risk, PhD | French living in Switzerland | ■ | ■ | ■ | ■ | ■ | ■ | ■ | ■ | ■ | ■ | ■ | ■ | ■ |
| 3 | Chemical Engineer | Irish living in UK | ■ | ■ | ■ | ■ | ■ | ■ | ■ | ■ | ■ | | ■ | ■ | ■ |
| 4 | Geophysicist | British living in UK | ■ | ■ | ■ | ■ | ■ | ■ | ■ | ■ | ■ | ■ | ■ | ■ | ■ |
|   |   |   |   | ■ |   |   |   |   |   |   |   |   |   |   |   |
| 5 | Engineer (MechEng) | British living in UK | ■ | ■ | ■ | ■ | ■ | ■ | ■ | ■ | ■ | ■ | ■ | ■ | ■ |
|   |   |   |   | ■ |   |   |   |   |   |   |   |   |   |   |   |
| 6 | Academic Biology, PhD | Spanish living in Spain | ■ | ■ | ■ | ■ | ■ | ■ | ■ | ■ | ■ | ■ | ■ | ■ | ■ |
| 7 | Gas utilisation and distribution engineer | British living in UK | ■ | ■ | ■ | ■ | ■ | ■ | ■ | ■ | ■ | ■ | ■ | ■ | ■ |
|   |   |   | ■ |   |   |   |   |   |   |   |   |   |   |   |   |
| 8 | Electronics / software | living in Ecuador | ■ | ■ | ■ | ■ | ■ | ■ | ■ | ■ | | ■ | ■ | ■ | ■ |
| 9 | Geologist / geophysicist | British living in Mexico | ■ | ■ | ■ | ■ | ■ | ■ | ■ | ■ | ■ | | ■ | ■ | ■ |
| 10 |   |   | ■ | ■ | ■ | ■ | ■ | ■ | ■ | ■ | ■ | ■ | ■ | ■ | ■ |
| 11 | Systems engineer / CEO nuclear | American living in USA | ■ | ■ | ■ | ■ | ■ | ■ | ■ | ■ | ■ | ■ | ■ | | ■ |
| 12 | Professor electrical engineering, PhD | American living in USA | ■ | ■ | ■ | ■ | ■ | ■ | ■ | ■ | ■ | | ■ | ■ | ■ |
| 13 | Pharmacist | French Reunion (Indian Ocean) | ■ | ■ | ■ | ■ | ■ | ■ | ■ | ■ | ■ | ■ | ■ | ■ | ■ |
| 14 | Nuclear engineer | British living in UK | ■ | ■ | ■ | ■ | ■ | ■ | ■ | ■ | ■ | ■ | ■ | | ■ |
|   |   |   |   |   |   | ■ |   |   |   |   |   |   |   |   |   |
| 15 | Physicist | German living in Germany | ■ | ■ | ■ | ■ | ■ | ■ | ■ | ■ | ■ | ■ | ■ | ■ | ■ |
|   |   |   |   |   |   |   |   |   |   |   | ■ |   |   |   |   |
| 16 | Investor / mathematics, Cambridge | British living in Malaysia | ■ | ■ | ■ | ■ | ■ | ■ | ■ | ■ | ■ | ■ | ■ | ■ | ■ |
| 17 | Asst. Prof. philosophy of science, PhD | German living in The Netherlands | ■ | ■ | ■ | ■ | ■ | ■ | ■ | ■ | ■ | ■ | ■ | ■ | ■ |
| 18 | Engineer / management consultant | Brit / German living in Germany | ■ | ■ | ■ | ■ | ■ | ■ | ■ | ■ | ■ | ■ | ■ | ■ | ■ |
| 19 | Electrical and petroleum engineer | American living in USA | ■ | ■ | ■ | ■ | ■ | ■ | ■ | ■ | ■ | ■ | ■ | ■ | ■ |

Figure 3 The 19 participants in our survey. Profession, qualifications, nationality and residence are also given where possible. The grey squares indicate which technologies the participants scored. Five individuals (unnumbered participants in the leftmost column) cast scores for only one or two technologies and are not included in the results.

For each criterion and for each technology, we calculated a mean score and summing the scores for 12 criteria across a technology provides a total mean score for that technology that is our benchmark for performance. For criterion specific scores, 1 = good / best and 10 = bad / worst. For technology specific total mean scores, the minimum possible = 12 (good / best) and the maximum = 120 (bad / worst).

Participants were asked to score the performance of a whole electricity supply system, including the supply chain, for the whole world. And so for example, when rating the environmental costs of coal, this had to include coal mining and transport, power generation and pollution. There are vast variations in the degree of impact in different parts of the world. It seems an impossible task but, for example for coal, our participants rated fatalities 6.8, chronic illness 7.9, external environmental costs 7.2 and $CO_2$ intensity 9.8. Coal stands out as by far the worst electricity technology on these health and environmental criteria. Coal, however, scored well on cost and grid categories. As we shall see in section 4, there is a high level of consistency between mean scores and "reality" for many but not all criteria. This and further checks presented below suggest that our expert judgment may convert to semi-quantitative measures in most instances (Section 4).

## 2.5 Participant selection

Participants self-selected from Energy Matters readership. The blog has ~200 regular commenters and the majority of, but not all, participants came from this group, already predisposed to online interaction. The project managers (authors of this paper), therefore, did not participate directly in selecting or recruiting the participants. All judgments made were included in the survey results. The biases of participants range from pro nuclear to pro renewables to pro coal. As discussed in Section 2.4, the blog comments are moderated, which excluded trolls from participation.

2.6 Data compilation and presentation

Scores were compiled by hand in an XL spread sheet as they were cast over a period of four weeks.

We display results on a spider diagram (MS XL "charts other" radar plot) that has 12 axes scaled 1 to 10 in the default format. Figure 4 shows the spider-diagram for nuclear with mean score ± 1 standard error. The axes are arranged according to categories in a way that tends to favor renewable energy in the top right quadrant and fossil based systems in the lower left quadrant.

The shape of the spider diagram creates a visual multi-criteria impact while the area is proportional to the technology specific total mean score. The spider diagrams for all 13 technologies are presented in section 3.3.

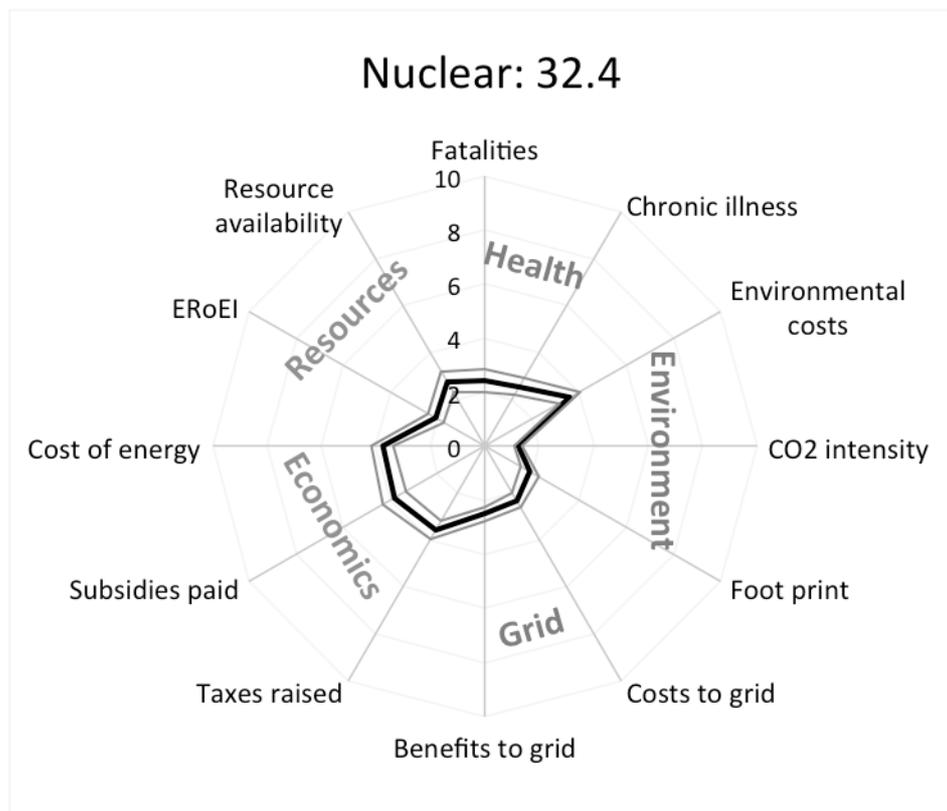

Figure 4 Representation of criterion-specific mean scores for nuclear power showing the means (heavy black line) ± 1 standard error (grey line). The total mean score for nuclear is 32.4. The criteria are represented by the axes, grouped into the 5 categories of health, environment, grid, economics and resources.

3. Results

## 3.1 Convergence on mean

We are alert to the possible criticism that the small size of our expert group may result in our mean scores or total mean scores not being representative. One advantage of the way results were compiled was that we updated spider diagrams as data came in over a period of hours to days. It became clear early on that once a handful of participants had cast their scores, the shape of the spider diagram was set. The convergence on mean score is illustrated in Figure 5 where it is shown that normally 5 participant scores was all that was needed to reach the final mean score recorded after 20 participant scores were cast. We infer that including more scores from this particular pool of experts would not result in any significant change to the results.

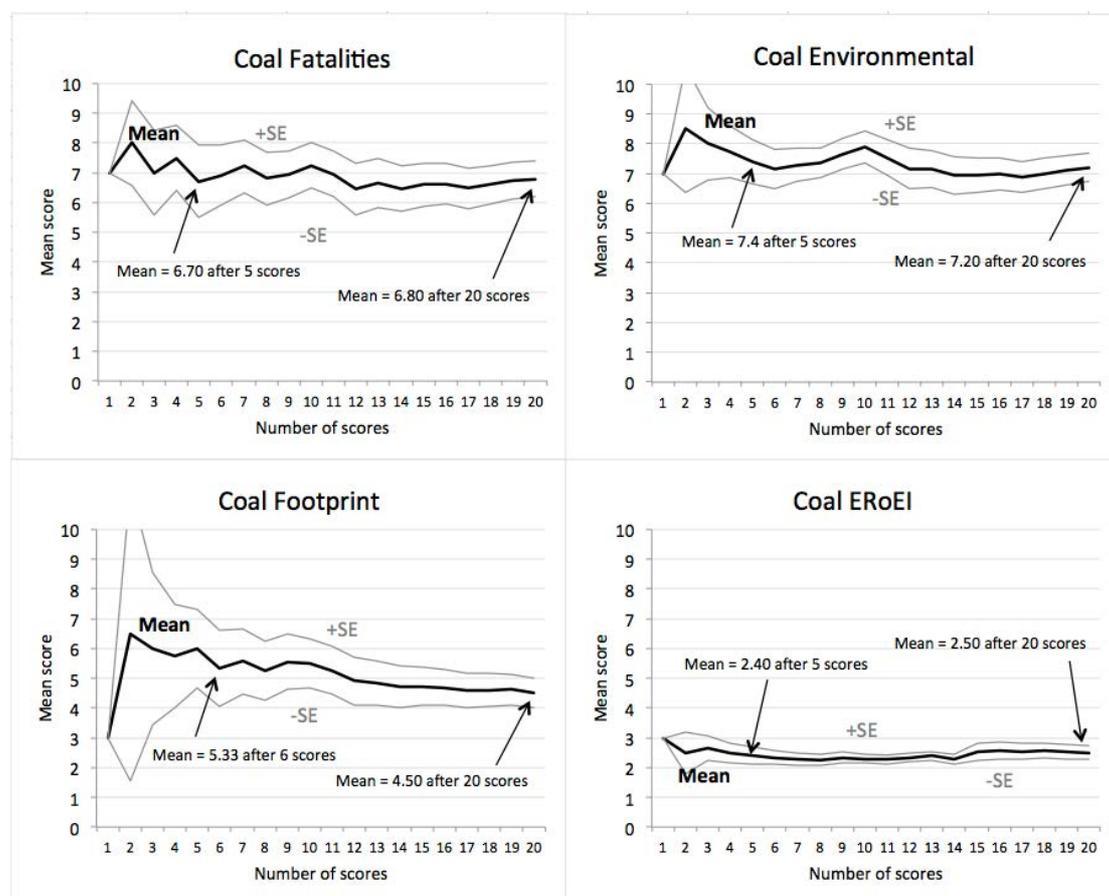

Figure 5 The charts show how the mean scores varied with the number of scores cast for 4 criteria applied to coal. In most cases, the final mean score is reached after ~ 5 scores are cast. Only for footprint is the mean still trending slowly up to 12 scores.

While there is rapid convergence on final score, this does not necessarily mean convergence on a truly quantitative score. It is unlikely but still possible that a different group of experts may produce scores quite different to ours.

## 3.2 Evaluation of weighting criteria

In conducting MCDA polls, it is common practice to weight criteria according to their importance in the minds of those casting scores [see for example 40]. This effectively means casting two scores per criterion; one for the relative importance of the criterion to a technology and one for the relative importance of one criterion relative to the others.

We conducted a pilot study on weighting criteria and considered three different methods:

1. Giving each participant 100 points to allocate as they saw fit to weight the 12 criterion,
2. Asking participants to allocate weights of 1-2-3 to each criterion
3. Asking participants to allocate weights of 1-2-5 to each criterion

With method 1, we found a tendency for most weight to fall on 2 or 3 criteria (different from participant to participant) and this defeated the objective of developing a multi-criteria system. With method 2, we found in experimentation that this made no difference to the un-weighted score ranks. And so we opted to test method three which offered a compromise between methods 1 and 2 providing greater leverage while still retaining some input for each criterion in the final total score. Participants were asked to allocate weights to the 12 criteria on April 20. The results are shown in Figures 6 and 7.

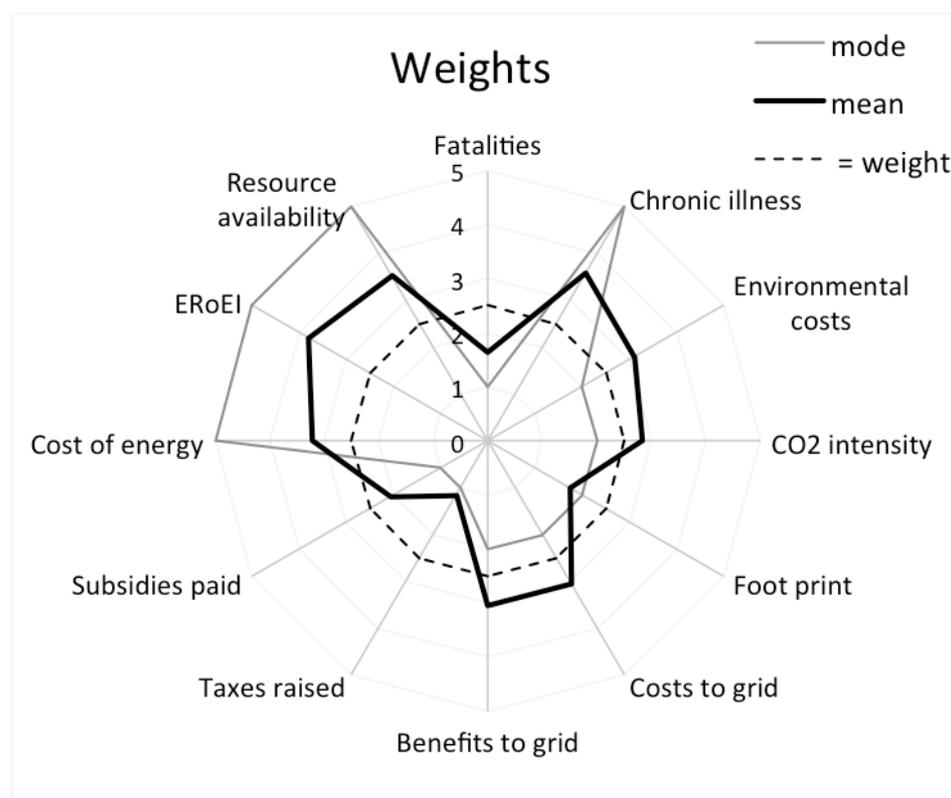

Figure 6 Spider diagram displaying the mean and mode weights assigned to the 12 criteria using the 1-2-5 weighting scheme.

It is important to appreciate that assigning weights to criteria is a different process to scoring criteria and electricity technologies. In the former, participants are expressing an informed judgment while in the latter they are expressing an opinion on which criteria are most important.

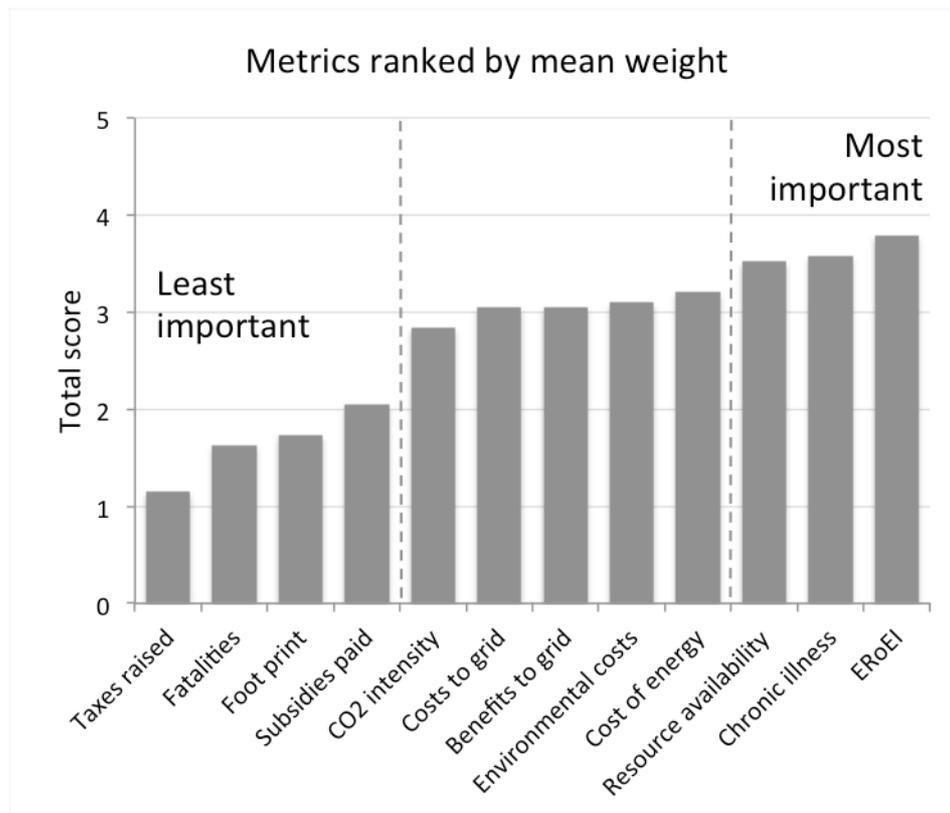

Figure 7 Rank-order for criteria weights based on participant opinion employing mean weight.

We find that using the mean or the mode effectively leads to the same conclusions. We judge that the mean probably provides the more equitable representation of group opinion that has produced a result that we find a little surprising. Using the mean weights, four criteria stand out as being least important – taxes raised, fatalities, footprint and subsidies, and three criteria stand out as most important – resource availability, chronic illness and ERoEI (Figures 6 and 7). Our expert group has tended to downgrade economic criteria (taxes, subsidies and cost) while placing ERoEI in top place. There is perhaps an element of quid-pro-quo where our experts have judged that a high ERoEI electricity source will automatically lead to low cost and high tax revenues and they have therefore discounted some double counting. Placing fatalities in second bottom place is also rather odd. We surmise that our expert group has rationalized that fatal accidents are not a major issue with any of the technologies assessed. In contrast, chronic illness (stemming from airborne pollution?) is perceived to be an important issue.

The weighting of results is discussed in section 3.4 and shows that applying weights of 1-2-5 using either the mean or the mode makes no to little difference to the outcome of our poll as measured by the rank order of technologies and so we opted to use the un-weighted results.

## 3.3 Summary of Scores for 13 Electricity Technologies

The total mean score for a technology equals the sum of the scores for the 12 criteria averaged between our 19 experts. Figure 8 shows the 13 electricity systems ranked by total mean score. We find three clear winners: nuclear, gas CCGT and hydroelectric power. These are clear winners but with very little variance in total mean score between these three technologies. We then have a second group of runners up in the form of geothermal, diesel and coal where there is somewhat greater differentiation between the scores for these three technologies. There is then another clear break to our group of 7 worst electricity technologies. In rank order: solar thermal, wind, tidal stream, wave, solar PV, biomass and tidal lagoon. With a gradual increase in score from left to right, there is little differentiation between these 7 technologies.

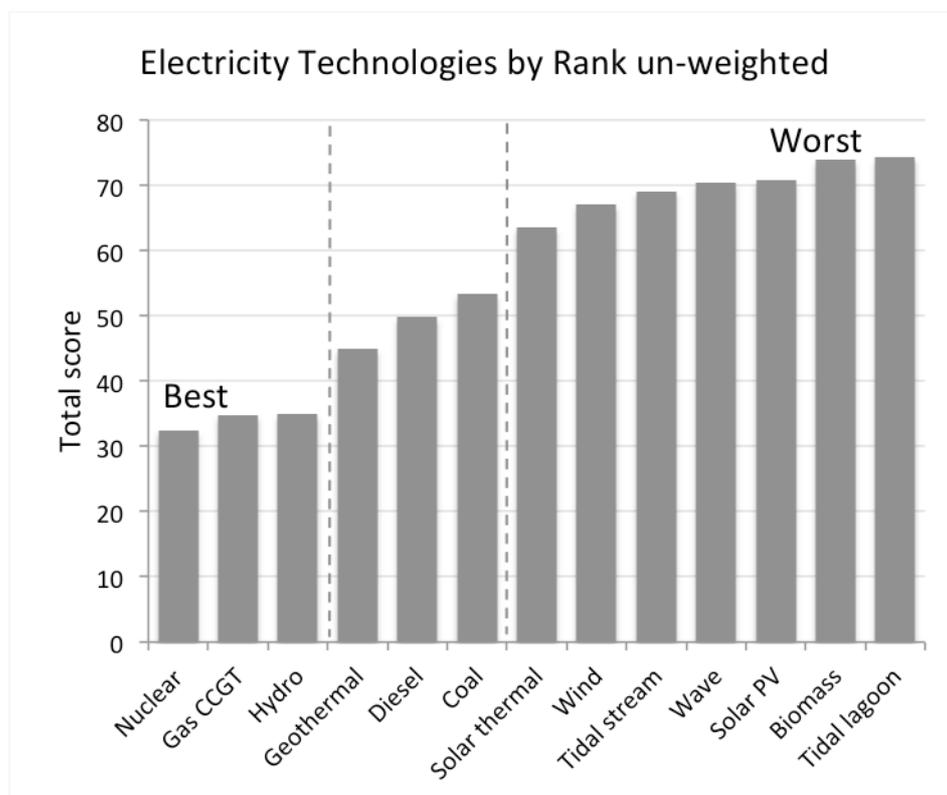

Figure 8 Thirteen electricity generation technologies ranked by total mean score.

The worst technology, tidal lagoon, has 2.3 times as many points as the best technology that is nuclear power. No one should be surprised at the three out right winners since these are all proven and already widely deployed technologies. What is surprising is that countries like France, Germany and Switzerland have chosen to shut down nuclear power stations (1st place) and to replace them with wind (8[th] place) and solar PV (11[th] place) technologies that according to our analysis are vastly inferior.

There now follows a discussion of each technology based on the distribution of scores for criteria as displayed on the spider diagrams.

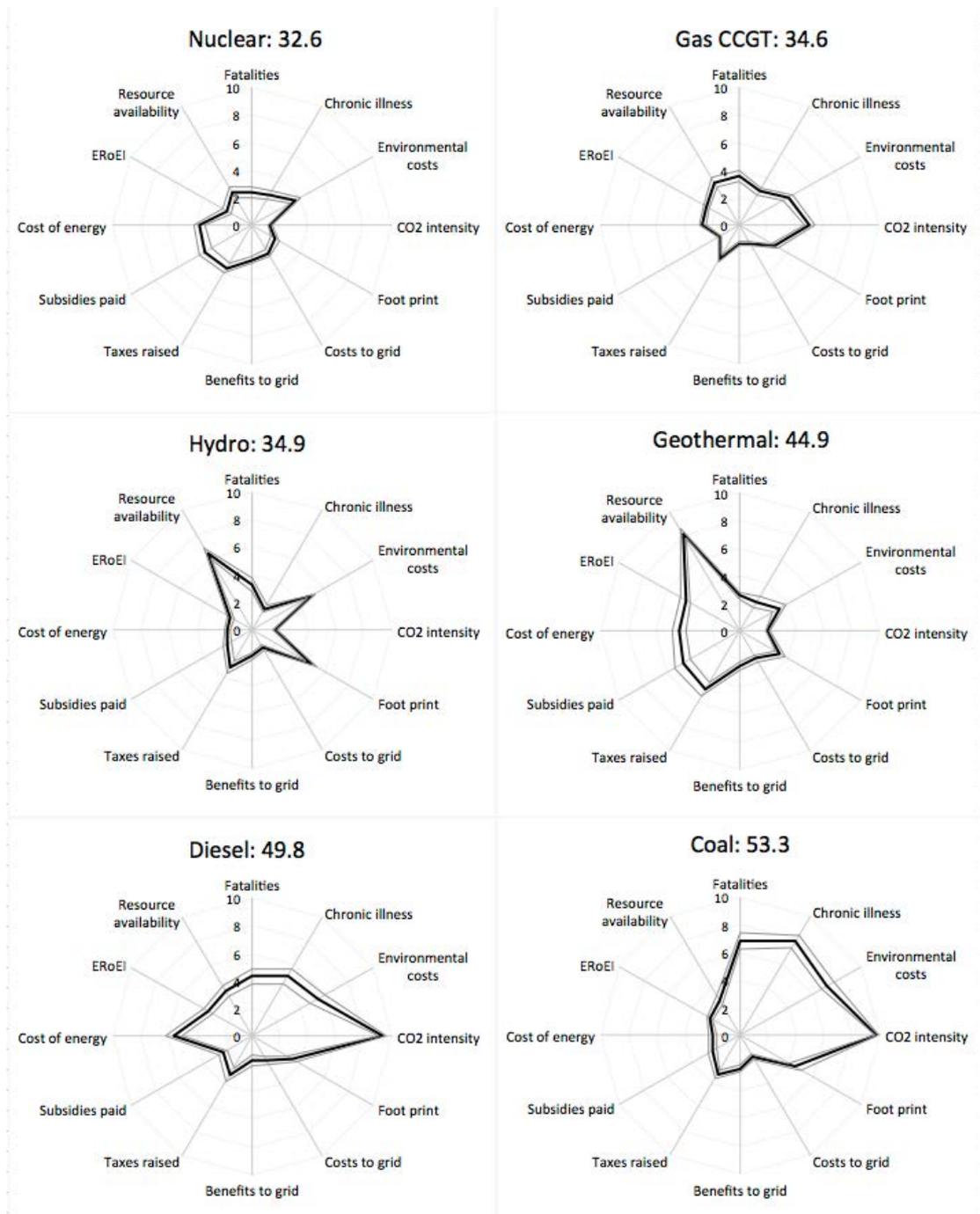

Figure 9 Spider diagrams for the 6 best technologies (i.e. with lowest scores) in rank order.

### 3.3.1 Nuclear Power
With a total mean score of 32.6, nuclear power was judged to be the highest quality electricity option. Nuclear scored well on all crtieria and particularly well on $CO_2$ intensity and ERoEI.

### 3.3.2 Combined Cycle Natural Gas
With a total mean score of 34.6, combined cycle gas turbines came in second place. Compared with nuclear, there is greater skew in the scores. Scores for costs and benefits to grid, taxes raised, subsidies paid, cost of energy and ERoEI are all particularly low. This is offset by poorer performance on all health and environmental criteria and on resource availability that scored 3.63, a little higher than nuclear power, but still a relatively low score. This signals that our expert group sees no immediate problems with natural gas supply.

### 3.3.3 Hydroelectric Power

With a total mean score of 34.9, hydroelectric power is just behind natural gas but with a rather different distribution of scores and an unusual spikey shape. Dispatchable hydroelectric power has low scores on costs and benefit to grid, cost of energy, ERoEI, chronic illness and $CO_2$ intensity. However, it has performed badly on environmental costs, footprint and resource availability. Flooding a valley of course totally obliterates the natural habitat. The three Gorges Dam lake in China covers an area of 1,084 km$^2$ flooding 13 cities, 140 towns and 1350 villages, displacing 1.24 million people [41]. This is environmental destruction on a grand scale [42]. On resource availability, our expert group sees that the majority of large dam sites are already used with little prospect for future expansion.

### 3.3.4 Geothermal

With a total mean score of 44.9, there is a significant step-wise decline in the quality of geothermal compared with the three leading technologies (Figure 8). Geothermal was allocated low scores on all health, environmental and grid criteria. On the latter, our expert group have taken the view that geothermal either provides uniform baseload or can be varied to help maintain supply matched to demand. On the economic and resource criteria, geothermal performed less well, particularly on resource availability. Our expert group saw geothermal as a niche application in volcanic regions where steam can be easily accessed and produced via shallow wells.

### 3.3.5 Diesel

Our participants were asked to score diesel generators that have been deployed in the UK as peaking plants. With a total mean score of 49.8, diesel is only just ahead of coal on 53.3, although once again the distribution of strengths and weaknesses is rather different. Diesel scored well on cost and benefit to grid, subsidies paid (largely unsubsidised), ERoEI and resource availability. Diesel performed badly on cost, seen as an expensive source. Some diesel engines won contracts in the UK capacity market auction [43]. Diesel engines have the advantage of being readily switched on and off, using little fuel, while gas and coal plants can take hours to become operational. Diesel performs badly on all Health and Environment criteria, especially $CO_2$ intensity.

### 3.3.6 Coal

With a total mean score of 53.3, coal has one of the most heavily skewed distributions. It was judged very well on all Grid, Economic and Resource criteria but scored very badly on all Health and Environmental criteria. Major health and environmental costs are counterbalanced by cheap, efficient and abundance benefits.

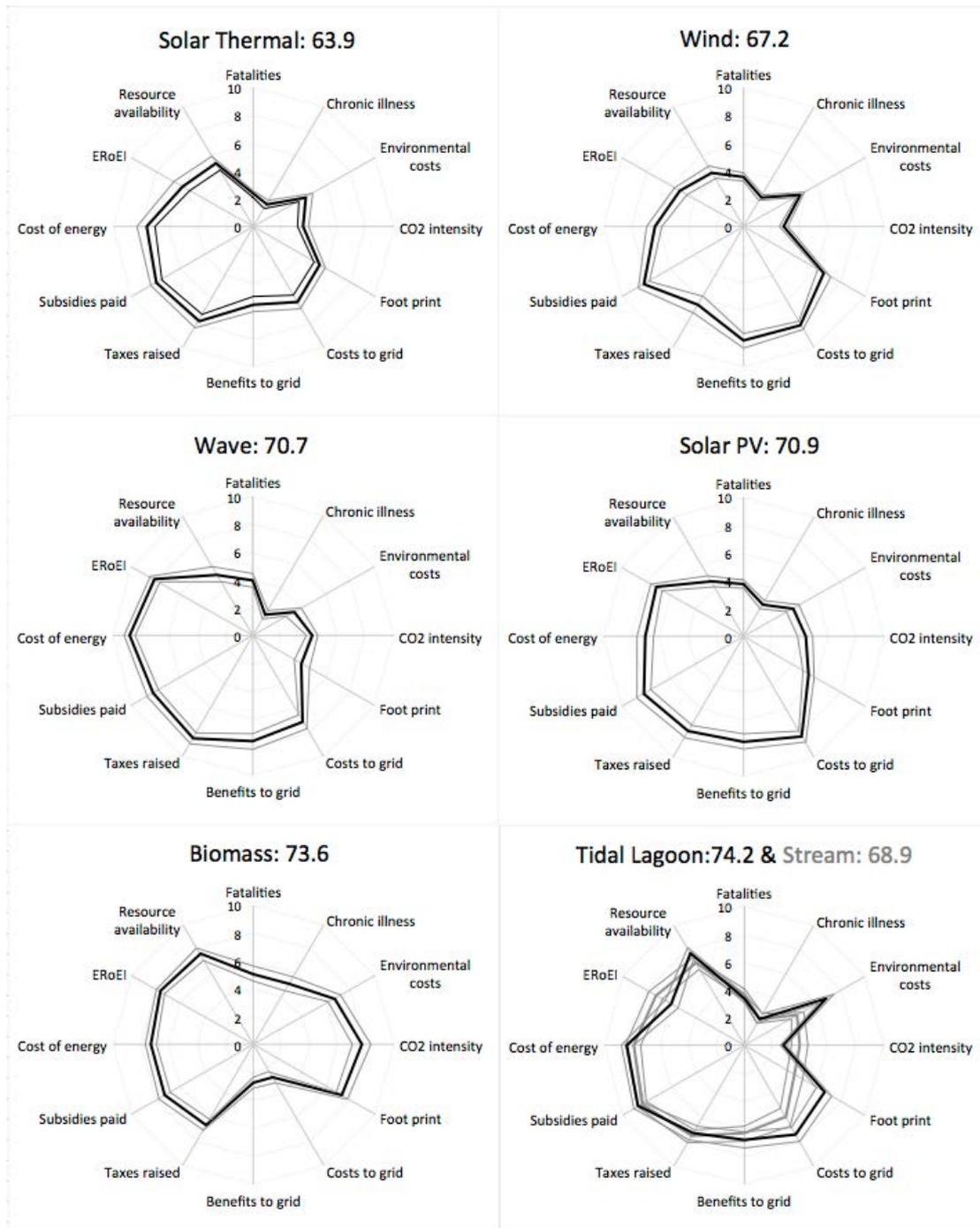

Figure 10 Spider diagrams for the six worst technologies (i.e., with the highest scores). Solar, wind, wave and tidal technologies all have similar shapes, tending to perform well on Health and Environment criteria and badly on Resource, Economy and Grid. For graphing convenience, tidal lagoon and tidal stream are plotted together.

### 3.3.7 Solar Thermal

With a total mean score of 63.9, there is once again a step-wise change going from coal to solar thermal (Figure 10). Solar thermal performed well on health and environmental criteria with the exception of footprint. Solar thermal power plants do occupy a lot of space. On environmental costs, solar thermal has an elevated score since solar thermal power plants tend to obliterate the natural habitat where they are built. It performed badly on Grid, Economy and Resource criteria. Notably, performance on Grid cost and benefit (6.2 and 5.6) is better than Solar PV (8.3 and 7.5). This is because solar thermal power plants should only be deployed in desert areas where sunshine is more continuous and seasonal variability is less than at temperate latitudes where much solar PV has been deployed. It may also have some thermal storage capacity providing a tiny amount of dispatch. On Economic metrics, solar thermal is seen as expensive and subsidised. On resource availability, it has a poor score for two reasons. First, desert areas are uncommon in proximity to population centers. Plans to transmit solar thermal electricity over long distances, for example Desertec [44], are not considered to be feasible by the expert group. The cost is simply prohibitive for an intermittent electricity source. Second, the resource is only available during the day and our group did not see any form of grid-scale electricity storage being economically viable for the foreseeable future.

### 3.3.8 Wind

In our survey we did not differentiate between onshore and offshore wind. The reason for this is that in our opinion there is no fundamental difference between these technologies. It is true that offshore wind tends to have higher capacity factors than onshore wind. For example, for onshore projects in the UK and Germany the large majority of average annualised capacity factors achieved by end-2019 were under 30% (70% in the case of the UK), while the vast majority of offshore projects have achieved well over 30%. This difference may be material to the project operator but the higher capacity factor offshore is offset by higher cost. Thus, the perceived advantage of higher capacity factor is in part offset by higher cost. It is often argued that more continuous offshore wind with widely dispersed installations will tend to smooth output and reduce the intermittency issue. This may be true but not in a material way. Our own unpublished analysis [73] of the combined wind output for Sweden, Denmark, France, The UK, Germany and Spain for the months of September and October 2015 showed that the combined capacity of 98GW on one occasion produced less than 3GW, a capacity factor of only 2.9%. This included many large offshore wind farms with a pan-European dispersion. A mean capacity factor offshore >30% is largely irrelevant to the grid and the consumer when collectively all pan-European wind farms can generate close to zero output during lulls caused by large areas of high pressure.

The total mean score of 67.2 for wind is approximately double that of our three leading technologies. This could mean that our expert group view wind as being only half as good quality, taking all metrics into account. Wind has not performed as well on health and environmental metrics as one might have expected. A score of 3.6 for fatalities reflects the fact that installing and maintaining wind turbines kills workers and members of the public on a regular basis [45]. Scores of 4.6 for environmental cost and 6.6 for footprint reflects the bats and birds killed and the sprawling blot on the landscape created by onshore wind farms. This highlights one of the great ironies of current energy policy direction. The single minded focus on $CO_2$ intensity ignores all other environmental factors and has led to environmentally damaging approaches being adopted. For example on Health and Environmental metrics, wind scores better than gas on $CO_2$ but has similar or worse scores on the remaining metrics.

|  | Wind | Gas |
|---|---|---|
| Fatalities | 3.6 | 3.6 |
| Chronic illness | 2.5 | 2.9 |
| Environmental costs | 4.6 | 4.0 |
| CO2 intensity | 2.8 | 5.1 |
| Footprint | 6.6 | 2.9 |

Because of variable intermittency, wind scored badly on costs and benefits to Grid. High system costs and current generous subsidies caused wind to perform badly on all Economic criteria. Scores on the resource criteria are intermediate. In summary, wind performed badly because it is viewed as a high system cost, unreliable source of electricity.

### 3.3.9 Wave

With a total mean score of 70.7, wave power has a heavily skewed distribution on the spider diagram, providing almost a mirror image of coal. Wave power scored well on all health and environment criteria but really badly on all Grid, Economy and Resource criteria. This exemplifies what we say in the introduction. Fixation on a single environmental criterion has led to a major distortion of holistic quality appraisal.

### 3.3.10 Solar PV
With a total mean score of 70.9, solar PV is ranked almost identically to wave power. Furthermore, the distribution of scores between criteria is also very similar (compare the shapes of the spider diagrams in Figure 10). Wave and solar have very different vectors of power generation, but despite this have been judged very equally. It is noted that solar PV is one of the most widely deployed new renewable technologies while wave power is one of the least. The small modular nature of solar PV makes it possible to install facilities of all sizes and costs while offshore wave farms have a much higher entrance level for cost and have yet to be proven to actually work [21].

### 3.3.11 Biomass
With a total mean score of 73.6, the distribution of scores among criteria for biomass is unique for the seven "new renewable" technologies. In our assessment, participants scored imported wood pellets from N America combusted in industrial – scale former coal-fired power plant in Europe. There is a long, energy intensive and hazardous supply chain [46] that ends in a power generation process that has many of the same health and environmental hazards of coal. Biomass therefore scores badly on the Health and Environment criteria. With a high score for $CO_2$ (7.7), our participants were not convinced that combusting biomass is an effective means to reduce emissions. Unlike all other new renewables, dispatchable biomass power scores well on Grid criteria (cost 2.65:benefit 2.75) compared with coal (1.8:2.4) and nuclear power (2.3:2.5). Biomass scores similar to nuclear power since in the UK at least it is run in base load mode. On all Economic and Resource criteria, biomass scores badly. While biomass has all the disadvantages of coal, it does not bring the benefits of low cost and abundance. In short, biomass brings us the worst of all worlds. Policy makers need to explain why they have encouraged its use via generous subsidies.

### 3.3.12 Tidal Stream and Tidal Lagoon

Tidal stream and tidal lagoon are plotted together for graphical convenience. While both utilise the energy of the tides, the applications are very different. Tidal stream is like an underwater wind farm, where propellers capture the energy from flowing tidal water. A lagoon is an artificial dam / impoundment where tidal range is captured and released and is more analogous to hydroelectric power.

With a score of 68.9, tidal stream is viewed as a better technology than tidal lagoon with a score of 74.2, tidal lagoon has come in last place. The underwater, out of sight, stream technology has scored better on footprint and environmental costs.

### 3.4 Applying weights

In section 3.2, we describe the different weighting methodologies we considered and how we settled on the 1-2-5 scheme. We now show how applying weights affects our results, focusing on rank order. In Figure 11, the left panel shows systems ranked using mean 1-2-5 weights and the right panels shows systems ranked by mode 1-2-5 weights. The key observation is that the three groups identified from the un-weighted results (Figure 8) are unchanged. In fact, the rank order is highly similar between un-weighted, weighted mean and weighted mode. In the vast majority of instances, technologies either appear in the same place in the rank order (difference = 0) or have moved by one place (difference = 1) (Figure 12).

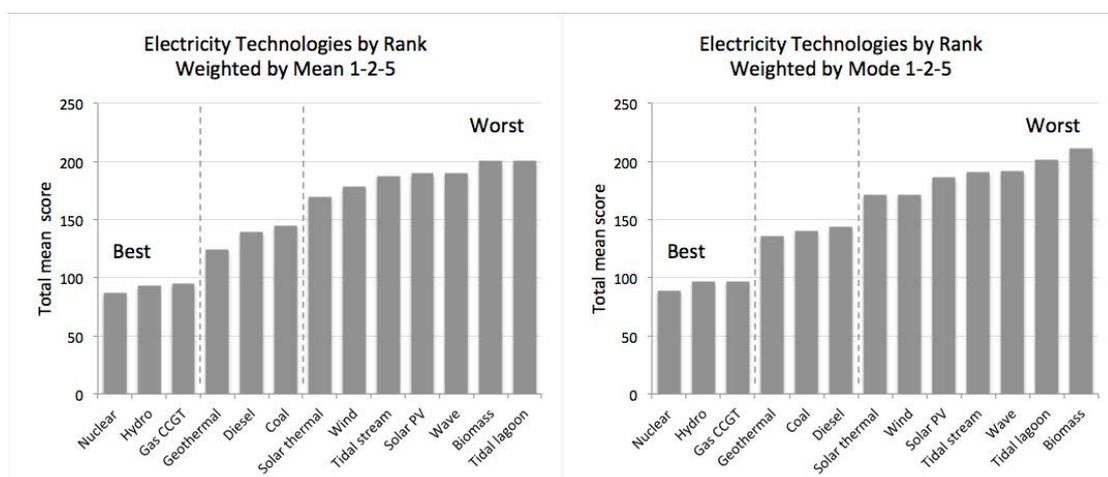

Figure 11 Weighting was applied to results by multiplying the total mean score by the weights cast by our participants (Figure 6) using the mean weight (left panel) and mode weight (right panel) both using the 1-2-5 weighting scheme.

From the results presented in Figures 11 and 12, it is clear that applying the 1-2-5 weighting method has made no material difference to the outcomes and we therefore elected to base our discussion around the un-weighted results.

| Un-weighted | Mean 125 | Mode 125 | Rank order difference | | |
|---|---|---|---|---|---|
| | | | Un v mean | Un v mode | mean v mode |
| Nuclear | Nuclear | Nuclear | 0 | 0 | 0 |
| Gas CCGT | Hydro | Hydro | 1 | 1 | 0 |
| Hydro | Gas CCGT | Gas CCGT | 1 | 1 | 0 |
| Geothermal | Geothermal | Geothermal | 0 | 0 | 0 |
| Diesel | Diesel | Coal | 0 | 1 | 1 |
| Coal | Coal | Diesel | 0 | 1 | 1 |
| Solar thermal | Solar thermal | Solar thermal | 0 | 0 | 0 |
| Wind | Wind | Wind | 0 | 0 | 0 |
| Tidal stream | Tidal stream | Solar PV | 0 | 1 | 1 |
| Wave | Solar PV | Tidal stream | 1 | 1 | 1 |
| Solar PV | Wave | Wave | 1 | 2 | 0 |
| Biomass | Biomass | Tidal lagoon | 0 | 1 | 1 |
| Tidal lagoon | Tidal lagoon | Biomass | 0 | 1 | 1 |
| | | Sum of rank order difference | 4 | 10 | 6 |

Figure 12 Electricity systems listed in rank order for un-weighted results, weighted results using the mean 1-2-5 weights and weighted results using the mode 1-2-5 weights. Summing the rank order difference gives a measure of similarity where the minimum possible sum is zero and the maximum possible is 84. The sum of difference between unweighted and weighted mean is only 4 indicating that weighting scores made little difference to the final outcome.

### 4. Validation of results

Validation of results can be viewed from three perspectives. The first is whether one agrees or not with our selection of 12 criteria to assess the holistic quality of an electricity generation system. There is no definitive answer to this judgment. The second is whether one agrees with the methodology of inviting experts to cast scores and using the sum of the means for each technology as a simple measure of quality. The third is whether or not the judgment of our expert group has any foundation in reality. It is this latter perspective that we seek to evaluate in this section employing three different approaches (section 2.4). In this section, we provide a summary of the validation process where a more full description and analysis are provided in Appendix C.

### 4.1 Pairwise correlations between scores for various criteria

We expect to find a correlation between pairs of scores for certain criteria. For example, cost of energy should correlate with ERoEI where high ERoEI sources are expected to provide cheap electricity and vice versa. We have tested 5 pairs of criteria where a correlation was expected and one pair of criteria where no correlation was expected as summarized below.

$R^2$
1. Cost of energy v ERoEI      0.84
2. Cost to grid v benefits to grid      0.97
3. Fatalities v Chronic illness      0.77
4. Cost of energy v Subsidies      0.83

5. $CO_2$ intensity v ERoEI             0.90 (non-combustion technologies)
   6. Fatalities v Resource availability   0.05 (non-correlating)

For the five pairs of criteria that are expected to correlate (1 to 5), the correlation coefficients are exceptionally high (0.77 to 0.97). This demonstrates that in aggregate the judgment of our expert group has been highly consistent. In the one instance where we would expect no correlation (between fatalities and resource availability), the correlation coefficient is 0.05. See appendix C for further discussion.

## 4.2 Correlations between scores and real data

The second way we validate our results is to compare our mean scores with quantitative data where this is available for five of our criteria – cost, deaths, illness, ERoEI and $CO_2$ intensity.

|  | $R^2$ |
|---|---|
| Cost of electricity versus Levelised Cost of Electricity [47] | 0.90 |
| Fatalities versus Deaths per TWh [48] | 0.71 |
| Fatalities versus Deaths per TWh (49) | 0.74 |
| Chronic illness versus serious illness [48] | 0.80 |
| ERoEI versus buffered ERoEI [50] | 0.92 |
| $CO_2$ intensity versus $CO_2$ intensity for 5 technologies [47] | 0.97 |

For cost, illness and EROEI, we find excellent correlations of 0.9, 0.8 and 0.92 respectively. We had difficulty finding reliable and reproducible data on fatalities and electrical power production. For example, the three sources we found yielded 10, 25 and 60 deaths / TWh for coal [64, 48 and 49 respectively]. Furthermore, there is not always exact equivalence in technology, for example [64] reports data for methanol and not biomass. We have tackled this problem by simply averaging the data from the three sources that produces a reasonable correlation with $R^2 = 0.71$. The high quality of this agreement surprised us and suggests that, given a group of diverse but true experts, expert judgment combined with MCDA may provide a near quantitative measure. This provides us with some confidence in the veracity of scores for criteria where it is less easy to find quantitative data at the global scale for, for example, external environmental costs, costs and benefits to grid.

## 4.3 Comparison between EMETS MCDA and NEEDS baseline costs

In the period 2004 to 2009, the EU majority funded (€7.6 of 11.7 million) a multi-institutional project called New Energy Externalities Development for Sustainability (NEEDS), with the aim of evaluating a number or electricity generation options for the future [19, 20]. The project had the specific aims of informing EU energy policy based on sustainability criteria in the future year of 2050. The project involved up to 67 institutions, mainly from Germany, Switzerland, France and Italy.

The NEEDS project was in two parts where 26 electricity technologies were evaluated using 1) conventional life cycle baseline cost analysis, including external costs, and 2) an elaborate on-line MCDA survey employing 36 Criteria where 159 "stakeholders" participated. We found the NEEDS MCDA analysis to be deeply flawed including 1) a heavy bias in participants towards academia, 2) the selection of numerous esoteric electricity systems for evaluation (see Figure 13), 3) many questions that were impossibly difficult to answer, 4) many criteria that were not universally applicable to all technologies imparting bias to results, 5) selection of criteria that were clearly biased against nuclear power and large thermal plant, 6) duplication of non-universal and biased criteria amplifying the bias of results, 7) evaluating 26 technologies against 36 criteria using a complex user interface (936 near impossible questions to answer) consumes excessive amounts of time and this may explain why only 159 out of 2848 invited experts participated in the survey (see Appendix B for a full critique).

The 26 technologies selected by NEEDS for analysis include many esoteric fossil fuel based systems and excludes common systems like hydroelectric power and geothermal (Figure 14). Nevertheless, there is an area of near overlap with 7 technologies common to both studies – nuclear power, coal, combined cycle gas, biomass, solar PV, solar thermal and wind - (Figure 13) and while in detail the EMETS and NEEDS are measuring different things (EMETS was measuring holistic quality today while NEEDS was measuring projected sustainability in 2050), we did expect to find a degree of coherency between the two sets of results for these seven overlapping technologies. Given the serious issues we find with the NEEDS MCDA approach, in this section, we will focus on comparing our results with the NEEDS baseline costs.

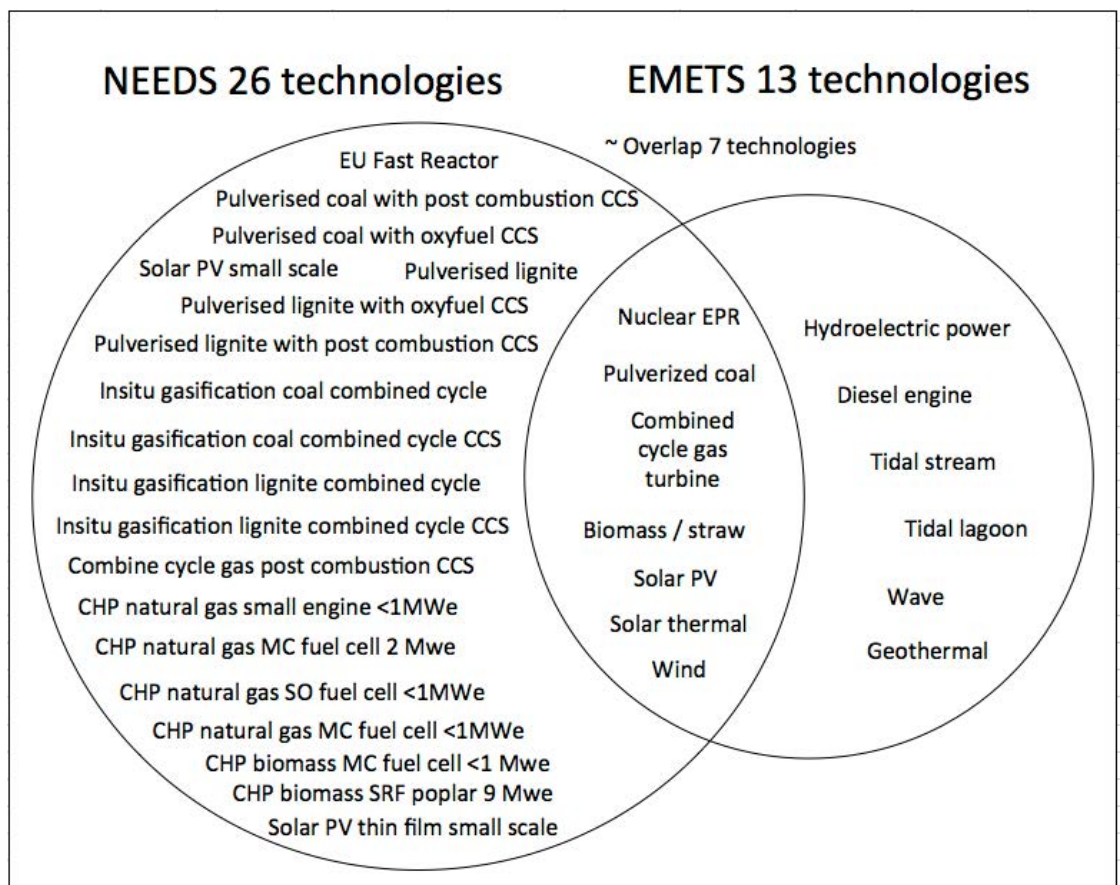

Figure 13 Venn diagram showing the 26 NEEDS and 13 EMETS technologies and the overlapping 7 technologies where in some instances, for example biomass, the overlap is approximate.

For the seven overlapping technologies, the EMETS total mean scores are cross-plotted against NEEDS baseline cost as taken from Table 6 in ref [19]. We find a reasonable correlation ($R^2$ = 0.57) for all seven technologies (Figure 14). Gas is an outlier and excluding it we find an excellent correlation with $R^2$ = 0.93. The fact that the regression does not pass close to the origin is not an issue as one cannot expect the best technology to have zero cost and the minimum possible score is 12 by construction. The EMETS participants ranked gas similar to nuclear power (total mean score = 34.6 v 32.6 respectively) while the NEEDS baseline costs rank gas at almost double the cost of nuclear in 2050. At the time the NEEDS survey was conducted (2004-2009), oil prices were rising fast leading to perceptions of global scarcity of petroleum [27, 28, 29]. The shale oil and gas revolution that has subsequently taken place has transformed perceptions of scarcity of oil and gas at the present time that could explain the diverging perceptions for natural gas.

While the coherency between the EMETS MCDA scores and NEEDS baseline costs is less good than the validation described in sections 4.1 and 4.2, it is sufficiently good to provide confidence that both methods are measuring similar things and supports the veracity of both approaches even although EMETS is rooted in the present day and NEEDS is aiming at 2050 (with insights also anchored in the present).

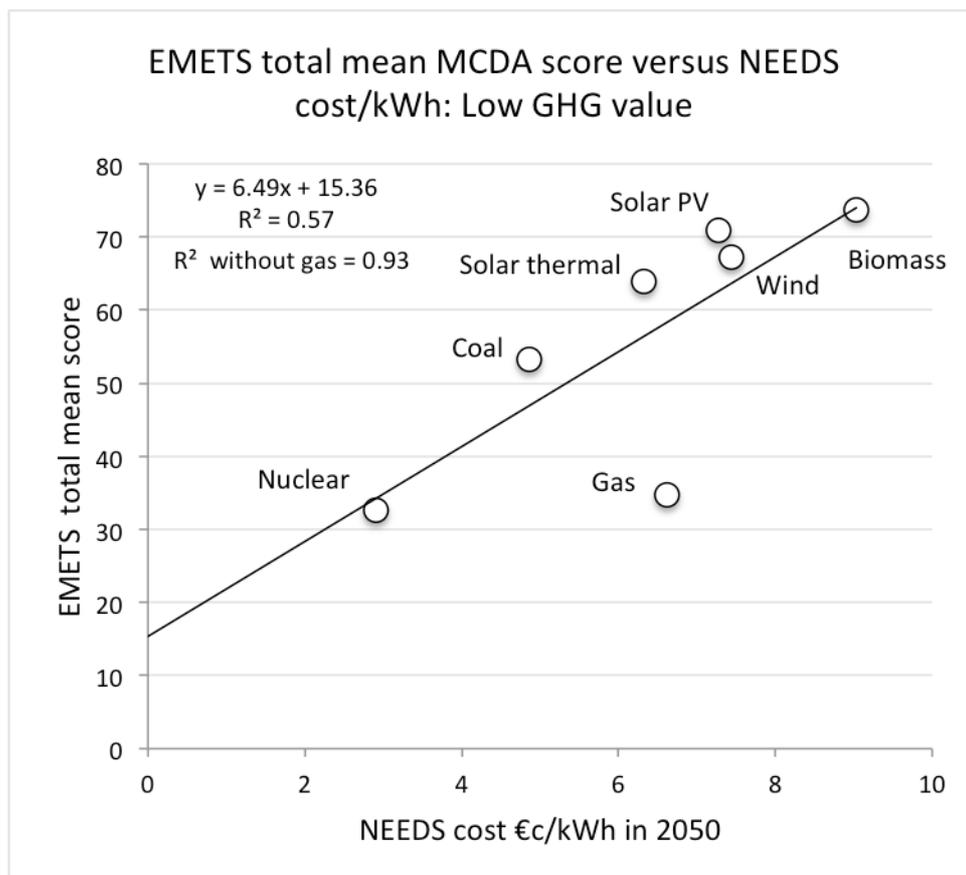

Figure 14 NEEDS baseline costs compared with the EMETS total mean scores for 7 technologies common to both surveys.

## 5. Conclusion and policy implications

*So the book was born out of anger and the feeling that humanity really does need to pay attention to arithmetic, and the laws of physics, and we actually need a plan that adds up.*

and…

*And the only reason that Solar got on the table was because of democracy, that the MPs wanted to have a Solar Feed-In Tariff…..You know, Britain's one of the darkest countries in the world.*

*The Late Professor David Mackay, author of [57] final interview 3 April 2016 [transcript at 58]*

International attitudes and actions aimed at averting catastrophic climate change by reducing and then eliminating $CO_2$ emissions from combusting fossil fuels varies enormously from one country to the next. The IPCC has now decreed that the world should aim for net zero emissions by ~2050 [1]. One needs to wonder if anyone in the IPCC has a plan that adds up? The European Union countries plus the UK, Canada and Australia have embraced this goal, which may work only if the rest of the world does follow suit. Many European countries have recorded falls in $CO_2$ emissions since the reference year of 1990 (Figure 15), giving the impression that various policy interventions have worked and will continue to do so. A number of factors have contributed to the declines in European emissions and understanding the mechanism is vital for the assessment of policy success.

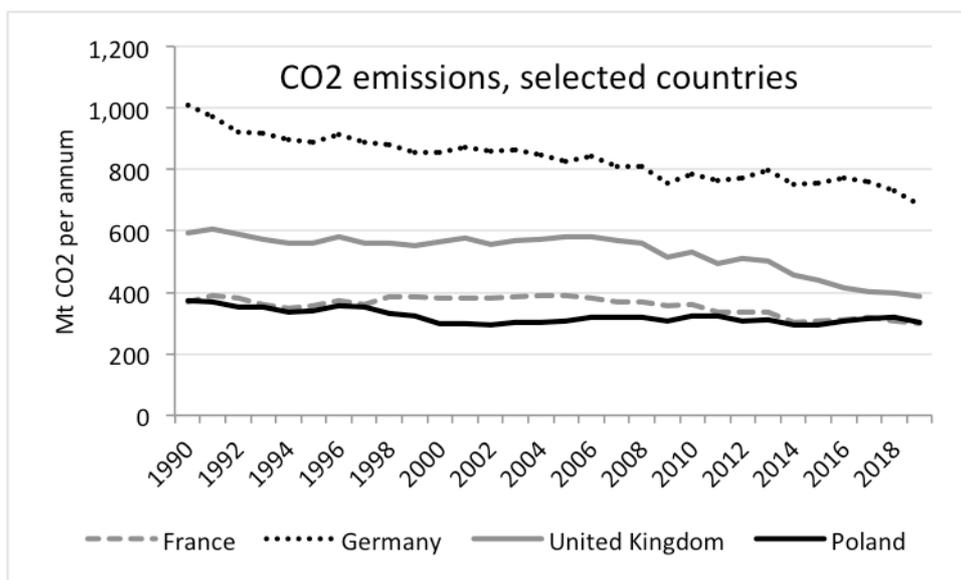

Figure 15    $CO_2$ emissions for selected countries. France, Germany and the UK all show a decline in emissions since the reference year of 1990. Data from BP 2020 [6]. This plot shows only emissions from fossil fuels consumed in country. In the case of the UK, these have fallen 37% since 1990. Including other categories of emissions, such as waste management, industrial processes and emissions embedded in manufactured imports and the decline is more modest at 23% (Figure 16, Appendix

A). Note that France has made little progress in reducing fossil fuel emissions, mainly because there are no emissions to eliminate from their nuclear-hydro energy mix, apart from aiming at other industrial sectors such as transportation, agriculture and so on.

In Appendix A, we use the UK as a case example of how and where greenhouse gas (GHG) emissions have been reduced (Figure 16). The UK Government claims that total reduction in MtCO2e (million tonnes carbon dioxide equivalent) has been 42% since 1990 giving the impression that the UK is well down the road to achieving net zero [15]. However, when net emissions embedded in imported goods are included the reduction is a more modest 23% leaving the UK with a mountain to climb. The UK has plucked some low hanging fruit with significant reductions in fluorine gasses in industrial processes, reduction of methane emanations in waste disposal and phasing out coal in power generation.  Coal has been substituted by wind and solar power, electricity imports and wood chips imported mainly from N America (Figure A4 in Appendix A).

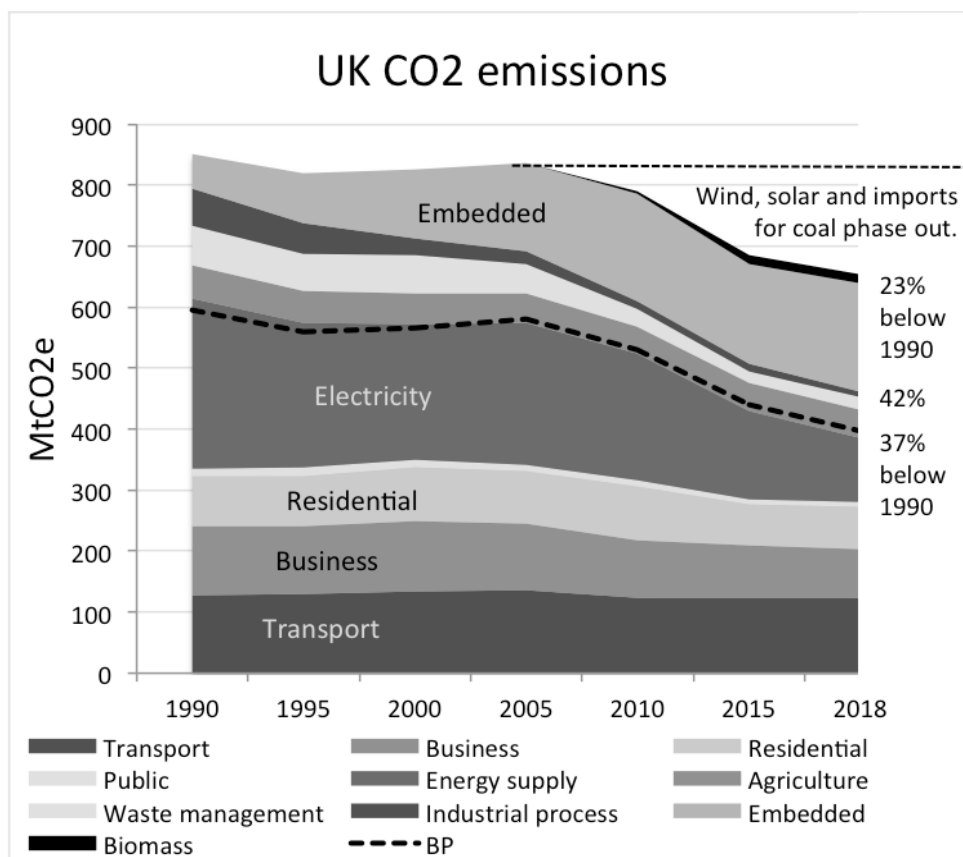

Figure 16 UK $CO_2$ emissions 1990 – 2018 as reported by the UK government [54]. Net emissions embedded in imported goods have been added [52]. Emissions from burning imported woodchips have also been added [55]. The emissions from categories transport, business, residential, public and electricity are dominated by the combustion of fossil fuels.

The challenge remaining for the UK as it stood in 2018 is summarized in Figure 17. Here we focus on four categories of 1) transport, 2) business (energy for industry), 3) residential and public heat and 4) energy supply. Categories 1-3 comprise mainly natural gas burned for space heating and liquid fuel combusted in transport. While distributed renewable energy may contribute to reducing domestic demand for gas, we posit that to eliminate emissions from categories 1-3 will require mass electrification of space heating and transport. The technology to achieve this exists by way of heat pumps, resistance heaters and electric vehicles. Electrification carries the significant advantage of being more energy efficient [57] and reducing air pollution. This is balanced on the negative side by currently higher capital costs. However, one thing is clear, opting for the electrification route will lead to significantly higher demand for electricity in a country like the UK. If power grids in the west are going to be completely rebuilt and expanded, this leads to the crucial question of how this should be done.

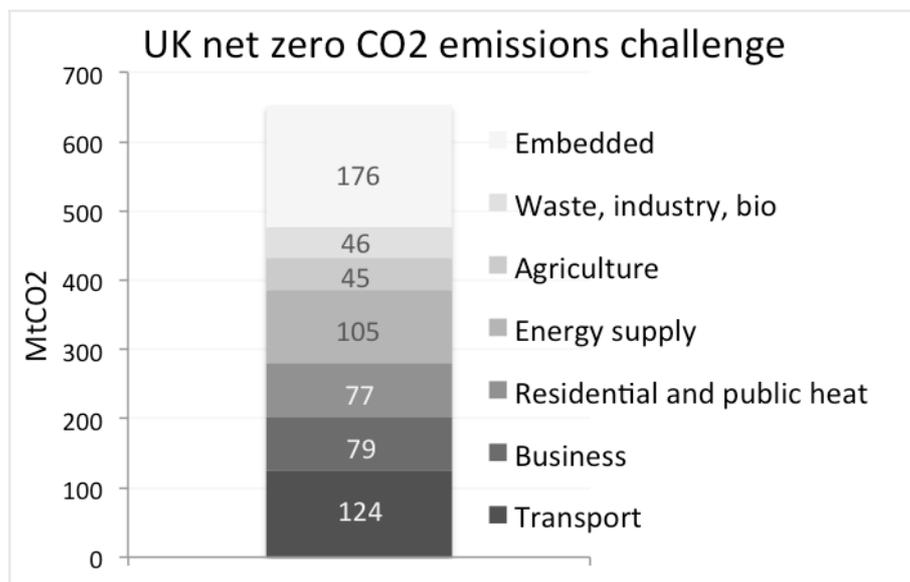

Figure 17 Summary of UK emissions as of 2018, extracted from Figure 17 based on [52] except for the embedded category that is based on [50].

With reference to the quote from Sir David MacKay [57,58] at the beginning of this section, it seems that energy policy is based more on hope and populism than on physics and economics. This led us to attempt to quantify the holistic quality of 13 electricity generating systems, most of which are in use today, based around 13 criteria arranged into 5 categories (health, environment, grid, economics and resources). Technologies were ranked on scores obtained via the judgment of an expert group within an MCDA framework. The survey produced 3 clear winners and 7 clear losers (Figure 8) (the numbers in brackets are the mean MCDA scores):

   1) Nuclear fission     (32.6)
   2) Gas CCGT         (34.6)
   3) Hydro electric     (34.9)

   4) Geothermal       (44.9)

5) Diesel             (49.8)
6) Coal               (53.3)

7) Solar thermal      (63.9)
8) Wind               (67.2)
9) Tidal stream       (68.9)
10) Wave              (70.7)
11) Solar PV          (70.9)
12) Biomass           (73.6)
13) Tidal lagoon      (74.2)

No one should be surprised at the three winning technologies that are tried and tested and already form the backbone of three European grids in France, Sweden and Finland. What becomes a real puzzle is why certain countries are planning to phase out their nuclear power, e.g. Germany and Switzerland, in favor of wind and solar judged by our expert group to be far inferior. The only way to currently back up and balance intermittent renewables is to use CCGT. And so replacing nuclear with renewables locks a country into a future with fossil fuels and higher emissions, the exact opposite of stated policy goals.

We note that the NEEDS MCDA study selected solar thermal as the most promising technology for European power generation in the year 2050 [19] (see appendix B). Our expert group scored solar thermal at the global scale and recognized, that in certain desert or low latitude countries, this is a technology with some potential. However, this technology can never work in the northern latitudes of Italy, France, Switzerland and Germany that was the geographic target of the NEEDS survey. It is already known to work badly in southern Spain where on cloudy days, it does not work at all. And of course it never works at night. The argument of providing some dispatch from stored heat is simply window dressing.

European policy makers and governments need to consider why they have staked so much on wind and solar PV power and why they ask their citizens to subsidise these technologies. These technologies deployed at scale harm the environment, they are not $CO_2$ free, are expensive and produce third-rate intermittent power. These facts are represented in the scores allocated by our expert judges.

Governments pursuing vigorous emissions reduction programs need to introduce technical and economic rigor to energy and economic accounting practices. This should include 1) $CO_2$ accounting through the whole supply chain per electricity generating technology (embedded emissions), 2) taking a system's approach to account for $CO_2$ emissions and other costs caused by intermittent technologies embedded in a grid, for example energy, emissions, other environmental and economic costs associated with energy storage and 3) account for net emissions embedded in traded goods.

To achieve the above, it is essential to take a multi criteria approach and we offer our 12 criteria scheme as a starting point, to be adapted in light of better scientific, economic, environmental and social understanding. The hierarchical multi criteria approach provides a holistic measure that balances costs against benefits over a range of critical criteria.

# Appendixes

# Appendix A   Failure of Global and UK energy policies

## A1     Introduction

The purpose of this appendix is to illustrate the harsh reality of how $CO_2$ abatement strategies have failed totally to stem the rise of atmospheric $CO_2$ thus far. We use the UK as a case study to provide some context for the reasons behind this failure. The UK government claims that $CO_2$ emissions have fallen by 42% since 1990 where our computation suggests the figure is closer 23%. This appendix aims to quantify the divergent arithmetic and logic behind these two scenarios and to highlight the serious consequences of following a wrong and delusional path.

Figure A1 left shows how atmospheric $CO_2$ has risen inexorably since the mid 19[th] century from a pre-industrial level of ~180 ppm to >400 ppm at the present day. An acceleration in the rate of increase is plain to see at around 1950. The data from both Law Dome and Siple ice cores from Antarctica overlap with the Mauna Loa atmospheric measurements, hence the dogleg is not an artifact of measurement source. This dogleg is coincident with an extraordinary post World-War II rise in energy consumption by the human species, argued by some to mark the beginning of the Anthropocene Epoch [53]. Human population growth and prosperity spreading through some very large countries, first the UAS and then China, India and Brazil has underpinned the phenomenal rise in energy consumption. It seems likely that these trends will continue for so long as human population and prosperity expands.

Figure A1-right shows human $CO_2$ emissions from the combustion of fossil fuels compared with atmospheric $CO_2$ as measured on Mauna Loa. While the rise in atmospheric $CO_2$ is uniform, the rise in emissions is not. It is punctuated by three pauses following oil price spikes in 1973 and 1979, that caused energy consumption to fall, and following the global financial crash of 2008 when the depression that followed caused energy use to temporarily dip down. We anticipate seeing a similar decline in 2020 and 2021 stemming from the plunge in energy consumption and prosperity caused by the Covid 19 pandemic. The general message here is that historic dips in energy use follow spikes in energy price that in turns causes economic slowdown, or conversely, economic slowdown that causes energy use to fall.

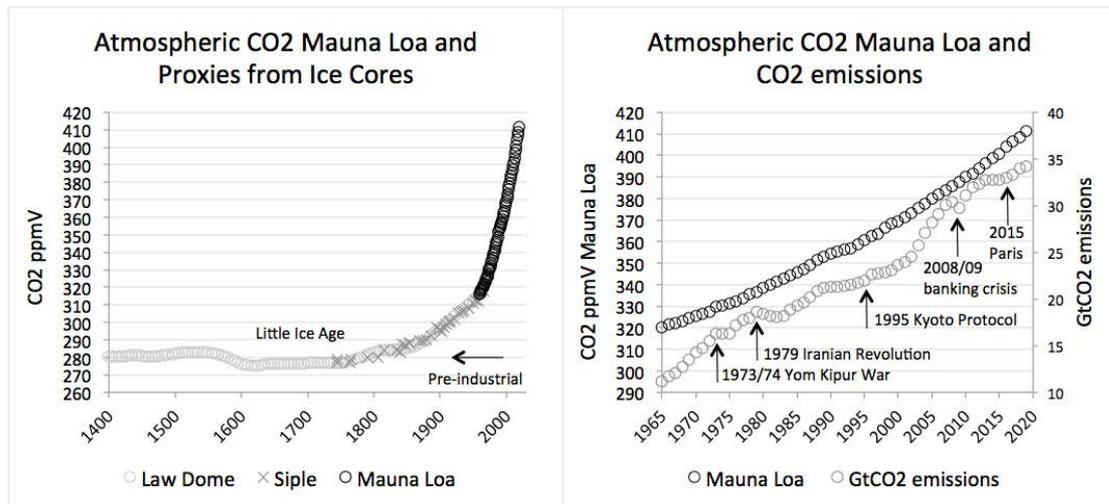

Figure A1 Left – reconstruction of atmospheric $CO_2$ based on ice cores from Law Dome [3] and Siple [4] (Antarctica) spliced with annual averages of actual atmospheric measurements made at Mauna Loa (Hawaii) [5]. Right – annual average $CO_2$ from Mauna Loa (Hawaii) [5] compared with human emissions as reported by BP [6]. The only detectable declines in emissions occur in 1973 and 1979 linked to spikes in oil price that led to a decline in oil consumption, and in 2008/09 linked to decline in energy consumption resulting from the global financial crisis. Climate treaties in 1995 and 2015 did not result in discernible declines in $CO_2$ emissions although it can still be argued that emissions may have risen more rapidly without those treaties. Irregularities in the emissions trend are not visible in the atmospheric $CO_2$ trend.

The 1995 Kyoto protocol had two commitment periods beginning in 2008 and 2012. The former is coincident with the finance crash and a massive spike in oil price that began in 2002. The latter is coincident with a flattening of the curve that has subsequently begun to rise again. The Paris Accord came into effect in 2016 but, as discussed in section 1, individual country's commitments fall well short of the $CO_2$ reductions required to meet the goals assuming that the physics behind the MAGIC C program, used to compute temperature rise for a given rise in CO2, is correct [14].

The reasons for the failure to abate $CO_2$ emissions are well understood. As already discussed rising global population and rising global prosperity are the main drivers. But there is a third reason and that is the mendacious nature of some western countries approach to $CO_2$ reduction strategies that are discussed below using the UK as a case study.

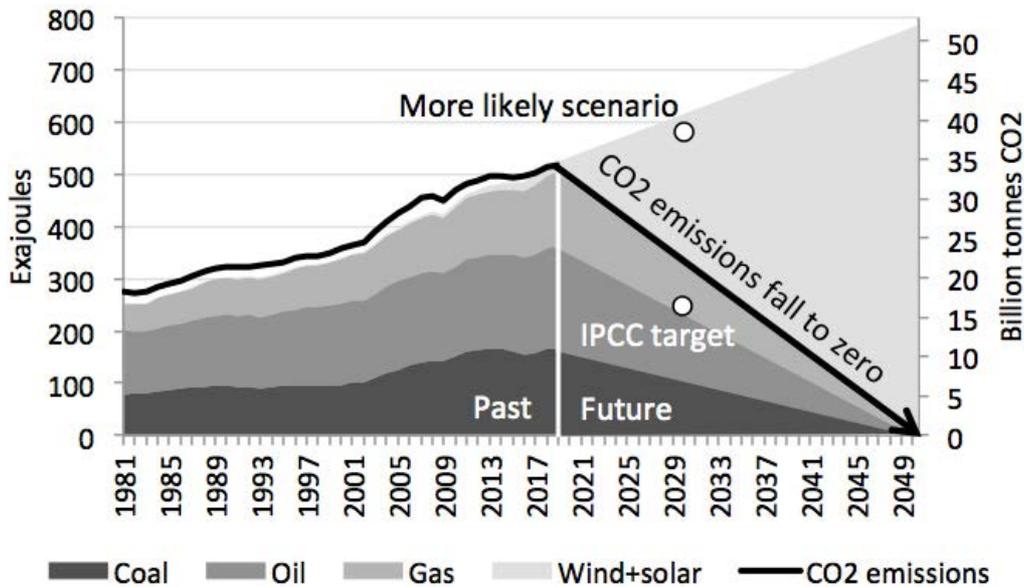

Figure A2  Global coal, oil, gas and wind+solar production from BP 2020 [6]. "IPCC target" equals the 45% fall of $CO_2$ emissions from 2010 levels by 2030 as described in [1]. The "more likely scenario" depicts business as usual and on-going growth in energy consumption, prosperity, global population combined with on-going policy failure.

In a special report published in 2018, disregarding past failure, the IPCC raised the bar on emissions reductions by declaring that dangerous climate change may occur if global temperatures rise more than 1.5°C above pre-industrial levels [1]. In general terms, the IPCC set a target for the world to reduce emissions to net zero by ~2050 with an interim target of 45% below 2010 levels by 2030. Net zero has subsequently become the new mantra of the western world's energy policies. Figure A2 illustrates the scale of failure to cut emissions to date and the scale of the future challenge where the IPCC target shows 17.1 GtCO2 per annum in 2030, being 45% lower than the 31.1 GtCO2 per annum reported by BP for 2010 [6]. The pale grey colour shows the barely visible contribution that wind and solar has made to the global energy budget to date and the massive expansion in wind and solar capacity required should global energy demand continue to rise on the historical trend and emissions fall to zero by 2050.

We now go on to illuminate some of the technical reasons for failure to cut $CO_2$ emissions using the UK as a case study. The UK aspires to be world leader in the fight against climate change. Labour Government legislation introduced in 2008 called the "Climate Change Act" enshrined in law the UK's commitment to reduce $CO_2$ emissions by 80% relative to the 1990 value by the year 2050 [56]. Part of the mechanism introduced to achieve this goal was to establish the independent Committee on Climate Change to advise the government on policy and interim targets aimed at achieving the legally binding goal. Surprisingly this legislation has cross party support and the current Conservative Government led by Boris Johnson seems to have even greater zeal to meet these goals manifest by the recent introduction of legislation aimed at reducing UK $CO_2$ emissions to zero by 2050.

An announcement by the Department for Business, Energy and Industrial Strategy sets the scene [15]:

*The UK has already reduced emissions by 42% while growing the economy by 72% and has put clean growth at the heart of our modern Industrial Strategy. This could see the number of "green collar jobs" grow to 2 million and the value of exports from the low carbon economy grow to £170 billion a year by 2030.*

Figure A3 shows how the 42% figure (in 2018) is derived using mainly government statistics from Table 3 in [54]. The government figures may be placed into three groups as follows:

1) Transport, Business, Residential and Public combined have fallen by 16% 1990-2018, and this group represents 60.8% of reported government emissions in 2018.
2) Energy supply (mainly electricity supply) has fallen by 62.3% 1990-2018 and now represents only 22.7 of reported government emissions in 2018.
3) Agriculture, waste management and industrial process have fallen by an equally impressive 57.7% and now represent only 16.5% of reported government emissions.

## A2   Net embedded emissions in imported goods

Embedded emissions reflect the energy used and $CO_2$ produced to manufacture goods. If goods are manufactured in China but then imported by the UK, the emissions associated with manufacture are borne by China while the goods are "consumed" in the UK. This is a straightforward and undisputed concept. For example, if the UK elects to close down a steel mill and to instead import steel from Germany, then UK emissions will fall with a corresponding increase in Germany. The outcome for the atmosphere is the same, this simply represents a way of the UK shifting a liability off its books.

One of the key reasons for constructing Figure A3 was to illustrate the impact on UK emissions of including the $CO_2$ emissions embedded in imported goods. For decades, the UK has pursued a strategy of deindustrialization and consequently has imported more and more goods from abroad, most notably from the European Union and from China [57]. It is widely acknowledged that significant amounts of $CO_2$ is embedded in traded goods and services, see for example [57] and the UK government recognizes that these do exist, see for example [59] and yet the international convention is to simply ignore embedded emissions in international $CO_2$ accounting, even though these can be substantial. For example in Figure A3, net embedded emissions amount to 176.4Mt$CO_2$ in 2018 or 27% of the total. Given the importance attached to emissions reduction and emissions accounting by the international community, it is extraordinary that such a significant anomaly should be allowed to exist.

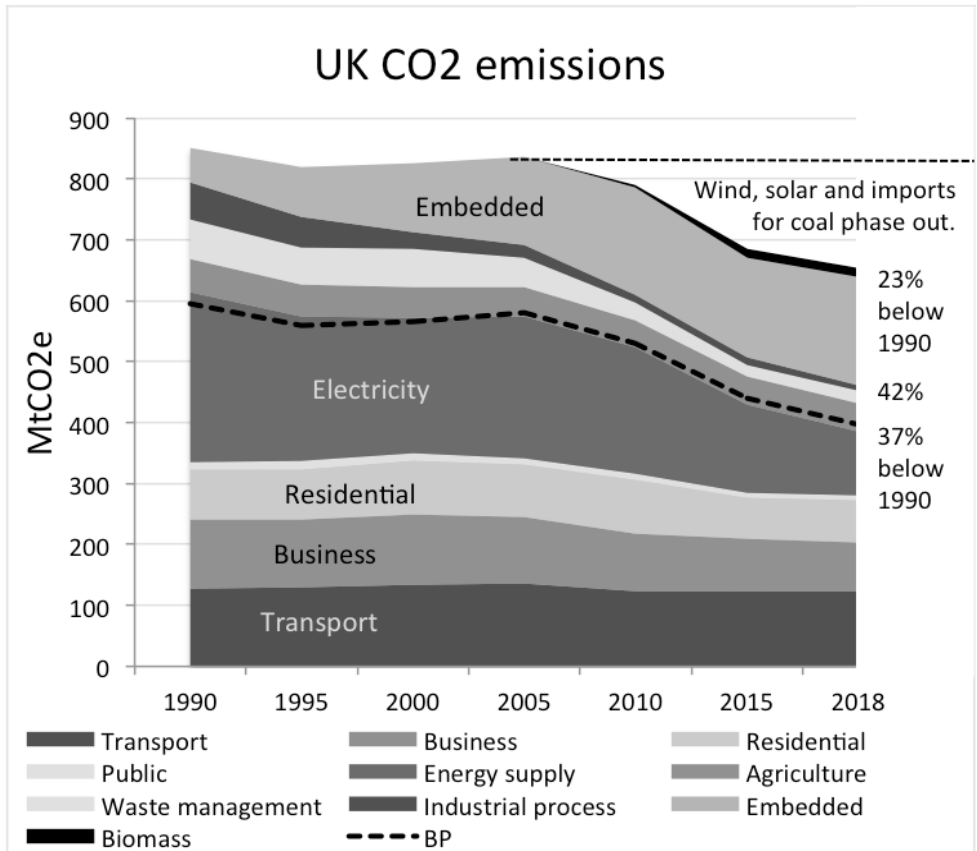

Figure A3 UK CO$_2$ emissions 1990 – 2018 as reported by the UK government [54]. Net emissions embedded in imported goods have been added [52]. Emissions from burning imported woodchips have also been added [55]. The emissions from categories transport, business, residential, public and electricity are dominated by the combustion of fossil fuels.

For convenience, we use tabulated data published by [52]. Other sources suggest that [52] could underestimate the scale of embedded emissions in the UK. For example [60] estimate 253 MtCO2e / annum for the UK in 2004 and UK government [59] reports 358 MtCO2 embedded in UK imports in 2017 compared with 176.4 reported by [52] for 2018. The wedge of embedded emissions has increased 3 fold between 1990 and 2018. Including embedded emissions and imported wood pellets used for power generation, the fall in UK emissions since 1990 is reduced to 23% from 42%.

This is important for a number of reasons. First, the UK government needs to be aware that the substantial investment made in reducing UK emissions has produced far lower returns than widely understood by government ministers and the public at large. Second, this lower return on investment emphasizes that aiming for net zero by 2050 is a far more formidable challenge than currently assumed. Midway between the 1990 to 2050 target period, the UK is only 23% of the way along the road with 77% of the effort lying ahead as opposed to the 42:58 % split that is currently assumed. And third, this accounting anomaly goes a long way to explain why international efforts to reduce global emissions have so far failed against expectations. Western countries are actually doing far less than claimed in this battle and seem content to pass the buck to developing nations like China.

Traded goods are not the only arena where western countries use sleight of hand to bring virtue to their climate policies. For example, European forests have increased in area by 9% in the period 1990 to 2014, some 13 million Ha, roughly the area of Greece [61]. This increase in forest cover has been achieved through importing a larger quantity of forest goods resulting in a reduction of forest area of roughly 11 million Ha elsewhere, most notably in Brazil and Indonesia [61]. Juvenile forests in Europe are replacing mature rainforests in these countries. Some of the most valuable habitats on earth are being destroyed so that Europe can look good on the world stage of climate mitigation policies.

### A3    $CO_2$ reduction mechanisms

### A3.1 Energy supply

In 1990, energy supply was by far the largest source of UK emissions (Figure A3) where $CO_2$ was the dominant gas emitted (94%). The main contribution to energy supply is the combustion of coal and natural gas in electricity supply but this category also includes emissions from other sectors such as coal mining. It is not clear if emissions from offshore oil and gas extraction are also included. In 1990, energy supply accounted for 33% of all emissions. This had fallen to 16% in 2018. This has been achieved via a number of mechanisms [54]:

1) The gradual closure of UK coal mining with the last deep mines closing in 2015 that reduced methane gas emissions associated with mining coal
2) The de-industrialisation of the UK economy
3) Increased dependency on electricity imports
4) A decline in electricity production since 2005
5) The phasing out of coal in electricity generation that occurred mainly in the period 2005 to 2018
6) Improved efficiency in gas powered generation with the introduction of combined cycle gas turbines (CCGT)
7) Increased production of renewable energy, mainly wind and solar

The underlying theme of these strategies is the progressive closing down of UK production that can hardly be viewed as a good thing. One strategy that may be viewed in positive light has been the progressive phasing out of coal-fired generation in the UK. Coal is both $CO_2$ intensive (twice the emissions of natural gas per TWh produced) and inherently dirty, producing large amounts of fly ash, not to mention the environmental harm caused by mining. Figure A4 illustrates how the phasing out of coal in UK power generation is balanced exactly by the increase in imports, production of electricity by burning imported wood and the expansion of wind and solar.

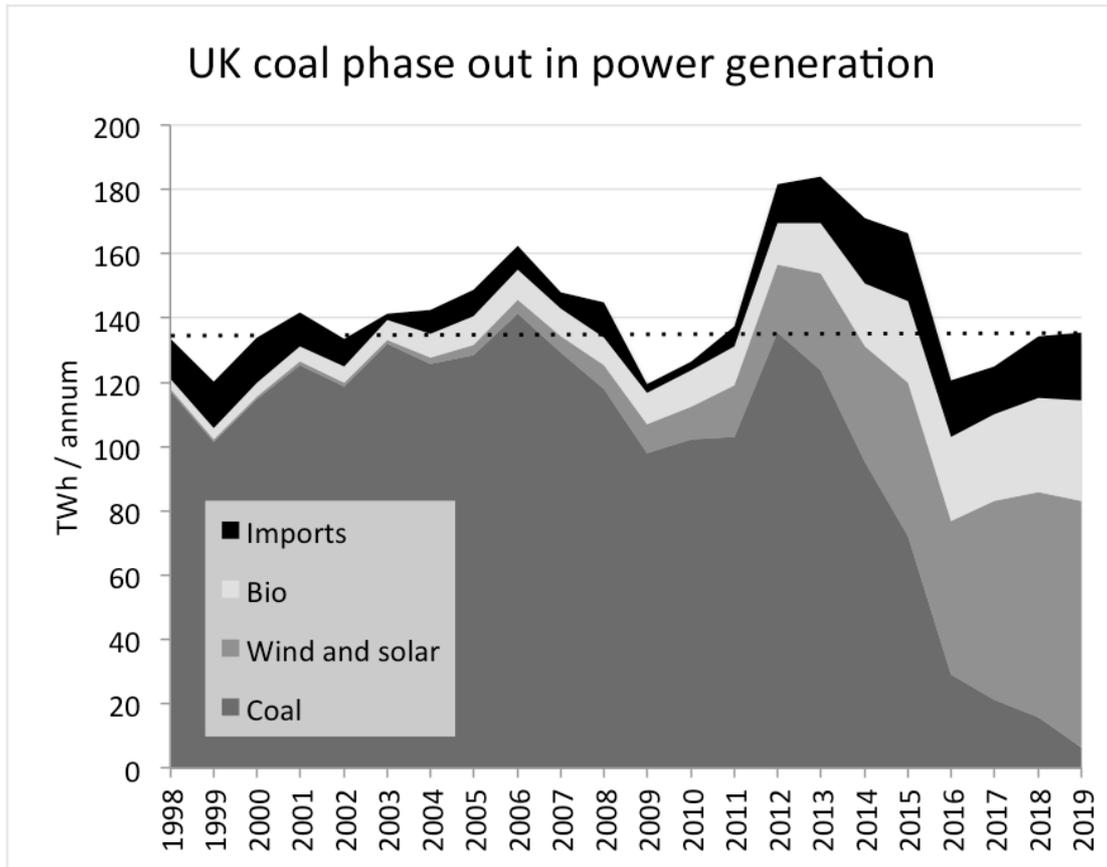

Figure A4  UK power generation by four sources [62] shows how the intentional phasing out of coal is replaced exactly by an increase in wind and solar, bio-mass and imports.

Imports come mainly from high nuclear France and may be regarded as having low $CO_2$ intensity. Wind and solar may also be viewed as largely emissions free so long as the $CO_2$ emissions embedded in imported solar panels and wind turbines are included in the accounting as described above. Burning wood pellets in former coal fired stations such as Drax is certainly not emissions free for two reasons. First, the supply chain for US wood pellets includes chain saws, trucks, a pellet mill, drying, ships and more trucks [46]. This represents a vast investment in steel and liquid fuel. It seems likely that as much energy will be consumed in the production of wood pellets as will ultimately be produced through their combustion [17]. Second, deforestation emits gasses from soil and the wood is burned emitting $CO_2$. It takes decades for the forest to re-generate.

A3.2   Transport, Business, Residential and Public (TBRP)

The emissions from transport (mainly cars, vans, trucks, trains and domestic air travel), business (manufacturing, refrigeration and air conditioning), residential (home heating) and public (heating schools, hospitals etc.) are dominated by combusting fossil fuels producing $CO_2$ [52]. Combined with Energy supply, these five sectors make up the lion's share of UK fossil fuel use. The CO2 emissions reported by BP exclusively relate to fossil fuel combustion [6] and Figure A3 shows a good match with the combined emission of these 5 sectors.

Emissions in the combined TBRP sectors have fallen 16% in the period 1990-2018 and there is a long way to go to decarbonize them. Electrification of space heating and transport is simultaneously straightforward and a huge challenge. Heat pumps and resistance heaters are an efficient but expensive way to heat space [58] and electric car technology is here but it too is expensive owing to the materials used and is limited by battery range. One of the greatest barriers to electrification is the need to generate a lot more carbon free electricity. Renewables cannot do this since they are and will remain dependent upon natural gas to provide load following service and to cover for windless and sunless periods. Nuclear fission power alone provides the only viable option for a country like the UK [12, 58].

A3.3   Agriculture, Waste Management and Industrial Processes (AWI)

Agricultural emissions include methane emanations from livestock (56%), nitrous oxides from soils (31%) and $CO_2$ from burning fossil fuel in tractors etc. (10%). Emissions from agriculture are relatively flat but have declined by ~ 16% since 1990. Emissions from waste management are dominated by methane from landfill (92%). These are deemed to have fallen by 69% since 1990 stemming from a program to send biodegradable waste to fermentation facilities where the methane is collected and used to generate electricity and by collecting landfill gas, also used to generate electricity. Methane emissions from waste therefore have been used to substitute for some fossil fuel methane. Furthermore, the methane is converted to $CO_2$, a less potent greenhouse gas.

In 1990, emissions from industrial processes comprised three roughly equal components: 1) $CO_2$ from cement production, 2) nitrous oxide from adipic acid production and 3) flourine gasses associated with the production of halocarbons (F gasses in [52]). $CO_2$ from cement production has remained relatively constant since 1990 while nitrous oxides and F – gases have all but been eliminated resulting in an 83% fall in this category. It is not clear whether the elimination of nitrous oxides and F gases is real, i.e. their production has been eliminated from the industrial process, or whether these have been offshored, i.e. they are still being produced, just not in the UK. [all information from 52]

Emissions from the AWI group have fallen by 57.7% since 1990.

A4     Discussion

The IPCC has decreed that, in order to limit global warming to <1.5°C, global emissions of greenhouse gasses should fall to zero by ~2050 [1]. Many western countries, such as the UK, have signed up to this challenge, enshrining in law the new net zero ambition. Governments like the UK have done so with a false picture of progress made to date in reducing emissions since 1990 that stems mainly from a failure to include the significant emissions embedded in imported goods in the accounting. In the UK, excluding embedded emissions, the computed fall from 2009 is 42% leaving a further 58% to be eliminated over the next 30 years. Including embedded emissions and the fall is a more modest 23% leaving a formidable 77% remaining. The scale of this challenge needs to be seen in the context that emissions reductions to date have emerged from plucking the low hanging fruit. The remaining 77% will be much harder to eliminate.

The low hanging fruit has included the closing down of coal-fired electricity generation and replacing this with a mixture of renewable energy, imports and wood pellets imported form N America. Renewable energy devices and wood pellets have significant embedded emissions and on-going failure to include these in the accounting, put bluntly, creates a false set of accounts. The significant reduction in methane emanations from landfill may be viewed as a major success but that fruit is already picked and little further progress can be made on that front. The elimination of F gases and nitrous oxides from industrial processes is also an achievement, but it needs to be verified that these emissions have not simply been offshored.

Figure A5 summarises the net zero challenge remaining for the UK, based on data as it stood in 2018. In order to chart the way ahead, in this discussion, we will simplify to three major components. First, $CO_2$ emissions in the categories of transport, business, and residential and public heat are all linked to combusting fossil fuels, on site, i.e. petroleum products. The technologies exist to substitute both natural gas (heat) and liquid fuel (transport) with electrical equivalents by way of heat pumps, resistance heaters and electric vehicles. Electrification of our energy system carries much virtue since electrical heating and electric cars are more energy efficient than their fossil counterparts [61] and tend to be far less polluting in cities although mining the many materials required to construct renewable devices may cause environmental harm elsewhere. These advantages are offset by considerably higher capital costs. The operational costs will vary according to the price of electricity used in their operation.

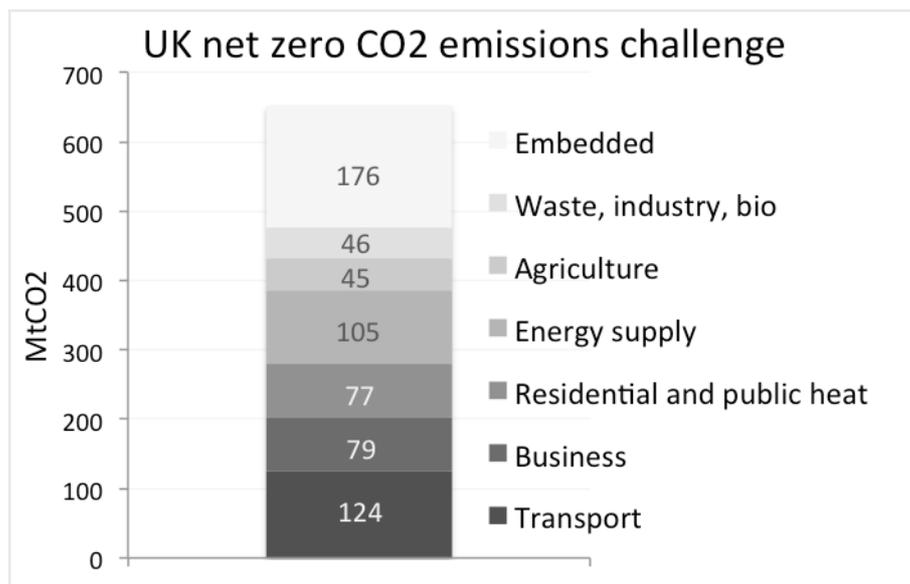

Figure A5    Summary of UK emissions as of 2018, extracted from Figure A3 based on [54] except for the embedded category that is based on [52] and biomass that is based on [55].

One inescapable fact flowing from electrification of society is that we will use a lot more electricity that brings us to our second discussion point. If the emissions associated with transport, business, residential and public heat are eliminated via electrification, then the burden is simply transferred to electricity supply. Here it is widely accepted that there are two viable routes of either renewable energies or nuclear fission. Since the whole decarbonistaion agenda tends to be driven by the environmental movement, who overwhelmingly tend also to be anti-nuclear, the current path tends to be driven by expansion of renewable energies.

It is these divergent pathways that prompted the energy quality MCDA study based on expert judgment that forms the main body of this paper. The study was designed to provide a holistic multi-criteria perspective of the quality of electricity delivery systems, as opposed to the one-dimensional, $CO_2$ reduction view taken by many policy makers today. Our panel of experts judged that, of 13 technologies reviewed, the leading three technologies in rank order were 1) nuclear fission, 2) CCGT and 3) hydro-electric power. Coming in last place were 11) solar PV, 12) biomass and 13) tidal lagoon. Wind power came in 8$^{th}$ place. Governments in Europe and the UK need to ask why they have chosen a pathway following expensive, subsidized, unreliable and environmentally destructive technologies.

We therefore question the claims that the 100% renewables pathway is either viable or desirable even although models have been published that suggest a 100% renewables future is attainable [9, 10]. It is beyond the scope of this Appendix / Paper to provide details, but put simply, to convert intermittent renewable energy (wind and solar) to dispatchable power requires vast investment, not only in renewables devices but in energy storage and inter-connectivity. The scale of storage is so large, it remains to be ascertained whether it is either desirable or attainable. The natural environment in areas with high renewable energy penetration has already been seriously degraded and we fail to see the argument that this is an environmentally friendly route to follow. We therefore strongly advocate that, if UK and OECD governments are really serious about zero carbon in 2050 and in environmental protection, nuclear fission is currently the only viable route. There are multiple nuclear fission technologies under development that further reduce the already vanishingly small risks [12].

Finally the largest slice of UK emissions that remain in 2018 are those embedded in imported goods. There are only two possible routes to follow if these are to be eliminated. The first is to ensure that the $CO_2$ reduction pathway of the countries manufacturing goods on our behalf parallels that in the UK. Much of the manufactured imports to the UK originate in east and Southeast Asia where environmental standards are often below those in the west. One reason for offshoring manufacturing in the first place was to access lower costs brought about in part by less regulation. The UK would have no control over the $CO_2$ intensity of imported goods. The only viable alternative therefore is to repatriate manufacturing to the UK. The Covid–19 pandemic has potential to reverse globalization, and so a plan to repatriate manufacturing would be aligned with the existing trend. This plan also has much to commend it by way of providing employment in the UK. One disadvantage of course is that energy (electricity) demand and prices would rise, but the UK would have control over the provenance of that electricity.

**Appendix B – Weaknesses of the NEEDS MCDA approach**

The European Union (EU) project called *New Energy Externalities Developments for Sustainability* (NEEDS) ran from 2004 to 2009 and had the objective of identifying the most sustainable advanced electricity generating technologies for France, Germany, Italy and Switzerland in the year 2050. The project employed two different methodological approaches, namely 1) the baseline cost approach that combined internal direct costs and indirect external costs and 2) Structured Multi-Criteria Decision Analysis (MCDA) based on soliciting judgments and preferences from an expert group. This project intrigued us for a number of reasons. First, in early 2018 we had access to the readership of a high quality energy policy blog called Energy Matters [38] and saw the possibility to conduct an analogous study using the judgment of the blog readership. Second, we were surprised by the fact that the MCDA approach gave results quite different to the traditional baseline costs approach. Indeed the MCDA approach placed CdTe solar PV in first place and Solar Thermal power in number 2 in the rank order. Since solar thermal power is barely functional today in Spain, we were intrigued by its perceived advantageous functionality in the year 2050, especially in countries like Germany and Switzerland that seldom have sufficient solar resource for a large solar thermal plant to function. Third, out of 7 common technologies (nuclear, gas, coal, biomass, wind, solar PV and solar thermal), the basic cost approach placed nuclear (not solar thermal) in top rank position while the MCDA approach placed nuclear in position 6 out of 7. In fact, the MCDA approach placed these 7 technologies virtually in the reverse rank order of the basic cost approach. Something seemed to be far wrong with this work that was meant to inform energy policy in Europe. This Appendix sets out our analysis of the weak methodology and multiple biases of the approach followed that seem to be intended to produce the outcomes reached.

In 2018, we asked Stephan Hirschberg of the Paul Scherrer Institute to review an earlier version of this manuscript and he responded with lengthy commentary that we found as difficult to follow as the methodology employed by NEEDS. For example:

> *This conclusion is based on a fundamental misunderstanding. The application of MCDA was motivated by criticism of the total cost approach as aggregated measure of sustainability. First, the scopes of the two types of evaluations are much different as among other factors the total cost approach in particular reflects a very limited treatment of the social dimension of sustainability while MCDA apart from having the capability to cover a broad range of social factors has also more flexibility in terms of accommodating some environmental and economic aspects that are not directly amenable to quantification. Second, the total cost approach assigns equal importance to the internal and external costs while MCDA is open to assignment of different preferences to the various related criteria/indicators. Thus, it was given from the beginning of NEEDS that the results obtained using the two methods will be quite different. The background to carrying out MCDA was that total costs (an approach absolutely necessary for cost-benefit analysis) lack general acceptance as a measure of sustainability, particularly in the case of nuclear energy. The*

> *main objection concerns the lack of addressing the social dimension of sustainability, certainly a widely debated issue in the nuclear context.*

From this, it is easy to develop the view that the traditional and accepted Total Costs approach gave results that were perceived to be faulty and that a new approach was required that gave results perceived to be correct. The key problem perceived with Total Costs was a lack of emphasis given to the Social Dimension. Looking to the accepted definition of sustainability:

> *Sustainability focuses on meeting the needs of the present without compromising the ability of future generations to meet their needs. The concept of sustainability is composed of three pillars: economic, environmental, and social.*

Stephen Hirschberg criticized our work for not giving sufficient weight to social issues, claiming that we had only one social metric in the way of health. In fact we had one category called health under which we had two criteria of fatalities and chronic illness and then several metrics that varied depending on the technology being evaluated. Scoring was done at the criteria level and with equal weight, direct social criteria accounted for 2/12 of our total score. However, important social criteria are embedded in other categories such as environment – the building of wind farms and nuclear power stations close to housing can cause social distress for current and future generations – and cost – high cost of electricity can be one of the socially most damaging characteristics of electricity technologies. Beyond that, we fail to see what social dimension we have missed that might compromise the ability of future generations to meet their needs.

The NEEDS reports of interest are organized into two parts called Survey 2 (or 2ask) [63] and Survey 3 [19]. Survey 2 was where potential stakeholders were asked about the proposed design of the survey to be carried out as Survey 3.

## B1   NEEDS – whose judgment was solicited and whose judgment was given?

The NEEDS MCDA analysis was conducted via a highly complex – state of the art – web based questionnaire that we will get to shortly. An initial questionnaire was sent to 2848 individuals referred to as stakeholders who were existing contacts of the NEEDS project team. The target stakeholders came mainly from 4 countries, namely France, Germany, Italy and Switzerland (Table 1).

| Country | # of individual stakeholders |
|---|---|
| France | 105 |
| Germany | 659 |
| Italy | 435 |
| Switzerland | 1120 |
| Other countries | 529 |
| **TOTAL** | **2848** |

The NEEDS project did target a diversity of stakeholders as follows:

- Energy supplier
- Energy consumer
- Non-governmental organization
- Government energy and environmental organisations
- Regulator / government authorities
- Associations
- Politicians
- Researcher / academia
- Consultants
- Other

The initial survey (Survey 2) was designed to get feedback from potential participants on the questionnaire design leading to a feedback loop where the questionnaire may be modified or redesigned.

Of the 2848 persons invited to participate, 660 visited the web site and 275 filled out the questionnaire representing a 9.7% response rate. The response rate was also heavily skewed towards academics with 61.5% of respondents coming from this group (Figure B1).

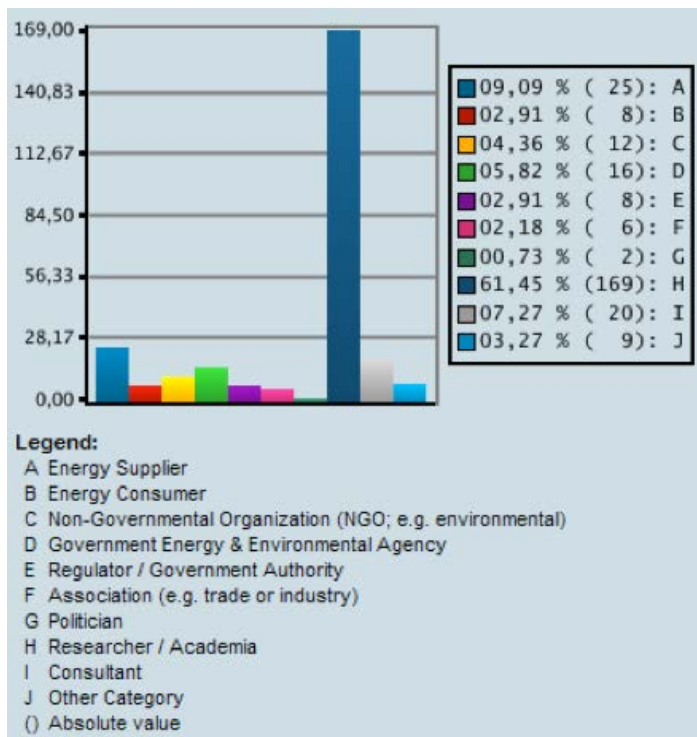

Figure B1 Category distribution of 275 out 2848 stakeholders invited to participate in the NEEDS questionnaire survey. The Y-axis shows the number of participants in each category. We do not know why this is quoted with two decimal precision nor why the axis interval is 28.17.

We can in part sympathize with the low response rate since we were also rather disappointed with the uptake of our own survey. But at this point, the managers of this multi-Euro flagship project should have realized that something was wrong and that they should not proceed with a stakeholder database so heavily skewed towards academics, who, as we discuss below, could be heavily biased in their outlook.

### B2     Why was the response rate so poor?

Bear in mind that, so far, stakeholders are only reviewing the questions to be asked and the questionnaire design. The difficult task of answering these questions as applied to 26 different energy technologies has yet to come. The questionnaire had 60 questions; questions 1 to 6 were about the individual's profile; questions 7 to 46 were the technical questions; 47 to 51 was general feedback on the questions; questions 52 to 55 were more personal information and 56 to 60 was feedback on the questionnaire.

While the NEEDS report claimed the approach had a high approval rating from the participants, it needs to be born in mind that >90% of those invited declined to participate. One published comment from a participant said this:

*The indicators must also be understood by "normal people", right?*

This alludes to the impossible complexity of some of the questions asked [19]. We give three examples below.

*Q9     Mineral resources (ores) This criterion quantifies the use of selected scarce metals used to produce 1 kWh of electricity.*

*It is based on the life cycle impact assessment method "CML 2001". The use of all single metals is expressed in antimony –equivalents, based on the scarcity of their ores relative to the reference ore (antimony). The indicator covers each complete electricity generation technology chain.*

*Indicator: Weighted total consumption of metallic ores [kg(Sb-eq)/kWh]*

How many respondents know that the most common antimony ore is a sulphide called stibnite and that most of the global production comes from a single mine in Hunan Province in China? It is quite simply impossible for even an expert to conceptualise the relative scarcity of all scarce metals, normalize these to antimony and to then work out how much scarce metal is used in, for example, a nuclear power plant as a proportion of the electricity produced over its life time. Of course, there may already be publications summarizing this information, but it cannot be reasonably expected that the range of stakeholders, who mainly declined to participate, should know this information.

*Q13     Impacts from normal operation / acidification and eutrophication*

*This criterion quantifies the loss of species (flora and fauna) due to acidification and eutrophication caused by pollution from production of 1 kWh of electricity. The "potentially damaged fraction" (pdf) of species is multiplied by the land area and years for each complete electricity generation technology chain.*

*Indicator: Impacts of air pollution on ecosystems [PDF*$M^2$a/kWh]*

This is one of a number of questions biased towards "normal operation" that excludes pollution caused by construction and installation of power stations where renewable technologies perform less well and this question seems specifically targeted at coal that is the only technology to produce significant quantities of noxious exhaust gas. But not all coal-fired power stations are born equal. In the developing world, untreated exhaust gasses may be released directly into the atmosphere causing great harm to health and to the surrounding environment. In the developed world, power stations may be equipped with a variety of exhaust scrubbers to remove $SO_2$ and NoX. Acidification may also be closely linked to the S content of coal. It is therefore impossible to answer this question without advance knowledge of the kind of coal technology, the chemistry of the coal and the environmental setting of the power station – desert versus virgin temperate forest.

Unknown to the respondents at this stage, they would go on to be asked to evaluate 9 different types of coal / lignite technology and this question may be helpful in discriminating between these. But without advance knowledge about which specific technology is to be judged, it is impossible to evaluate this question.

*Q34    Expert-based risk estimates for normal operation / non-fatal illnesses due to normal operation.*

*This criterion is based on the increased rate of sickness or morbidity due to normal operation of the electricity generation technology and its associated energy chain. It is measured in the years of life affected by disabilities (disability affected life years or DALY) suffered by the entire population, compared to their expected health without the technology in question.*

*Indicator: Morbidity due to normal operation [DALY / kWh].*

This is another question biased towards normal operation that skips over morbidity in poorly regulated mining operations for rare-earth elements and cobalt, essential to manufacture of certain renewable technologies. It asks specifically about associated energy chain, which may include coal mining, but excludes supply chain that would include the aforementioned special metals required in the manufacture of actuators and batteries.

The question requires the subject to know, compute or estimate the duration of unspecified disabilities for "the entire population" – and does not specify which population? A kWh is a tiny unit of electricity suggesting that perhaps disability should be measured in hours or days per unit that may aggregate to months and years for larger quantities such as MWh and GWh. Once again, unless a subject has recently read a report on this topic, the question is quite impossible to answer. Was it anticipated that ordinary consumers and politicians should know and understand such details?

B3      Soliciting Political Opinion

The NEEDS management, perhaps disappointed with the minuscule participation of politicians in Survey II (Figure B1), organized a special targeting of politicians via an organization called Globe Europe who solicited participation from 1500 parliamentarians from 27 EU member states and states like Norway, Iceland and Turkey. 37 of these visited the web site and 3 completed the questionnaire (0.2% of the original 1500).

The NEEDS management team's appraisal:

> *The reasons for this very low response rate may be manifold, including the large number of requests to which politicians are expected to provide their opinion. As a consequence, no separate evaluation of Survey II results for European politicians was undertaken.*

The NEEDS team seem oblivious to the fact that it is quite impossible for experts, let alone lay people, to evaluate their impossibly difficult questions.

B4      Bias of stakeholder responders

While the NEEDS team no doubt set out with honorable intentions to get a broad base of participants in their survey, the fact remains that 61.5% of eventual participants were academics. It seems likely that this came about because academics feel they have more free time to participate in a lengthy survey and perhaps feel better equipped to tackle the impossibly difficult questions. Those who did respond in the main thought the level of difficulty was appropriate! Among the academic responders, the following disciplines were represented:

| | |
|---|---|
| Energy systems analysis | 19.3% |
| Renewables | 9.5% |
| Nuclear | 11.6% |
| Energy other | 6.2% |
| Non-energy | 11.3% |
| | |
| Total | 57.9% |

We do not know why this does not add up to 100% (or to 61.5%) and we cannot fathom that non-energy disciplines could begin to understand the impossible questions.

One issue with academic bias is a trend in recent decades towards energy researchers being recruited to ubiquitous sustainability departments that by their very nature assume at the outset that renewable energy is the only sustainable option. Thirty years ago, energy researchers in academia may have been heavily skewed towards petroleum exploration and production, and indeed, in petroleum rich nations such departments still exist. However, since neither Germany, Switzerland or France ever had significant petroleum resources, this important group of energy experts are excluded from the mix – perhaps included in "energy other". This is an extraordinary bias given that fossil fuels still provide 84.5 % of primary energy consumed on Earth in 2019 [6]. New renewables account for only 5.0%. The big drive for new renewables began around 1985 and it has therefore taken 35 years to reach 5%. Those academics and policy makers now driving for net zero carbon by 2050 founded on renewable energy need to take stock of these realities.

### B5    Bias of questionnaire

One of the respondents offered this comment at the end of his questionnaire:

> *The questionnaire is one-sided in favor of renewable energies. Even fusion energy would be rated very poorly with these criteria.*

For reasons given below, we come to the conclusion that the questionnaire is extremely biased and designed to produce the specific outcome it produced, reversing the order of conventional cost based analysis, moving nuclear from the top to the middle of the rank order. This raises serious questions on the quality of this research and on the justification for the use of a large sum of EU funds.

The 40 difficult to near-impossible questions were to be applied to 26 existing and emerging energy technologies (see below), giving 1040 responses in total. The majority of the emerging technologies are barely known among energy experts, making it doubly difficult to answer questions in a meaningful way. For example, how is a layman supposed to evaluate the antimony equivalent use of scarce metals (Q9) to a Molten Carbonate Fuel Cell using Natural Gas 0.25 MW (Technology 16)?

#### Nuclear
1 European Pressurized Reactor
2 European Fast Reactor

#### Coal and lignite
3 Pulverized Coal
4 Pulverized Coal with post combustion Carbon Capture and Storage
5 Pulverized Coal with oxyfuel combustion and Carbon Capture and Storage
6 Pulverized Lignite
7 Pulverized Lignite with post combustion Carbon Capture and Storage
8 Pulverized Lignite with oxyfuel combustion and Carbon Capture and Storage
9 Integrated Gasification Combined Cycle coal
10 Integrated Gasification Combined Cycle coal with Carbon Capture and Storage
11 Integrated Gasification Combined Cycle lignite
12 Integrated Gasification Combined Cycle lignite with Carbon Capture and Storage

#### Natural gas and biogas

13 Gas Turbine Combined Cycle
14 Gas Turbine Combined Cycle with Carbon Capture and Storage
15 Internal Combustion Combined Heat and Power
16 Molten Carbonate Fuel Cells using Natural Gas 0.25 MW
17 Molten Carbonate Fuel Cell using wood derived gas 0.25 MW
18 Molten Carbonate Fuel Cells using Natural Gas 2MW
19 Solid Oxide Fuel Cells using Natural Gas 0.3 MW

Bio mass
20 Combined Heat and Power using short rotation coppiced poplar
21 Combined Heat and Power using straw

Solar
22 Photovoltaic, ribbon crystalline Silicon - power plant
23 Photovoltaic, ribbon crystalline Silicon - building integrated
24 Photovoltaic Cadmium Telluride – building integrated
25 Concentrating thermal – power plant

Wind
26 Offshore Wind

We now provide a list of all 35 technical questions / preferences in the 2ask questionnaire (the same questions as in Survey 3 that comprised the actual analysis) in order to illustrate how these questions are biased against fossil and nuclear and in favor of renewables. The superscripts refer to the type of bias that are listed at the end of the questions.

Q7    *Fossil fuels: This criterion measures the total primary energy in the fossil resources used for the production of 1 kWh of electricity. It includes the total coal, natural gas and crude oil used for each complete electricity generation technology chain.*[1]

Q8    *Uranium: This criterion quantifies the primary energy from uranium resources used to produce 1 kWh of electricity, It includes the total use of uranium for each complete electricity generation technology chain.*[1, 2, 4]

Q9    *Metal ore: This criterion quantifies the use of selected scarce metals used to produce 1 kWh of electricity.*

Q10    *$CO_2$ emissions: This criterion includes the total of all greenhouse gasses expressed in kg $CO_2$ equivalent to produce 1 kWh of electricity.*

Q11    *Biodiversity: This criterion quantifies the loss of species (flora and fauna) due to the land used to produce 1 kWh of electricity.*

Q12    *Ecotoxicity: This criterion quantifies the loss of species (flora and fauna) due to ecotoxic substances released to air water and soil to produce 1 kWh of electricity.*

*Q13     Air pollution: This criterion quantifies the loss of species (flora and fauna) due to acidification and eutrophication caused by pollution from production of 1 kWh of electricity.*

*Q14     Hydrocarbons: This criterion quantifies large accidental spills of hydrocarbons (at least 10,000 tonnes) which can potentially damage ecosystems.* [1, 3, 9]

*Q15     Land contamination: This criterion quantifies land contaminated due to accidents releasing radioactive isotopes.* [1, 2]

*Q16     Chemical waste: This criterion quantifies the total mass of special chemical wastes stored in underground repositories due to the production of 1 kWh of electricity.*

*Q17     Radioactive waste: This criterion quantifies the volume of medium and high level radioactive wastes stored in underground repositories due to the production of 1 kWh of electricity.* [1, 2]

*Q18     Generation cost: This criterion gives the average generation cost per kWh. It includes the capital cost of the plant, the fuel and operation and maintenance costs.*

*Q19     Direct jobs: This criterion gives the amount of employment directly related to building and operating the generating technology, including the direct labor involved in extracting or harvesting and transporting fuels. Indirect labor is not included. Unit = person-years / GWh.* [4]

*Q20     Fuel autonomy: Electricity output may be vulnerable to interruptions in service if imported fuels are unavailable due to economic or political problems related to energy resource availability.* [1, 9]

*Q21     Finance risk: Utility companies can face a considerable financial risk if the total cost of new electricity generating plant is very large compared to the size of the company. It may be necessary to form partnerships with other utilities or raise capital through financial markets.* [5, 9]

*Q22     Fuel sensitivity: The fraction of fuel cost to overall generation cost can range from zero (solar PV) to low (nuclear power) to high (gas turbines). This fraction therefore, indicates how sensitive the generation costs would be to a change in fuel prices.*

*Q23     Construction time: Once a utility has started building a plant it is vulnerable to public opposition, resulting in delays and other problems. This indicator therefore gives the expected plant construction time in years. Planning and approval time is not included.* [5, 9]

*Q24    Marginal cost: Generating companies "dispatch" or order their plants into operation according to their variable cost, starting with the lowest cost base-load plants up to the highest cost plants at peak load periods. This variable (or dispatch) cost is the cost to run the plant. [6]*

*Q25    Flexibility: Utilities need forecasts of generation they cannot control (renewable resources like wind and solar), and the necessary start-up and shut-down times required for the plants they can control. This indicator combines these two measures of planning flexibility, based on expert judgment. [7]*

*Q26    Availability: All technologies can have plant outages or partial outages (less than full generation), due to either equipment failures (forced outages) or due to maintenance (unforced or planned outages). This indicator tells the fraction of the time that the generating plant is available to generate power. [7]*

*Q27    Secure supply: Market concentration of energy suppliers in each primary energy sector that could lead to economic or political disruption. [7, 9]*

*Q28    Waste repository: The possibility that storage facilities will not be available in time to take deliveries of waste materials from whole life cycle.*

*Q29    Adaptability: Technical characteristics of each technology that may make it flexible in implementing technical progress and innovations. [8]*

*Q30    Conflict: Refers to conflicts that are based on historical evidence. It is related to the characteristics of energy systems that trigger conflicts. [9]*

*Q31    Participation: Certain types of technologies require public, participative decision-making progress, especially for construction or operating permits. [9]*

*Q32    Mortality:  Years of life lost (YOLL) by the entire population due to normal operation compared to without the technology.*

*Q33    Morbidity: Disability adjusted life years (Daly) suffered by the whole population due to normal operation compared to without the technology.*

*Q34    Accident mortality: Number of fatalities expected for each kWh of electricity that occurs in severe accidents with 5 or more deaths per accident. [5, 9]*

*Q35    Maximum fatalities: Based on the reasonably credible maximum number of fatalities for a single accident for an electricity generation loading technology chain. [5, 9]*

*Q36    Normal operation: Citizen's fear of negative health effects due to the normal operation of the electricity generation technology. [10]*

*Q37 Perceived acc: Citizens perception of risk characteristics, personal control over it, scale of potential damage, and their familiarity with the risk.* [5, 10]

*Q38 Terror-potential: Potential for a successful terrorist attack on a technology. Based on its vulnerability, potential damage and public perception of risk.* [2, 5, 7, 9]

*Q39 Terror effects: Potential maximum consequences of a successful terrorist attack. Specifically for low-probability high-consequence accidents.* [2, 5, 7, 9]

*Q40 Proliferation: Potential for misuse of technologies or substances present in the nuclear electricity generation technology chain.* [1, 2]

*Q41 Landscape: Overall functional and aesthetic impact on the landscape of the entire technology and fuel chain. Note: Excludes traffic.*

*Q42 Noise: This criterion is based on the amount of noise caused by the generation plant, as well as transport of materials to and from the plant.*

1. Not applicable to all technologies
2. Applicable to only nuclear
3. Applicable only to oil and gas
4. Unclear if high or low value is good or bad
5. Biased against large plant
6. Inaccurate and incomplete
7. Plain biased formulation
8. Mainly applicable to defective technologies that don't work well. This turns deficiency into merit.
9. Repetitive and duplication
10. When asked about public opinion, risk that stakeholder will substitute his / her views for those of the public

In our reading, the whole questionnaire is loaded against nuclear power and fossil fuels and in favor of renewables and it is therefore not surprising that nuclear fared badly and renewables fared well as a result. One further blanket criticism is that 1 kWh is used as the reference point for many questions. This is a tiny amount of energy. For example, a 1 GW nuclear power station may generate 1*24*365.25*40*0.9 = 316 TWh of electricity over its life (GW*hrs*days*years*mean load factor). To answer Q34, we may assume that, in the unlikely case where a nuclear plant has an accident that kills 5 people once in its 40 year life, this works out at $5/(316 \times 10^9) = 1.6 \times 10^{-11}$ death/kWh. This is an impossible scale to work with.

## B6 Answering the questions / recording preferences

At this juncture, we have to confess that we find it difficult to fathom exactly how the NEEDS Survey 3 worked. It seems that many of the stakeholders (i.e. participants) had the same problem:

> *In order to obtain sustainability preference information from as many*

> *stakeholders as possible, it was decided early in the NEEDS project to use an interactive, web-based MCDA application for Survey 3 (also called the MCDA Survey). The purpose was also to create a MCDA tool that would not just solicit the stakeholder's preferences on sustainability criteria, but also provide back the stakeholder's personal technology ranking results in an interactive way that would provide an iterative learning process for the user. The hope was therefore that the feedback of individualized results would provide an incentive or reward for the stakeholder to participate, since the complexity of the survey posed a learning curve that proved to be a barrier for many participants. [in 19, section 3.6]*

Stakeholders answer two questions per criteria per technology. The first is a ranking from best to worst and, from this, it becomes evident that stakeholders do not actually need to perform deterministic calculations to answer the impossibly difficult questions but may simply rank a technology based more on "gut feel". This part of the process, therefore, is rather similar to the methodology we used. The second question is to state a "preference" that is a measure of how important a criteria is. This is effectively a weight and a highly aggressive point scale was used ranging from 1/16 through 1/4, 1/2, 1, 2, 4, and ending with 16. This provides leverage with a dynamic range of 256 where the stakeholder has the discretion to pour all the weight on a handful of criteria. In fact, it is an eight point scale since a zero importance option was also available, which, if selected, excludes that criteria for the relevant technology. The following passage therefore concerns us:

> *The decision-maker can see the consequences of his preferences, see whether or not the results agree with any preconceived ideas of the desired outcome, and hopefully reconcile any differences to a more consistent understanding. (in [19] section 2.3).*

This procedure has pros and cons. On the one hand, as it seemed intended by the organisers of the study, each stakeholder can see the consequences of their answers in terms of how they lead to a final ranking of the different technologies and whether this ranking is **compatible with their "preconceived ideas of the desired outcome".** The stakeholder is thus empowered to modify the weights at their discretion to converge to a ranking outcome that they feel comfortable with, i.e., which represents most faithfully their understanding and appreciation of the different technologies. On the other hand, this procedure is likely to anchor the answers of the stakeholder on their biases, which are explicitly elicited through this "reconciliation" process. Rather than providing a series of compartmentalized answers for each of the quantitative and technical question and let the designers independently analyze them [67, 68], the stakeholder is enabled to modify the weight of each answer in reference to the whole "desired" output. This effective grouping has the danger of producing a "carryover" effect among the responses to the multiple questions that are aggregated as a final ranking [69]. This might artificially inflate the biases by making the stakeholder effectively both judge and jury. This could be construed as fiddling with the buttons until any preconceived **preference is produced.** It seems that a more streamlined approach might have been to simply ask the stakeholders to arrange the technologies in a rank order that matched their preconceived preferences.

Figure B2 illustrates the state-of-the-art user interface. On the criteria chart, the various technologies are located along the bar according to whether they are worst or best (see figure B5). On the User Preferences panel, the various criteria are weighted according to their importance. In figure B3, fossil fuels are weighted "6", much more important than average that has a weight of 4. Using fossil fuels is arbitrarily deemed to be bad (worst), thus most fossil technologies will fall to the left in Figure B5 and since this is said to be "much more important than average", most fossil technologies will be significantly downgraded in the assessment, apparently fully neglecting that coal and natural gas offer some of the cheapest and most reliable electricity. It should be clear that we are not supporting coal or natural gas specifically and we have no agenda. Here, we just outline some issues that we deem important in assessing the reliability of the conclusions of the NEEDS project.

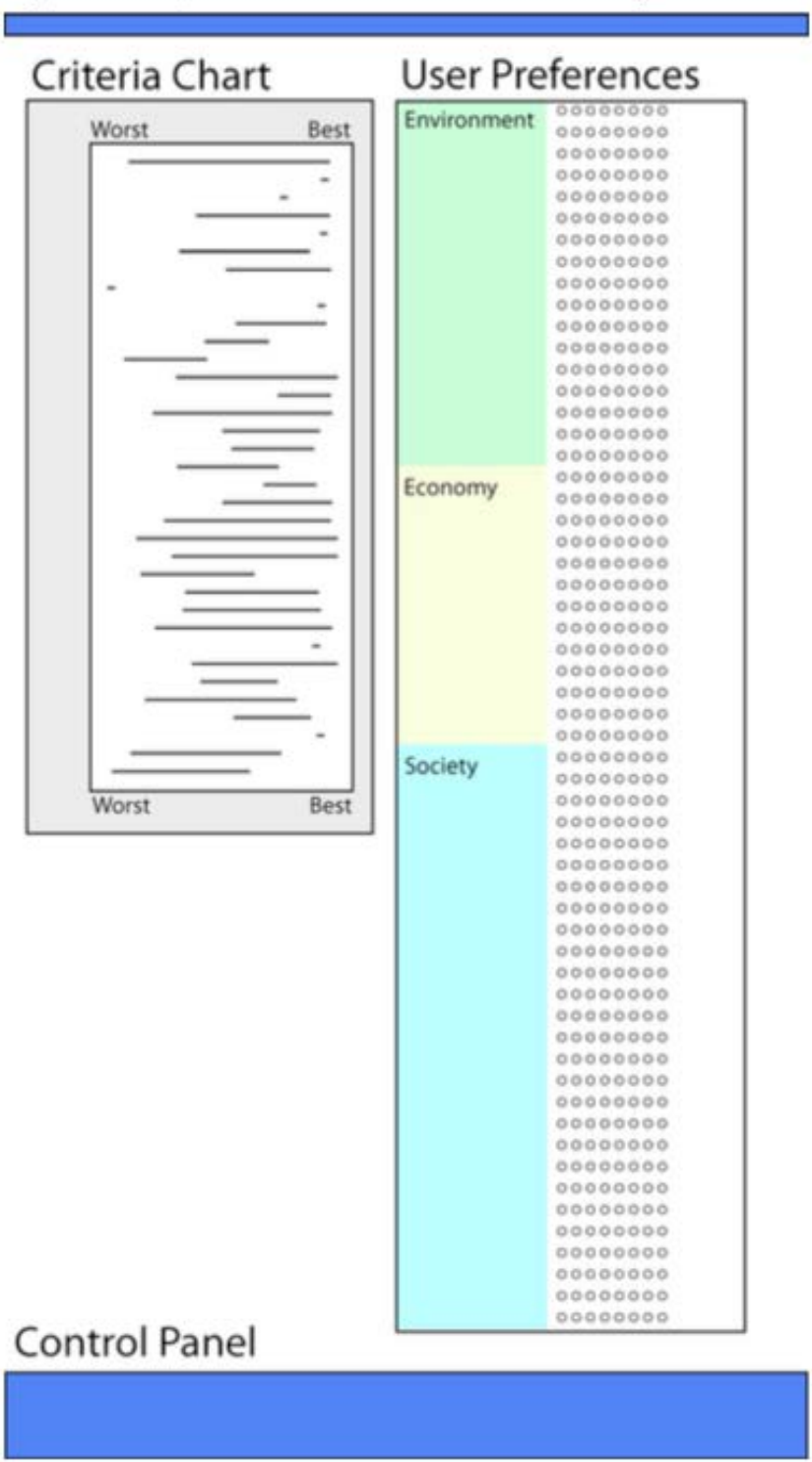

Figure B2: Figure 5 in [19]. The Survey 3 user interface. It seems that all technologies are entered into a single line on the Criteria chart where each line represents one criterion. This allows the user to sort these into a rank order (see figure B5).

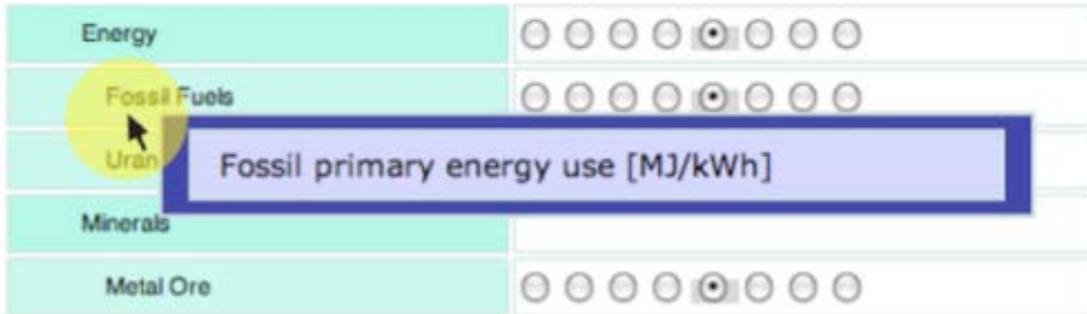

Figure B3 Figure 7 in [19] shows how the amount of fossil fuel used to generate a kWh of electricity is weighted in the Preferences table. In this case, position 5 = a weight of 1.

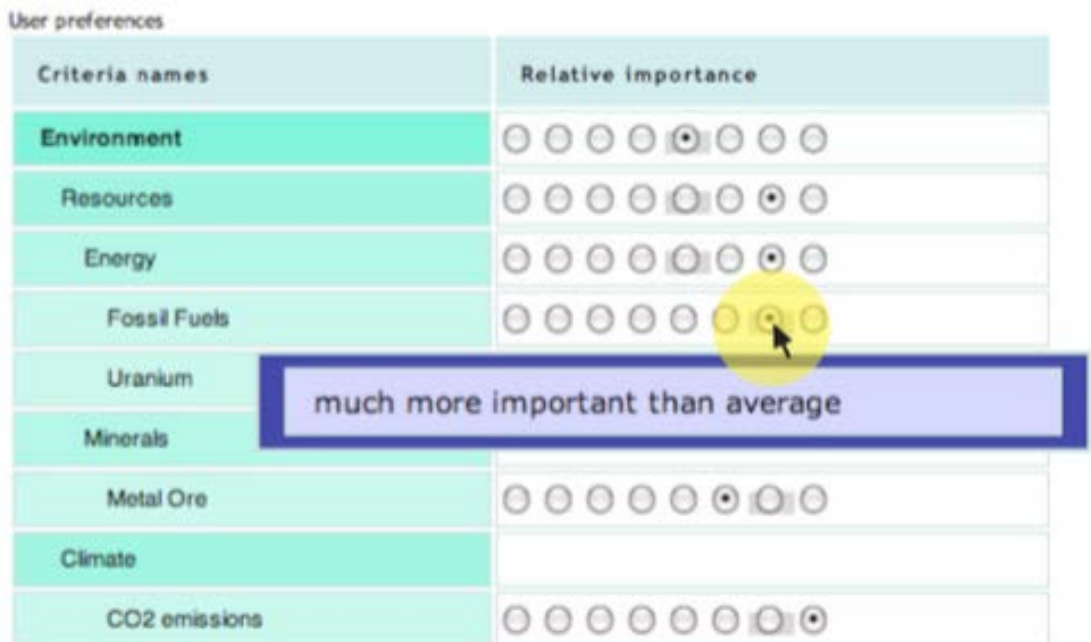

Figure B4 figure 8 in [19] shows a second pop up saying that position 7 is much more important than average. The seven buttons range from "vastly less" to "vastly more" important than average and have relative weights of 1/16, 1/4, 1/2, 1, 2, 4, and 16. The left most button (button 8) has a weight of zero, and if selected, excludes that criteria for the relevant technology.

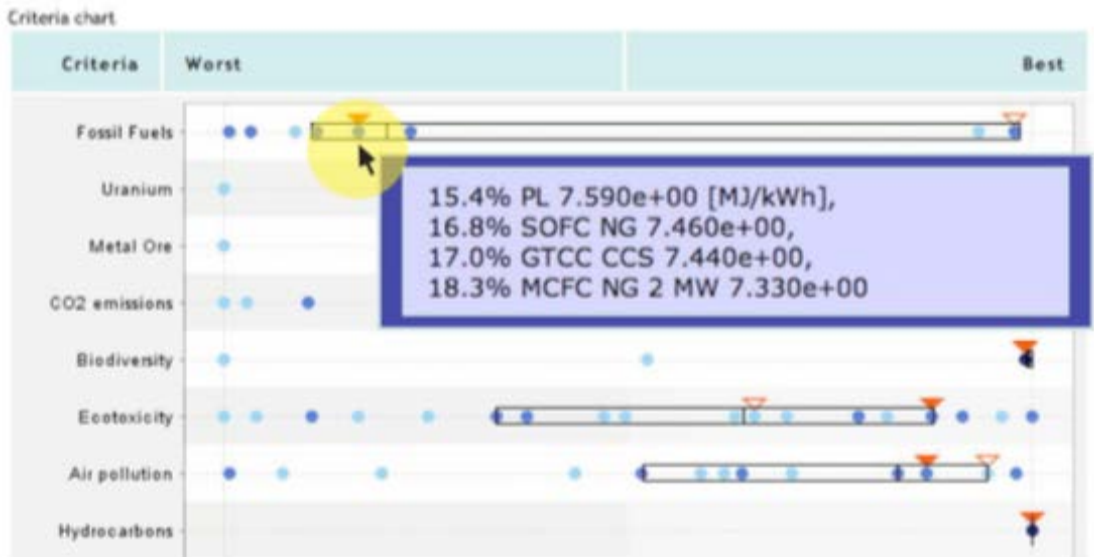

Figure B5 Figure 9 in [19] shows the criteria chart where the various technologies are placed in different locations relative to each other using either calculation or expert judgment. In this case, it is arbitrarily deemed that low fossil fuel use is best, which is a puzzle since coal provides by far the cheapest electricity and economy is one of three main pillars in the overall project design.

## B7    NEEDS MCDA and Baseline costs compared

In Figure 14 of the main text, we compare our EMETS total mean scores based on MCDA and expert judgment with the NEEDS total baseline costs where we had broadly comparable overlap of 7 electricity technologies. Gas CCGT was an outlier but the remaining 6 technologies lay on a line with a positive correlation coefficient $R^2$ = 0.93 (main text Figure14).

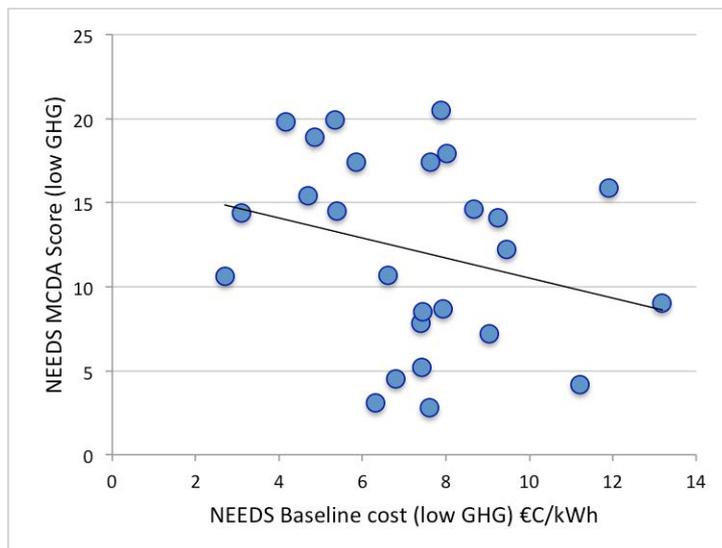

Figure B6 NEEDS total baseline costs compared with NEEDS MCDA results. Based on Table 6 in [19]. The correlation coefficient is <0.1.

Figure B6 compares the NEEDS MCDA with NEEDS Baseline costs, and shows that there is low correlation with R^2 <0.1. However, whatever correlation there is, is negative. The NEEDS MCDA methodology has successfully turned quantitative scientific and economic reality upside-down.

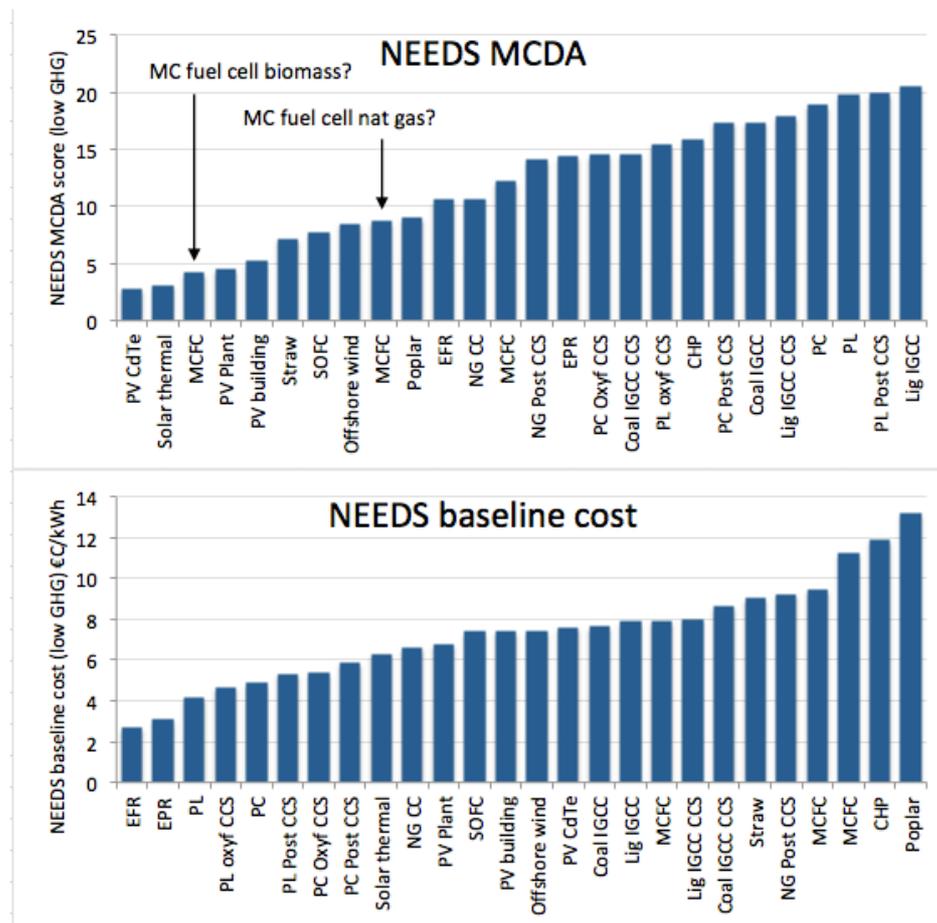

Figure B7 Rank order of NEEDS MCDA and NEEDS baseline cost scores. Based on Table 6 in [19].

## B8     Discussion

In their discussion, the NEEDS Final Report authors [19] say this:

> *The total cost approach has the advantage of being conceptually simple, and produces technology rankings that are unambiguous. Its merits for cost-benefit assessment are undisputable.*

We agree with this summation. But this leaves the question of why NEEDS should overturn the results of a methodology described as unambiguous and undisputable? Their Total Cost approach placed the European Fast Reactor in first place and the European Pressurised Reactor in second place. Unabated pulverized lignite came third and unabated pulverized coal came fifth. Solar thermal came in ninth. Cadmium-Tellurium PV came well down the rankings in fifteenth place. The baseline cost approach placed unabated natural gas CC in tenth place. This lowly placing must surely indicate the assumption of a very high gas price in 2050, influenced by the very high prices at the time when the project was on-going. Very different developments in the last decade underline the danger of extrapolating recent trends. Indeed, this assumption of a very high gas price in 2050 may not come to pass, highlighting the impossibility of trying to predict technology outcomes what was 40 years in the future when the NEEDS work was done.

In contrast to these eminently sensible rankings, the NEEDS MCDA study placed cadmium-tellurium PV (more commonly known as thin film solar) in first place. Cadmium is highly toxic and tellurium is about as rare as platinum. Without vast energy storage or back up from a third party electricity supply, Europe cannot run on indigenous solar power. The cost of adding storage or load following backup has clearly not been factored into the MCDA costs. Placing this technology first in the rank order is questionable to say the least, absent a major breakthrough in energy storage.

Second in the rank order is solar thermal. Unlike roof top solar PV that may be widely distributed across a country reducing the impact of uneven cloud cover on continuity of supply, solar thermal power stations depend on continuous solar energy at a single location. The only place in Europe with commercial solar thermal plant is Spain where it can be easily demonstrated that the power plants do not work when there is local cloud cover [70, 71]. Like solar PV, solar thermal requires either vast amounts of energy storage or conventional thermal backup in order to properly service a grid. Thermal storage does enable solar thermal to deliver some power at night but this is a small fraction of daily production and currently does not come close to providing reliable dispatch in southern Spain [70]. It is relatively safe to say that solar thermal could never function north of a line running through Marseilles and San Marino. This excludes virtually all of France, Germany, Switzerland and Northern Italy. The fact that the NEEDS MCDA study selected this as the second best option to power these countries in 2050 raises serious questions on the veracity of this study.

The EMETS MCDA study placed solar thermal in the middle of the rank order. However, this study was taking a global perspective and recognized that this technology could become useful in hot desert environments that are conspicuously absent through most of Europe. There is perennial talk about Europe drawing power from solar thermal plants located in N Africa (for example see [44]). This simply adds the cost of HVDC transmission over hundreds to thousands of kilometers, introduces the risk of terrorist attacks on power plants and power lines, introduces the risk to energy security by relying on a third party and adds to the energy import bill for a technology that does not work at night.

Third in the NEEDS MCDA rank order is a molten carbonate fuel cell running on either bio-gas or natural gas (in their table 6 [19], MCFC appears three times but the fuel source is not specified). This is largely an experimental technology. Some of the key features are 1) high operating temperature >600˚C, 2) a corrosive internal environment, 3) use of specialized materials such as $LiAlO_2$ ceramic, 4) use of other rare metals such as Ni and Cr, 5) currently small size, NEEDS scored units of 1 or 2 MW. Approximately 3000 such units would be required to substitute for a single EPR. On the positive side, the technology is said to have >60% energy efficiency. Without access to data on costs and durability, it is impossible for us to evaluate this experimental technology.

Fourth, fifth and sixth in the NEEDS MCDA rank order are large-scale solar PV power plant, roof - top solar and straw. Thus 4 of the 6 top ranked technologies are based on solar accompanied by a high-tech experimental fuel cell and agricultural waste.

## B9    Summary

The NEEDS final report recognizes that the merits of the total baseline cost approach are undisputable. And yet, at the outset, the aim of the NEEDS MCDA study appears to have been to overturn the outcome of sound scientific and economic analysis, citing social criteria to justify this approach. Two fundamentally important social criteria for electricity supply are cost and reliability. The latter is not directly addressed in the project, in fact Q25 appears to turn this important aspect on its head in a way that disadvantages nuclear. Of 35 questions, only one deals with cost of generation and this can be eliminated from the pool at the discretion of individual stakeholders.

While the project managers have likely set out with good intentions to access a broad spectrum of participants, they ended up with a majority of academics (62%) that, by the nature European Universities are now organized in the energy sphere, tend to be anti-fossil fuel, anti-nuclear and pro-renewables. Since participants were able to heavily change the weight of their responses until they achieved the desired result, we are concerned that the used methodology has not followed the best practice applicable to surveys to avoid cognitive, psychological as well as social biases. In particular, we are concerned with the introduction of the criterion of "social acceptance", which is taken as a causal independent factor (in the language of statistical regressions), while we argue that it should be a derived dependent factor. In other words, social acceptance should derive from scientific and technical facts that are patiently pedagogically presented to the public. Indeed, social acceptance is not scientifically and objectively derived, it is the results of myriads of factors that have shaped the mindset of the population. As a vast literature on the dynamics of the psychology of social groups attests, one cannot trust social perception to drive a rational decision process. Social perception, like fashions or financial bubbles, may come and go, and are often decoupled from physics, engineering and economics, until eventually reality comes back with a fury.

The way questions were parsed made them impossibly difficult to answer, although in reality participants were only asked to place technologies in a weighted rank order per criterion.

The questions are clearly biased against nuclear power, again this is borne out by the survey results. For example:

*Q8      Uranium: This criterion quantifies the primary energy from uranium resources used to produce 1 kWh of electricity, It includes the total use of uranium for each complete electricity generation technology chain.*[1, 2, 4]

*Q15     Land contamination: This criterion quantifies land contaminated due to accidents releasing radioactive isotopes.* [1, 2]

*Q17     Radioactive waste: This criterion quantifies the volume of medium and high level radioactive wastes stored in underground repositories due to the production of 1 kWh of electricity.* [1, 2]

*Q40     Proliferation: Potential for misuse of technologies or substances present in the nuclear electricity generation technology chain.* [1, 2]

These questions are all blatantly anti-nuclear and the polarity of any response may be dependent upon the outlook of the responder. A pro-nuclear individual may rank nuclear as "best " on Q8 and would then go on to weight Q15, Q17 and Q40 as zero (irrelevant). An anti-nuclear response would answer Q8, Q15, Q17 and Q40 as "worst" and give them all high weight in the preferences panel ensuring that nuclear technologies fair badly. There is no sound balance involved in this process that is simply founded on the social opinion of the stakeholder. The bias is further amplified by the absence of environmental questions that may disadvantage renewable technologies, for example linked to rare earth element mining in China and cobalt mining in the Democratic Republic of Congo.

This study took 54 months to complete and received €7.6 million in support from the European Union. The mechanics of the survey questionnaire empowered the participants to manipulate variables to produce a result consistent with their pre-conceived notions. The project might as well have asked the participants 3 simple questions, 1) are you pro or anti-nuclear, 2) are you pro or anti-fossil fuels, 3) are you pro or anti-renewables. This would have produced a similar outcome and saved everyone a vast amount of time and money.

### Appendix C: Validation of EMETS MCDA scores

Validation of results can be viewed from three perspectives. The first is whether one agrees or not with our selection of 12 criteria to assess the holistic quality of an electricity generation system. There is no definitive answer to this judgment. The second is whether one agrees with the methodology of inviting experts to cast scores and using the sum of the means for each technology as a simple measure of quality. The third is whether or not the judgment of our expert group has any foundation in reality. It is this latter perspective that we seek to evaluate in this Appendix employing three different approaches. First we perform a pairwise comparison (correlation) of criteria MCDA scores where we may expect a correlation to exist, for example between ERoEI and cost of energy. Second, we compare the mean MCDA scores with real data where this exists, for example between MCDA score for cost versus published levelised costs of energy. Third, we compare the mean MCDA scores with the baseline costs as reported by NEEDS [19].

#### C1    Validation of results I. Pairwise comparison of MCDA scores

In this section, we present cross plots for 5 pairs of criteria where we expected to see a correlation and for 1 pair of criteria where no correlation was expected, as a control. The results are summarised below and discussed in greater detail in the following sections.

|   | $R^2$ |
|---|---|
| Cost of energy v ERoEI | 0.84 |
| Cost to grid v benefits to grid | 0.97 |
| Fatalities v Chronic illness | 0.77 |
| Cost of energy v Subsidies | 0.83 |
| $CO_2$ intensity v ERoEI | 0.90 (non-combustion technologies) |
| Fatalities v Resource availability | 0.05 (non-correlating) |

It is clear that the correlation for the 5 correlating pairs is very good. This in itself does not validate results and bestow quantitative meaning but it does show that in aggregate our participants have shown great consistency in estimating scores. Where no correlation was expected between resource availability and fatalities, $R^2$ is effectively zero.

#### C1.1 Cost of energy versus ERoEI

ERoEI is a measure of the efficiency of energy production (see for example [36, 37]) and it is to be expected that efficient systems will produce lower cost electricity than inefficient systems and hence that Cost and ERoEI should be correlated. The best systems have low cost and high ERoEI and occur at the lower left, namely hydroelectric, coal, gas and nuclear power. Note that high ERoEI systems (good) were given low scores. Participants have used their knowledge of costs and ERoEI to apply scores but there must also be a high level of conjecture since the ERoEI of systems like solar thermal and tidal stream are not actually known with sufficient reliability. Herein lies the strength of the MCDA approach that allows values to be allocated to technologies where actual data does not exist.

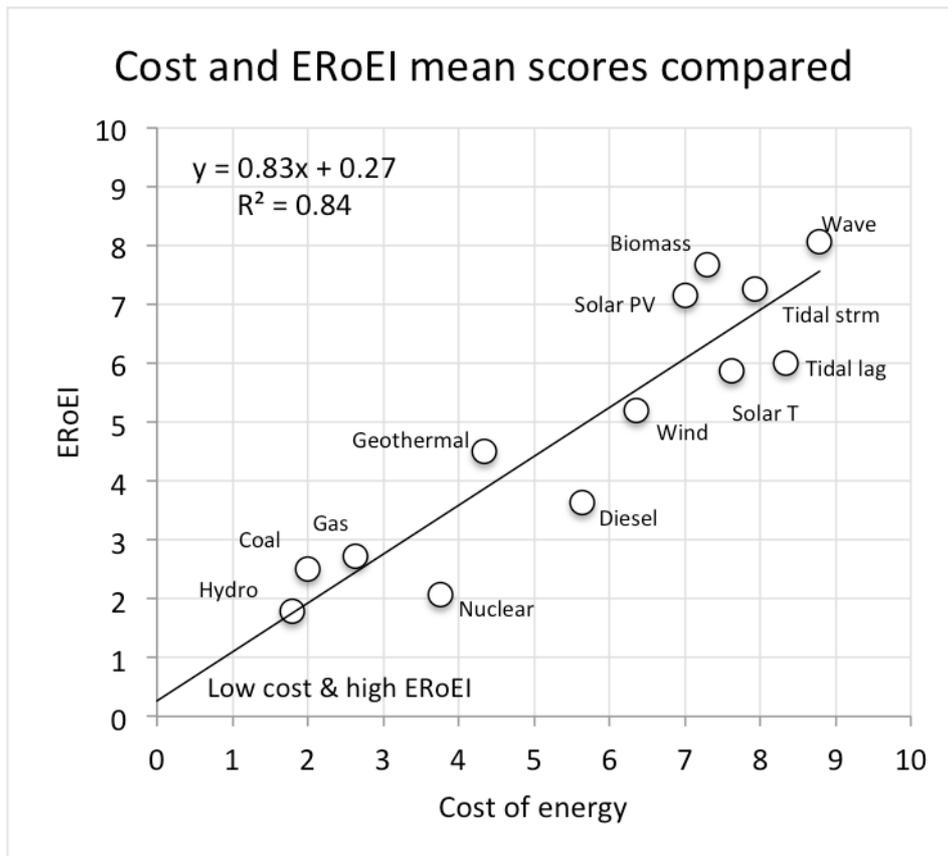

Figure C1 Comparison of the MCDA scores for cost of energy versus ERoEI

### C1.2 Cost versus benefit to grid

As discussed in the main text, certain electricity technologies are provide balancing, load following and backup to maintain grid stability, while other technologies are polluting the grid with noise and instabilities. These are opposing sides of the electricity dispatch coin. Participants have shown a very high level of consistency in applying grid scores with R2=0.97, a gradient of 0.93 and an intercept close to the origin. Two clear groups emerge, the virtuous dispatchable systems at bottom left and the bad intermittent systems at top right. The systems providing greatest benefit to the grid by providing load following, load balancing and backup are gas, hydro, diesel and coal. Nuclear, geothermal and biomass perform less well since they normally run in continuous base load mode (note that biomass stations could load-follow but, in the UK, opt to operate in base-load mode [46]). Base-load is still a service to the grid but less valuable than load following. There is also some skill in the way scores have been applied to the intermittent group. Solar thermal is better than solar PV because it is normally installed at lower latitude in desert environments where sunshine is more continuous (less cloud) and seasonal variability is less pronounced. It also has the capacity to provide some nighttime dispatch from thermal storage. Tidal stream is also less intermittent than tidal lagoon since the former is close to continuous but with fluctuating output while the latter as configured for the now stalled Swansea Bay scheme [23] was planned for 4 short bursts of generation per day.

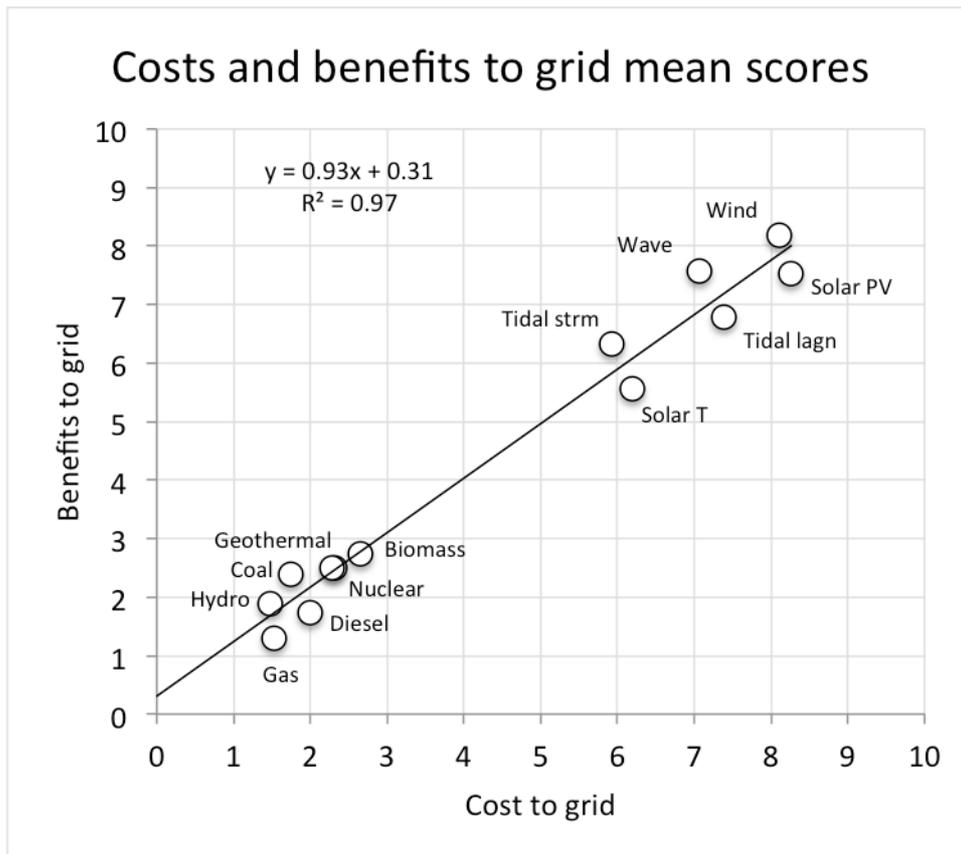

Figure C2 Comparison of MCDA scores for costs versus benefits to grid.

### C1.3 Fatalities versus chronic illness

In as far as chronic illness can lead to death, a link is expected between fatalities and chronic illness. A dangerous working environment like a coal mine that may lead to chronic illness may also be associated with higher fatal accident rates.

The way participants cast scores has led to three distinct groups. At the extreme lower left, solar thermal, nuclear power and geothermal are assessed to be the safest form of power production. To the right of this group, viewed as slightly more hazardous from the fatalities perspective lies hydroelectric power, solar PV, wave, tidal lagoon and stream, wind and gas. There is very little differentiation within this group.

To the top right are a group of three technologies that are assessed to pose significantly higher risks to health and life, namely coal, biomass and diesel. This stems from loss of life obtaining solid fuel in the case of coal and biomass and health risks stemming from combustion – particulates, NOX etc. combined with the production of ash. Notably, natural gas, also a combustion technology, is assessed to be significantly better than coal, biomass and diesel. That is because there is less risk associated with acquiring gas and its combustion is clean, producing only $CO_2$ and $H_2O$.

Coal came in a respectable 6th place in our assessment, scoring well on criteria like cost, grid, resource availability and ERoEI. It is an interesting question whether the poor performance on Health and Environmental criteria should be allowed to over-rule the perceived benefits. This is a decision not easily made by MCDA – weighted or unweighted. However, those calling for coal to be phased out globally should consider the benefits that need to be weighed against the costs.

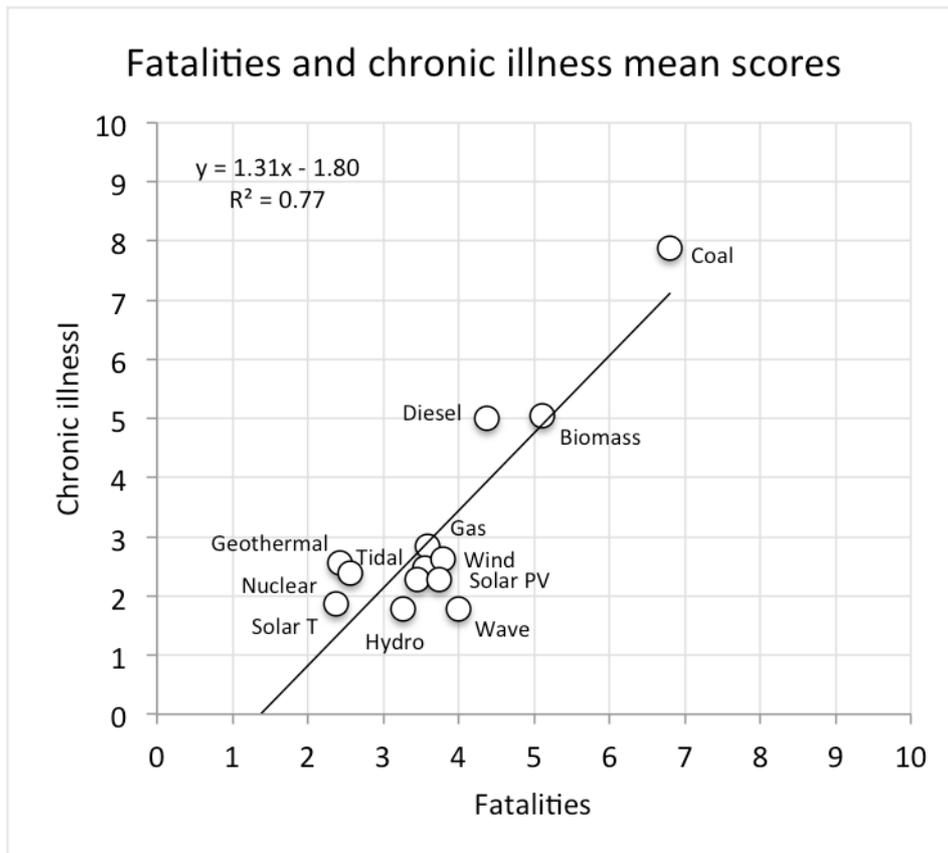

Figure C3 Comparison of MCDA scores for fatalities versus chronic illness.

As a general observation, with most technologies clustering towards the lower left, our expert group has judged that the majority of technologies do not represent significant health risks.

### C1.4 Cost of energy versus subsidies paid

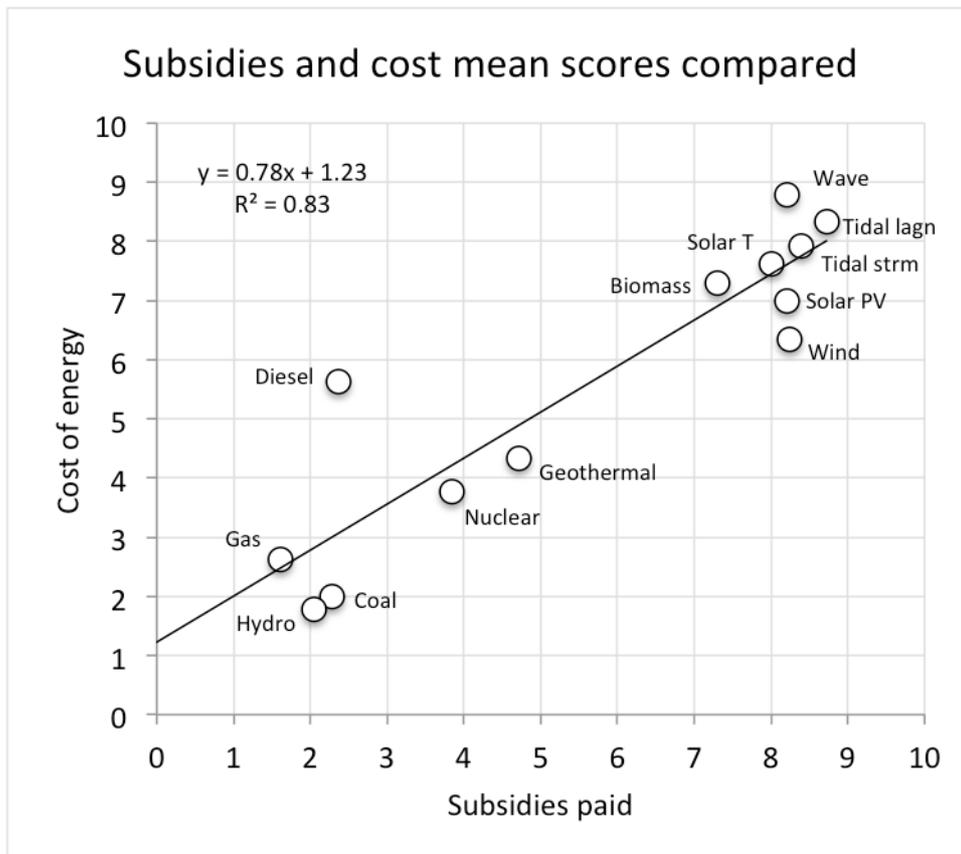

Figure C4 Comparison of MCDA scores for cost of energy versus subsidies paid.

On these criteria, there are three groups and an outlier (diesel). At the lower left are hydroelectric power, coal and natural gas assessed as cheap and unsubsidized. Moving up the regression line is nuclear power, viewed as more expensive and receiving subsidies. This stems from the view that national governments may subsidize the husbandry of nuclear waste and the financing of Hinkley C nuclear power station in the UK was done via a generous contract for difference (CfD), a mechanism now also used to subsidize renewable projects. The group to the top right, comprises all of the new intermittent renewables and are assessed to be expensive and heavily subsidized. Participants were asked to score system cost that builds in the cost of mitigating for intermittency.

Diesel is an interesting outlier assessed similar to hydro, coal and gas on subsidies (i.e. unsubsidized) but much more expensive. The answer here probably lies in the peaking-plant market where diesel generators are deployed. This niche market can command high price, which is not the same as high cost. It is possible our expert group have got this judgment wrong.

### C1.5 CO2 intensity versus ERoEI

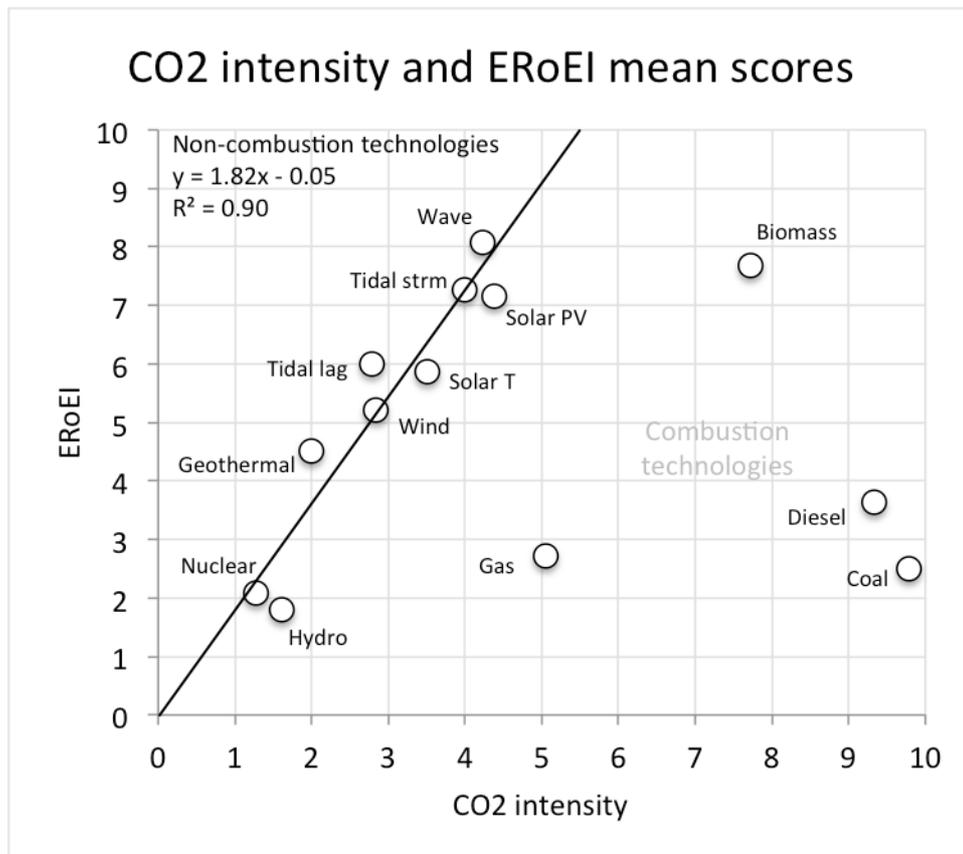

Figure C5 Comparison of MCDA scores for $CO_2$ intensity versus ERoEI. Note that a low ERoEI score equates with a high ERoEI and vice versa.

Figure C5 of $CO_2$ versus ERoEI shows two domains, one for the combustion technologies and one for the non-combustion technologies. The latter show a good correlation with $R^2=0.90$. The $CO_2$ intensity of non-combustion technologies is associated with the embedded emissions produced from materials mining, manufacture, maintenance and decommissioning. Surprisingly, our expert group view solar PV as almost $CO_2$ intense as natural gas. When one factors in emissions associated with providing back up and load following services with the low ERoEI, it is possible that this could be correct.

Biomass (wood chips) is an interesting outlier scoring badly on both $CO_2$ intensity and ERoEI. Biomass burners in the UK do receive large subsidies reported to run at over £1 million / day at the Drax plant in England [64]. The high $CO_2$ intensity reflects two opinions. The first is burning wood does produce $CO_2$ that will not be gone for many decades. The second is the large amount of embedded fossil fuels involved in harvesting, drying, grinding, chipping and transporting wood from N America to Europe [46].

The $CO_2$ intensity scores for the combustion technologies make a lot of sense. It is well established that the CO2 intensity of Gas combined cycle is about half that of coal. Diesel is viewed as more CO2 intense than gas because it has a higher proportion of C-C bonds and because the diesel engine is viewed as less efficient compared with the combined cycle gas turbine.

### C1.6 Fatalities versus resource availability

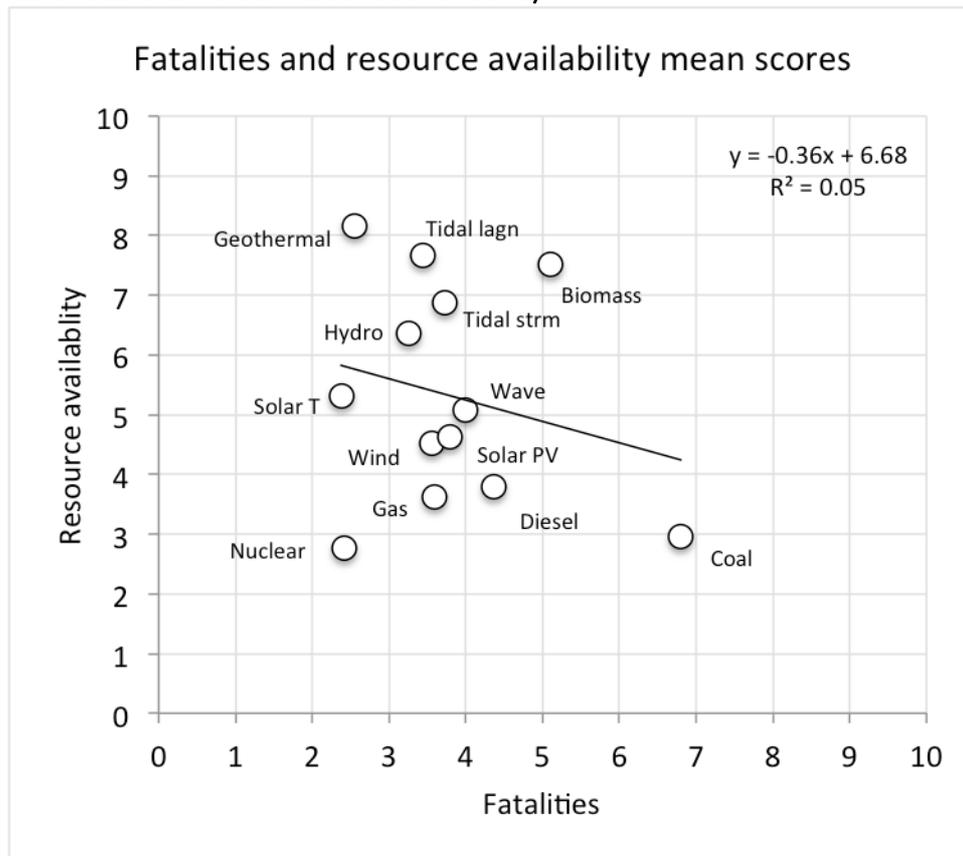

Figure C6 Comparison of MCDA scores for fatalities versus resource availability.

On the back of very good correlations for the five preceding pairs of criteria, we wanted to check this was not down to some systematic bias in our approach. And so here we cross plot two criteria where no correlation was to be expected. We can think of no reason why fatalities should correlate with resource availability. With R2=0.05, there is no correlation and we conclude that the fine correlations described above is down to the skill of our expert group who found correlations where they are expected to exist.

### C2      Validation of results II. Correlations between MCDA scores and real data

True validation of our MCDA poll is only possible through comparison with real data. This desirable approach is hampered in two ways. First, real data for our criteria are in themselves estimates and, second, real data are in short supply or non-existent and frequently only partial. For example, for ERoEI, we can only find reliable data for 6 of our 11 criteria (partial) and, for external environmental costs, real data is not available at all. It is a fundamentally interesting question to ask if MCDA scores can be used as a proxy in the absence of real data?

Regressions of real data versus MCDA scores for four pairs of criteria are summarized below:

$R^2$

| | |
|---|---|
| Cost of electricity versus Levelised Cost of Electricity [47] | 0.90 |
| Fatalities versus Deaths per TWh [48] | 0.64 |
| Fatalities versus Deaths per TWh [49] | 0.74 |
| Chronic illness versus serious illness [48] | 0.80 |
| ERoEI versus buffered ERoEI [50] | 0.92 |

Where we have managed to compare the MCDA with real data, we have a decent level of correspondence, albeit where certain outliers are omitted in some cases, sufficient for us to suggest that our MCDA scores are rooted in the real world and may perhaps be used in a semi-quantitative way.

### C2.1 Cost of electricity

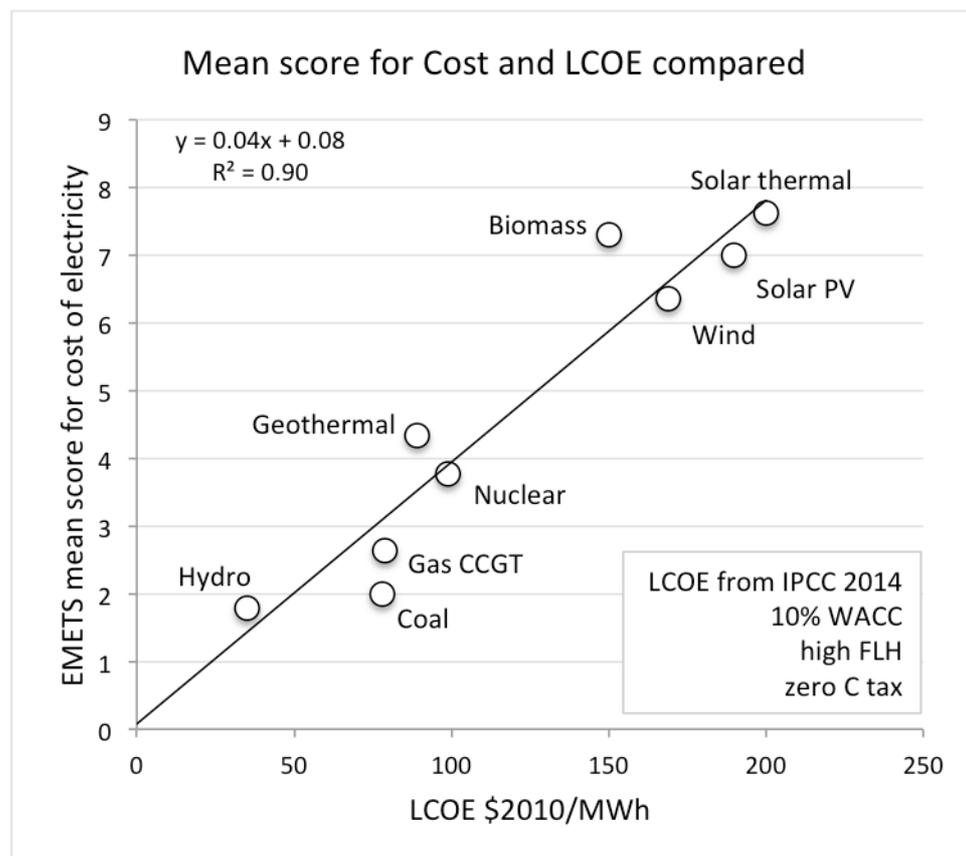

Figure C7 Comparison of LCOE [47] and the MCDA scores for cost of Electricity.

In comparing EMETS MCDA with quantitative data, the comparison between Cost of Energy and LCOE provides the best and most convincing correlation. What this tells us is that our expert group had good awareness of costs in advance and possesses skill at translating that awareness into an accurate score.

### C2.2 Fatalities

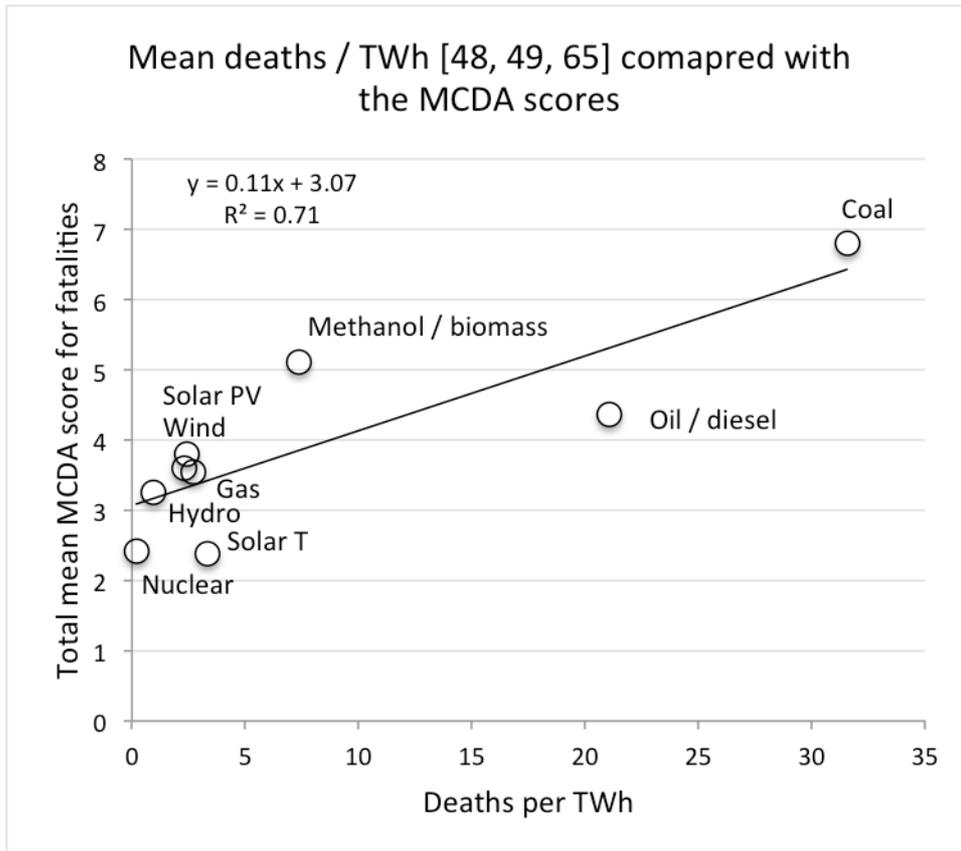

Figure C8 Fatalities expressed as deaths per TWh for three different sources [64, 48, 49] versus MCDA mean scores.

We had difficulty finding reliable and reproducible data on fatalities and electrical power production. For example, the three sources we found yielded 10, 25 and 60 deaths / TWh for coal [64, 48 and 49 respectively]. Furthermore, there is not always exact equivalence in technology, for example [64] reports data for methanol and not biomass. We have tackled this problem by simply averaging the data from the three sources that produces a reasonable correlation with $R^2 = 0.71$ (Figure C8). Our expert group has more or less got the rank order correct but have failed to capture the dynamic range of fatalities properly, assigning scores that were perhaps too high for the low mortality cluster.

### C2.3 Chronic Illness

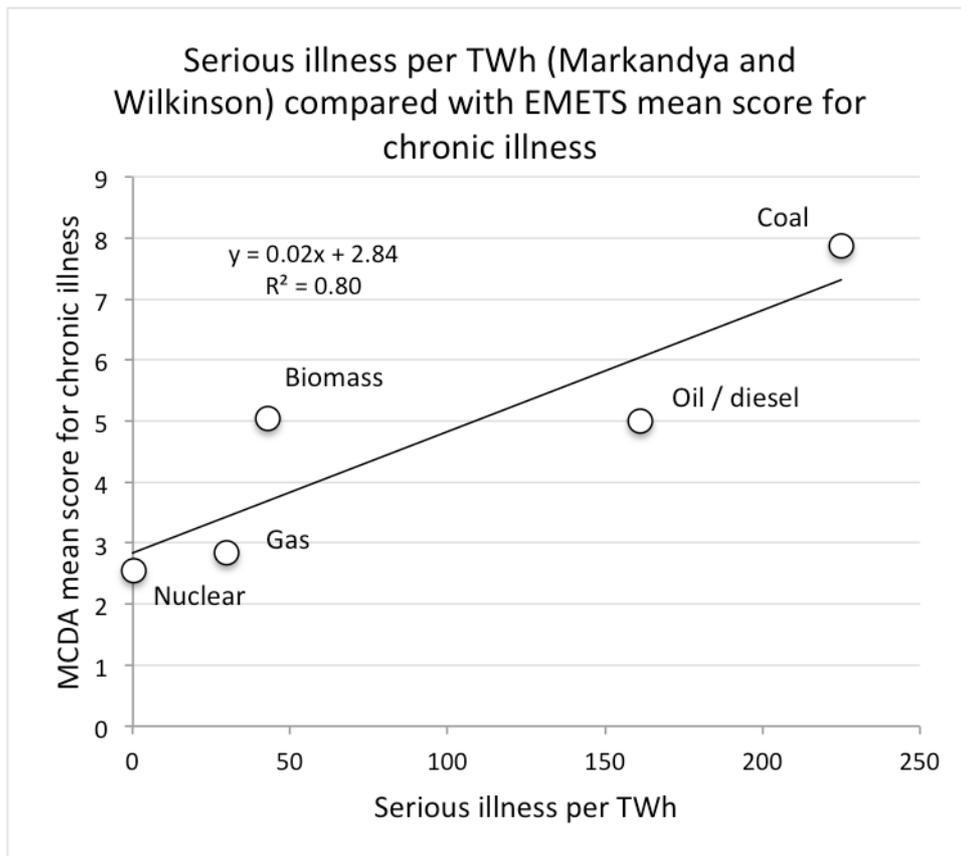

Figure C9 MCDA chronic illness mean scores compared with incidence of serious illness per TWh as reported by [48].

For technologies common to our MCDA survey and data reported by [48], we find the rank order for the MCDA scores and the measured data to be virtually the same and R2 = 0.8. This is another satisfactory result validating our approach.

C2.4 ERoEI

The correlation with ERoEI is negative because high ERoEI = good and is consequently given a low score in our scheme and vice versa. We plot our MCDA scores against buffered ERoEI reported by [50], as shown on their Figure 3. [50] report data for 8 technologies, all of which are in our sample set. Only seven are plotted here since nuclear was a major outlier with reported ERoEI of 75 that is well off the scale of our chart (see discussion below). There are many studies that report data for ERoEI, we selected [50] since it reports "buffered" data that means the ERoEI has been adjusted to include the energy cost of intermittency. This is equivalent to the way our scores were cast. The correlation coefficient of 0.91 once again provides good validation for our approach.

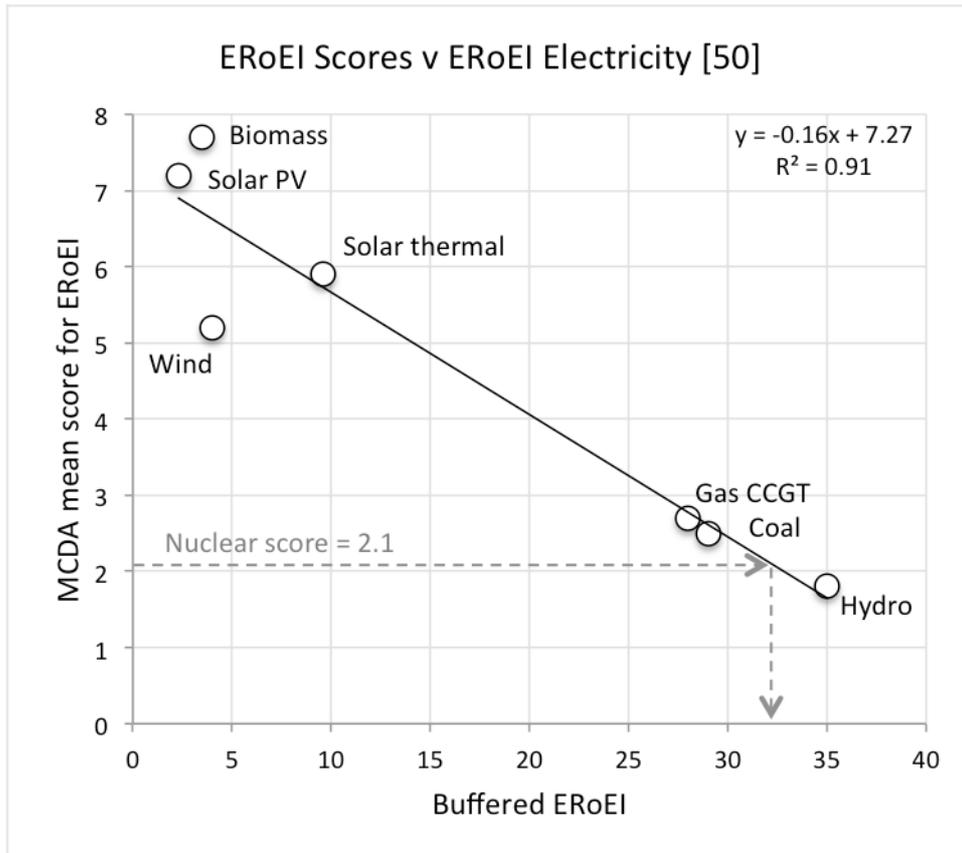

Figure C10 Correlation between buffered ERoEI [50] and MCDA mean scores for ERoEI.

The ERoEI of nuclear power is a controversial and highly uncertain issue. [50] report a value of 75 that is high compared to other sources. For example, [37] report a range of 5 to 15, at the other end of the extreme spectrum. Our expert group mean score for nuclear was 2.1 and using the relationship shown in Figure C10 as a calibration, this translates to an ERoEI ~ 32. A recent paper [66] report an ERoEI for nuclear of 32.

One issue with the value reported in [50] is that it is based on assumptions that include 8000 full load hours / year equal to a capacity factor of 91% and a 60 year operational life. This may be relevant to future Gen III nuclear but not to today's Gen II power stations. Assuming 80% capacity and 40 year life reduces the ERoEI reported by [50] to a value of 44. This is within range of our distribution and would be more representative of Gen II nuclear power stations.

### C2.5 $CO_2$ intensity

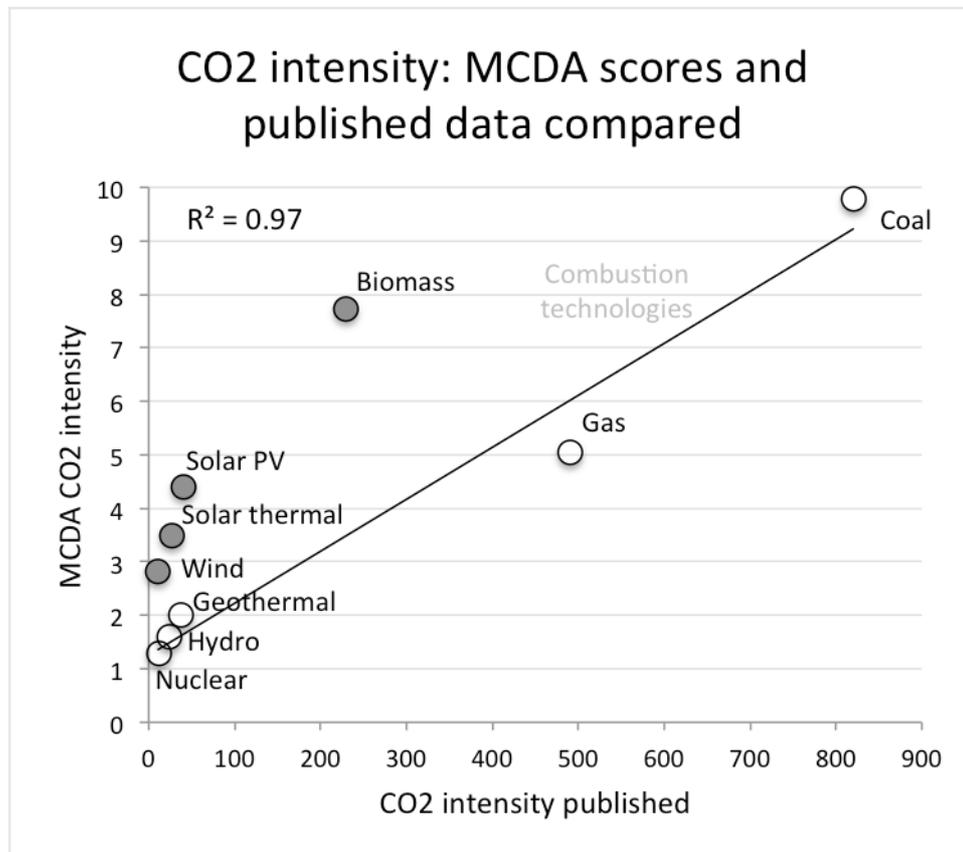

Figure C11 MCDA scores for CO2 intensity compared with published data [47].

There is no straightforward correlation between the MCDA CO2 intensity scores and CO2 intensities published by the IPCC [47] (Figure C11). The 5 "traditional" technologies – nuclear, hydro, geothermal, gas and coal, have an excellent correlation with $R^2$=0.97. The new technologies – wind, solar thermal, solar PV and biomass, however, have poor to no correlation. The IPCC [47] do make allowance for emissions embedded in the devices.

According to the MCDA scores, the IPCC [47] does not recognize all emissions associated with new renewables. A likely explanation lies in the fact that our expert group allocated system scores as opposed to isolated device scores. The system scores include emissions associated with the provision of balancing and backup services that, in Europe, are dominated by burning natural gas. Having renewables on the grid does reduce the amount of gas burned, as reflected by the MCDA scores, but it does not virtually eliminate it as suggested by the IPCC [47] data.

C3. Validation of results III: Comparison between EMETS and NEEDS

In the period 2004 to 2009, the EU majority funded (€7.6 of 11.7 million) a multi-institutional project called New Energy Externalities Development for Sustainability (NEEDS) with the aim of evaluating a number or electricity generation options for the future [19, 20] (see Appendix B). The project had the specific aims of informing EU energy policy based on sustainability criteria in the future year of 2050. The project involved up to 67 institutions, mainly from Germany, Switzerland, France and Italy and was managed by the Paul Scherrer Institute in Switzerland.

On the face of it, the NEEDS project was much larger and more complex than the EMETS. It certainly had a much larger budget, over 2 orders of magnitude larger. The NEEDS project was in two parts where 26 electricity technologies were evaluated using 1) conventional life cycle baseline cost analysis, including external costs, and 2) an elaborate on-line MCDA survey employing 36 Criteria where 259 "stakeholders" participated.

We wanted to compare the MCDA scores from our study with the NEEDS baseline costs. The 26 technologies selected by NEEDS for analysis include many esoteric fossil fuel based systems and excludes common systems like hydroelectric power and geothermal. It was considered unlikely that new hydroelectric or geothermal plant would be installed in Europe in the coming decades. Nevertheless, there is an area of overlap with 7 technologies common to both studies – nuclear power, coal, combined cycle gas, biomass, solar PV, solar thermal and wind - (Figure C12) and while in detail EMETS and NEEDS are measuring different things, we did expect to find a degree of coherency between the two sets of results for these seven overlapping technologies.

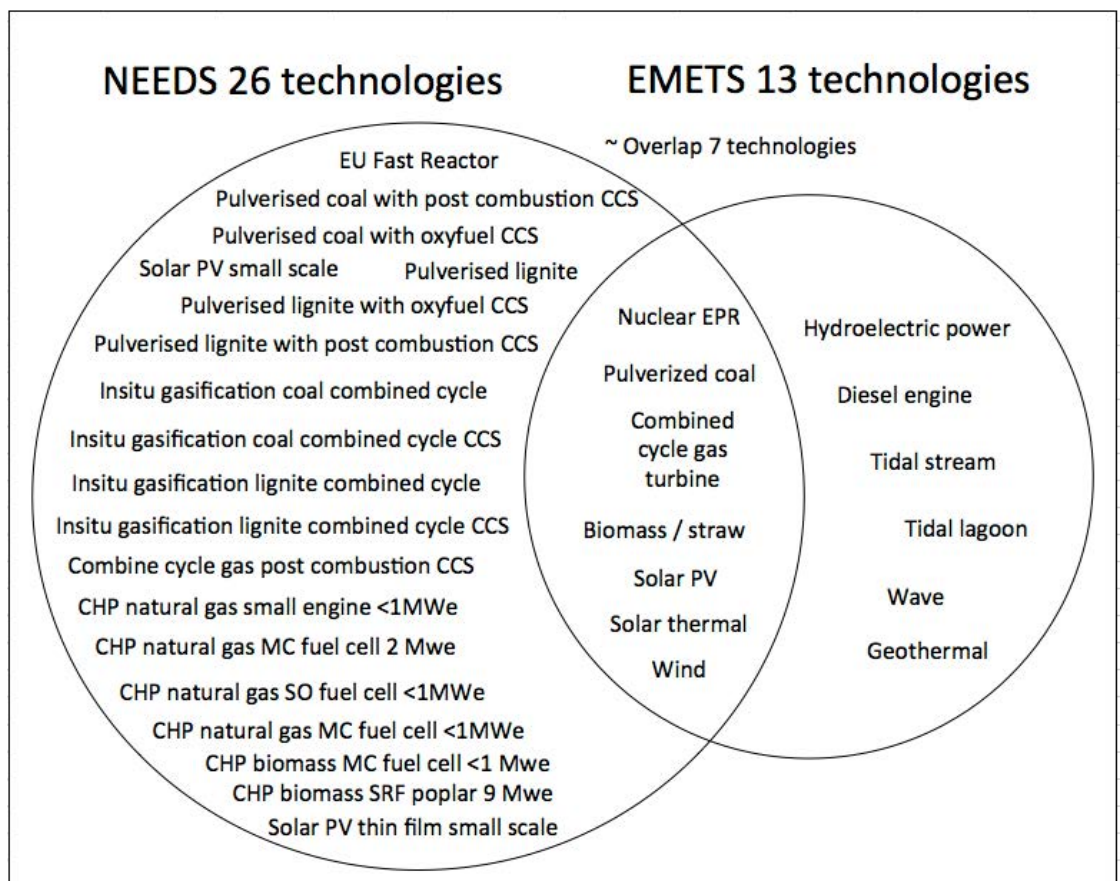

Figure C12 Venn diagram showing the 26 NEEDS and 13 EMETS technologies and the overlapping 7 technologies where in some instances, for example biomass, the overlap is approximate.

In section C3.1, we compare the EMETS MCDA with the NEEDS baseline costs. The comparison between EMETS MCDA and NEEDS MCDA are dealt with in Appendix B.

C3.1 Comparison of EMETS MCDA with NEEDS baseline costs

For the seven overlapping technologies, the EMETS total mean scores are cross-plotted against NEEDS baseline cost as taken from Table 6 in [19]. As expected, we find a reasonable correlation ($R^2$ = 0.57) for all seven technologies. Gas is an outlier and, excluding it, we find an excellent correlation with $R^2$ = 0.93. The EMETS participants ranked gas similar to nuclear power (total mean score = 34.6 v 32.6 respectively) while the NEEDS baseline costs rank gas among one of the most expensive technologies in 2050. At the time the NEEDS survey was conducted (2004-2009), oil prices were going up to historical high levels, giving rise to perceptions of global scarcity of fossil fuel [27, 28, 29]. The shale oil and gas revolution that has subsequently taken place has transformed this perception [32].

The coherency between the EMETS MCDA scores and NEEDS baseline costs tends to provide mutual confirmation that both methods are measuring similar things and supports the veracity of both approaches even though EMETS is rooted in the present day and NEEDS is rooted in 2050.

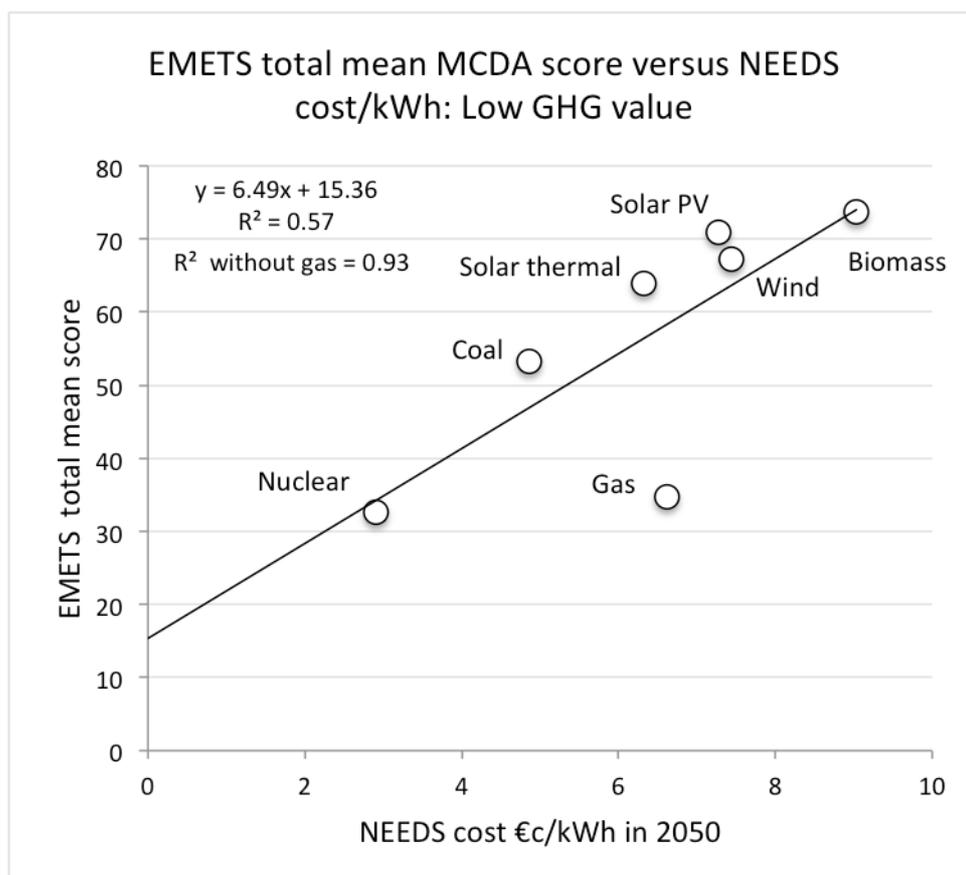

Figure C13 NEEDS baseline costs compared with EMETS total mean scores for 7 overlapping technologies.

C4. Summary and conclusions

In this Appendix, we set out to validate our MCDA scores using three different approaches:

1. Pairwise correlations between scores for various criteria
2. Correlations between scores and real data
3. Comparison between EMETS MCDA and NEEDS baseline costs

In some instances, comparison was hampered by a lack of exact equivalence between data available on different technologies and what we tried to measure, but in general we find excellent correlations where correlation is expected to exist.

### C4.1 Pairwise correlations between scores for various criteria

|   |   | $R^2$ |
|---|---|---|
| 1. | Cost of energy v ERoEI | 0.84 |
| 2. | Cost to grid v benefits to grid | 0.97 |
| 3. | Fatalities v Chronic illness | 0.77 |
| 4. | Cost of energy v Subsidies | 0.83 |
| 5. | $CO_2$ intensity v ERoEI | 0.90 (non-combustion technologies) |
| 6. | Fatalities v Resource availability | 0.05 (non-correlating) |

For the five pairs of criteria that are expected to correlate (1 to 5), the correlation coefficients are exceptionally high (0.77 to 0.97). This demonstrates that in aggregate the judgment of our expert group has been highly consistent. In the one instance where we would expect no correlation (between fatalities and resource availability), the correlation coefficient is 0.05.

### C4.2 Correlations between scores and real data

The second way we validate our results is to compare our mean scores with quantitative data where this is available for five of our criteria – cost, deaths, illness, ERoEI and $CO_2$ intensity.

|   | $R^2$ |
|---|---|
| Cost of electricity versus Levelised Cost of Electricity [47] | 0.90 |
| Fatalities versus Deaths per TWh [48, 49, 64] | 0.71 |
| Chronic illness versus serious illness [48] | 0.80 |
| ERoEI versus buffered ERoEI [50] | 0.92 |
| $CO_2$ intensity versus $CO_2$ intensity for 5 technologies [47] | 0.97 |

For cost, illness and EROEI, we find excellent correlations of 0.9, 0.8 and 0.92 respectively. We had difficulty finding reliable and reproducible data on fatalities and electrical power production. For example, the three sources we found yielded 10, 25 and 60 deaths / TWh for coal [64, 48 and 49 respectively]. Furthermore, there is not always exact equivalence in technology, for example [64] reports data for methanol and not biomass. We have tackled this problem by simply averaging the data from the three sources that produces a reasonable correlation with $R^2 = 0.71$. The high quality of this agreement surprised us and suggests that, given a group of diverse but true experts, expert judgment combined with MCDA may provide a near quantitative measure. This provides us with some confidence in the veracity of scores for criteria where it is less easy to find quantitative data at the global scale for, for example, external environmental costs, costs and benefits to grid.

### C4.3 Comparison between EMETS MCDA and NEEDS baseline costs

The 26 technologies selected by NEEDS for analysis include many esoteric fossil fuel based systems and excludes common systems like hydroelectric power and geothermal (Figure C12). Nevertheless, there is an area of near overlap with 7 technologies common to both studies – nuclear power, coal, combined cycle gas, biomass, solar PV, solar thermal and wind - (Figure C12) and while in detail the EMETS and NEEDS are measuring different things (EMETS was measuring holistic quality today while NEEDS was measuring projected sustainability in 2050), we did expect to find a degree of coherency between the two sets of results for these seven overlapping technologies. In this appendix, we focused on comparing our results with the NEEDS baseline costs.

| | $R^2$ |
|---|---|
| EMETS MCDA versus NEEDS baseline costs (6 of 7 technologies) | 0.93 |

Natural gas was an outlier, probably because NEEDS assumed a high gas price in 2050. Excluding gas, we find an excellent correlation between the NEEDS costs and EMETS MCDA scores.

### C4.4 Conclusion

There is a high degree of consistency between the EMETS MCDA scores for energy technologies based on the mean judgment of an expert group and other measures where we expect to find a correlation. In particular, we have good to excellent correlations between the MCDA scores and real world data. This provides us with some confidence that the MCDA scores may be used as proxies for energy quality where data is lacking, for example external environmental costs and footprint. Or where data is lacking for a technology, for example the ERoEI of tidal lagoon or tidal stream. The MCDA scores may be used to guide energy policy.